\documentclass[twocolumn, times]{aastex61}
\usepackage{amsmath}

\shorttitle{ISM Evolution in Post-Starburst Galaxies}
\shortauthors{Li et al.}

\begin{document}
\title{The Evolution of the Interstellar Medium in Post-Starburst Galaxies}

\correspondingauthor{Zhihui Li}
\email{zhihui@caltech.edu}

\author{Zhihui Li}
\affiliation{Cahill Center for Astrophysics, California Institute of Technology, 1200 East California Boulevard, Pasadena, CA 91125, USA}
\affiliation{Kavli Institute for Astronomy and Astrophysics, Peking University, Beijing 100871, China}
\affiliation{Department of Astronomy, School of Physics, Peking University, Beijing 100871, China}
\affiliation{Steward Observatory, University of Arizona, 933 North Cherry Avenue, Tucson AZ 85721}

\author{K. Decker French}
\affil{Steward Observatory, University of Arizona, 933 North Cherry Avenue, Tucson AZ 85721}
\affil{Observatories of the Carnegie Institution for Science, 813 Santa Barbara Street, Pasadena CA 91101}

\author{Ann I. Zabludoff}
\affiliation{Steward Observatory, University of Arizona, 933 North Cherry Avenue, Tucson AZ 85721}

\author{Luis C. Ho}
\affiliation{Kavli Institute for Astronomy and Astrophysics, Peking University, Beijing 100871, China}
\affiliation{Department of Astronomy, School of Physics, Peking University, Beijing 100871, China}

\begin{abstract}
We derive dust masses (\emph{M}$_{\rm dust}$) from the spectral energy distributions of 58 post-starburst galaxies (PSBs). There is an anticorrelation between specific dust mass (\emph{M}$_{\rm dust}$/\emph{M}$_{\star}$) and the time elapsed since the starburst ended, indicating that dust was either destroyed, expelled, or rendered undetectable over the $\sim$1 Gyr after the burst. The \emph{M}$_{\rm dust}$/\emph{M}$_{\star}$ depletion timescale, 205$^{+58}_{-37}$ Myr, is consistent with that of the CO-traced \emph{M}$_{\rm H_2}$/\emph{M}$_{\star}$, suggesting that dust and gas are altered via the same process. Extrapolating these trends leads to the \emph{M}$_{\rm dust}$/\emph{M}$_{\star}$ and \emph{M}$_{\rm H_2}$/\emph{M}$_{\star}$ values of early-type galaxies (ETGs) within 1-2 Gyr, a timescale consistent with the evolution of other PSB properties into ETGs. Comparing \emph{M}$_{\rm dust}$ and \emph{M}$_{\rm H_2}$ for PSBs yields a calibration, log $M_{\rm H_2}$ = 0.45 log $M_{\rm dust}$ + 6.02, that allows us to place 33 PSBs on the Kennicutt-Schmidt (KS) plane, $\Sigma \rm SFR-\Sigma M_{\rm H_2}$. Over the first $\sim$200-300 Myr, the PSBs evolve down and off of the KS relation, as their star formation rate (SFR) decreases more rapidly than \emph{M}$_{\rm H_2}$. Afterwards, $M_{\rm H_2}$ continues to decline whereas the SFR levels off. These trends suggest that the star-formation efficiency bottoms out at 10$^{-11}\ \rm yr^{-1}$ and will rise to ETG levels within 0.5-1.1 Gyr afterwards. The SFR decline after the burst is likely due to the absence of gas denser than the CO-traced H$_2$. The mechanism of the \emph{M}$_{\rm dust}$/\emph{M}$_{\star}$ and \emph{M}$_{\rm H_2}$/\emph{M}$_{\star}$ decline, whose timescale suggests active galactic nucleus (AGN) or low-ionization nuclear emission-line region (LINER) feedback, may also be preventing the large CO-traced molecular gas reservoirs from collapsing and forming denser star forming clouds.
\end{abstract}


\section{Introduction} \label{sec:intro}
Lying in the `green valley' \citep{2012MNRAS.420.1684W} of the galaxy color-magnitude diagram, post-starburst (PSB) galaxies \citep{1983ApJ...270....7D} is a transitioning phase between star-forming spirals and gas-poor quiescent galaxies (\citealt{2004ApJ...607..258Y, 2008ApJ...688..945Y}, and references therein). The absence of significant nebular emission lines (e.g., [O II], H$\alpha$) is indicative of little-to-no ongoing star formation. However, the presence of strong Balmer absorption reveals young and recently formed A-stars \citep{1983ApJ...270....7D, 1987MNRAS.229..423C}. These signatures indicate a recent starburst within the last $\sim$ Gyr. Although PSBs are a rare species at almost all redshifts ($<$1\% by \emph{z} $\sim$ 0.5, \citealt{2016MNRAS.463..832W}), the shortness of the PSB phase suggests that a large fraction (25-40\%) of field galaxies at \emph{z} $<$ 1 may have passed through it (\citealt{1996ApJ...466..104Z, 2004ApJ...609..683T, 2011ApJ...741...77S}). Thus, PSBs are important to understanding the evolutionary path from star-forming galaxies to early-type galaxies (ETGs).

Over the past decade, many attempts have been made to study the interstellar medium (ISM) properties of PSBs. The existence of atomic gas (HI) has been confirmed in several small samples of PSBs (e.g., \citealt{2001AJ....121.1965C, 2006ApJ...649..163B, 2013MNRAS.432..492Z}). Recent work has also revealed the existence of large molecular gas (H$_2$) reservoirs in PSBs (e.g., \citealt{2015ApJ...801....1F, 2015MNRAS.448..258R}). \citet{2018ApJ...862....2F} even discovered that the CO-traced H$_2$ declines with post-burst age over a timescale that would lead to ETG levels in 0.7-1.5 Gyr. However, as obtaining gas masses requires large amounts of radio telescope time, these studies are limited to small sample sizes. Alternatively, dust mass (\emph{M}$_{\rm dust}$) can be used to track the ISM evolution for a larger sample of PSBs, as it is more easily measured, i.e., by fitting the galaxy's spectral energy distributions (SED) over mid-infrared (MIR) to far-infrared (FIR) wavelengths. Archival data are now available for a statistically significant PSB sample, making it possible to calibrate the relation between \emph{M}$_{\rm dust}$ and CO-traced \emph{M}$_{\rm H_{2}}$ for the first time, as well as to examine the evolution of \emph{M}$_{\rm dust}$ over a wide range of post-burst ages.

\citet{2018ApJ...855...51S} derived \emph{M}$_{\rm dust}$ for 33 PSBs with CO detections from \citet{2015ApJ...801....1F} and investigated the evolution of their ISM properties in detail. In contrast, here we search all available archival IR data for three large PSB samples from \citet{2018ApJ...862....2F}, \citet{2016ApJS..224...38A}, and \citet{2015MNRAS.448..258R} and derive \emph{M}$_{\rm dust}$ for those 58 PSBs with sufficient IR data. Thus, we study the evolution of \emph{M}$_{\rm dust}$ with a larger sample size and a wider age baseline.

As \citet{1998ApJ...498..541K} points out, there is a universal correlation between the surface density of gas and star formation rate (SFR) for local normal star-forming galaxies and starburst galaxies (the Kennicutt-Schmidt relation, or the KS relation). For a sample of PSBs, \citet{2015ApJ...801....1F} observed a significant offset from the KS relation. But what is the evolutionary track for PSBs in the KS plane? The tight correlation between gas and dust that we observe here makes it possible for us to map this evolution for the first time and to connect it to changes in the star formation efficiency (SFE).

In this paper, we derive \emph{M}$_{\rm dust}$ for 58 PSBs by performing ultraviolet (UV) to FIR SED fitting. We study the evolution of \emph{M}$_{\rm dust}$ and SFE after the burst ends. We also investigate the dust-derived KS relation. In Section \ref{sec:sample}, we summarize our PSB sample selection criteria. In Section \ref{sec:data}, we describe the archival fluxes and errors used to construct full SED of our sample. In Section \ref{sec:fitting}, we discuss the CIGALE-based (Code Investigating GALaxy Emission; \citealt{2009A&A...507.1793N, 2019A&A...622A.103B}) SED fitting procedure and present the results. In Section \ref{sec:results}, we consider the evolution of \emph{M}$_{\rm dust}$, the position on the KS plane, and the SFE of our PSB sample. Section \ref{sec:summary} lists our conclusions. Throughout this paper we adopt a flat $\Lambda$CDM cosmology with $\Omega_{m}$ = 0.308, $\Omega_{\Lambda}$ = 0.692, and $H_{0}$ = 67.8 km s$^{-1}$ Mpc$^{-1}$ \citep{2016A&A...594A..13P}.

\section{Sample Selection} \label{sec:sample}
In this work, we combine three well-studied PSB samples from \citet{2018ApJ...862....2F}, \citet{2016ApJS..224...38A}, and \citet{2015MNRAS.448..258R} to make the time baseline since the starburst ends as wide as possible. Our combined PSB sample ranges in post-burst age from $\sim$-100 to $\sim$800 Myr, which enables us to sample any significant trends. The general idea of constructing PSB samples is requiring strong Balmer absorption lines (suggesting recent starbursts), and weak nebular emission (indicating little ongoing star formation). Specifically, \citet{2018ApJ...862....2F} use H$\delta_{A}$ - $\sigma$(H$\delta_{A}$) $>$ 4 \AA\ (where $\sigma$(H$\delta_{A}$) is the measurement error of the H$\delta_{A}$ index) and H$\alpha$ rest-frame equivalent width, EW(H$\alpha$) $<$ 3 \AA\ as their selection criteria, which yield a sample of real \emph{post}-starburst galaxies. \citet{2016ApJS..224...38A} allow for emission lines from shocks and use H$\delta_{A}$ $> 5$ \AA\ after emission-line correction; as this emission also may arise from star formation, their sample could still have ongoing starbursts. These objects turn out to be at earlier PSB stages, while some even have negative post-burst ages, signifying an ongoing burst \citep{2018ApJ...862....2F}. Thus, they serve as crucial links between  the cessation of the recent burst and the subsequent decline in star formation. \citet{2015MNRAS.448..258R} use a Principal Component Analysis (PCA) technique at 3175 -- 4150 \AA, which essentially requires strong Balmer absorption and weak 4000 \AA\ break strength; such a selection focuses on young stellar ages and thus, like the \citet{2016ApJS..224...38A} sample, includes transitioning galaxies from a starbursting to PSB phase.

One of the primary goals of our work here is to derive \emph{M}$_{\rm dust}$. As previous studies have shown that FIR ($\lambda \geqslant\ $40$\mu$m) photometry is crucial \citep{2008MNRAS.388.1595D, 2012ApJ...745...95D}, we define our sample as those galaxies among the aforementioned three samples of PSBs with archival FIR data. In addition to utilizing the processed \emph{WISE} and \emph{Herschel} data from \citet{2018ApJ...855...51S} for 33 PSBs from \citet{2015ApJ...801....1F}, we searched for \emph{Herschel} observations of other galaxies in these three samples in the PACS Point Source Catalog \citep{2017arXiv170505693M} and SPIRE Point Source Catalog \citep{2017arXiv170600448S}, available at the NASA/IPAC Infrared Science Archive (IRSA)\footnote{http://irsa.ipac.caltech.edu/.}. As a result, 37 PSBs from \citet{2018ApJ...862....2F}, 12 PSBs from \citet{2016ApJS..224...38A}, and 11 PSBs from \citet{2015MNRAS.448..258R} have detections in $\geqslant$3 \emph{Herschel} bands\footnote{We will justify the necessity of using at least three \emph{Herschel} bands in Appendix \ref{sec:Herschelbands}.}, constituting our final sample of 58 PSBs in total\footnote{Two objects (R8/A11 and R11/A12) are in both the \citet{2016ApJS..224...38A} and \citet{2015MNRAS.448..258R} samples, and we refer to them as A11 and A12 in this paper.}. The 37 PSBs from \citet{2018ApJ...862....2F} 
are labeled EAH01-EAH18 and EAS01-EAS15, consistent with the nomenclature in \citet{2015ApJ...801....1F} and \citet{2018ApJ...855...51S}, and F34-F37, for those without previous names. Their redshifts are 0.02 $<$ \emph{z} $<$ 0.11. The 12 PSBs from \citet{2016ApJS..224...38A} are A1-A12, with 0.02 $<$ \emph{z} $<$ 0.18. The 11 PSBs from \citet{2015MNRAS.448..258R} are R1-R11, with 0.03 $<$ \emph{z} $<$ 0.05.

\section{Multiwavelength Data} \label{sec:data}
We establish UV to FIR SEDs for our PSB sample. We incorporate the processed \emph{WISE} and \emph{Herschel} data from \citet{2018ApJ...855...51S} for 33 PSBs from \citet{2015ApJ...801....1F} and compile other data from different catalogs. In addition to compiling the \emph{Herschel} data from PACS/SPIRE Point Source Catalogs, we utilize the archival photometric data from \emph{GALEX}, SDSS, 2MASS, and \emph{WISE}. We have also calculated the flux uncertainties by combining in quadrature the cataloged measurement uncertainties with different systematic uncertainties in each band, which are described individually in the following paragraphs.

For \emph{GALEX} data, we search for NUV and FUV detections from the \emph{GALEX} All-Sky Survey Source Catalog (GASC) and the Medium Imaging Survey Catalog (GMSC)\footnote{http://galex.stsci.edu/galexview/.}. We use the \texttt{mag\_FUV} and \texttt{mag\_NUV} magnitudes, which should be representative of the total galaxy flux. We further add zero-point calibration errors of 0.052 and 0.026 mag to the FUV and NUV photometry errors, respectively \citep{2007ApJS..173..682M}.

For SDSS data, we search for \emph{ugriz} photometry in the \texttt{Photoobjall} catalog of the SDSS 14th Data Release (DR14, \citealt{2018ApJS..235...42A}). We adopt the \texttt{modelmag} magnitudes, as they provide reliable colors and represent the total light of our sources\footnote{http://www.sdss.org/dr14/algorithms/magnitudes/.}. To ensure all the magnitudes are on the AB system, we add -0.04 to measured \emph{u}-band magnitudes and 0.02 to \emph{z}-band magnitudes\footnote{http://www.sdss.org/dr14/algorithms/fluxcal/.}. Zero-point calibration errors of  5\%, 2\%, 2\%, 2\%, and 3\% are added to the photometry errors of \emph{ugriz} bands, respectively \citep{2007AJ....133..734B}.

For 2MASS data, we search for \emph{JHK}$_{s}$ photometry from the 2MASS Point Source Catalog (PSC; \citealt{2006AJ....131.1163S}) and Extended Source Catalog (XSC; \citealt{2000AJ....119.2498J}). If the source is cataloged in PSC, we adopt the standard aperture, which is measured in a 4$\arcsec$ radius aperture, but has already been corrected to an infinite aperture. If the source is cataloged in XSC, we choose the extrapolated total magnitude \texttt{x\_m\_ext}, which should represent the total flux\footnote{https://www.ipac.caltech.edu/2mass/releases/allsky/doc/sec4\_5e.html.}. We convolve a 5\% calibration error \citep{2009ApJ...703..517D} with the photometry error in quadrature.

For \emph{WISE} data, we search for \emph{W}1-\emph{W}4 photometry in the ALL\emph{WISE} Source Catalog \citep{2011ApJ...731...53M}. We use the magnitudes measured with profile-fitting photometry (\texttt{wxmpro}) for point sources (defined with \texttt{ext\_flg} = 0). For extended sources (\texttt{ext\_flg} $>$ 0), we follow the instructions from the \emph{WISE} official website\footnote{http://wise2.ipac.caltech.edu/docs/release/allsky/faq.html.}: when \texttt{ext\_flg} = 5, we adopt the \texttt{wxgmag} measured with an elliptical aperture; when 0 $<$ \texttt{ext\_flg} $<$ 5, we choose the circular aperture magnitude that best matches the extrapolated total radius \texttt{r\_ext} provided in 2MASS. We correct zero-point errors by adding 0.03, 0.04, 0.03, and -0.03 mag to the measured \emph{WISE} \emph{W}1-\emph{W}4 bands, respectively \citep{2012AJ....144...68J}. We add an overall 6\% calibration error for the \emph{W}1-\emph{W}4 bands to the photometry error \citep{2015nwis.rept....1C}.

For \emph{Herschel} data, we adopt the quantity \texttt{flux} from the PACS/SPIRE Point Source Catalogs. The PACS flux uncertainties are derived by convolving the \texttt{snrnoise} (including sky confusion and instrumental error) and the background \texttt{rms}. The SPIRE `total' flux uncertainties \texttt{flux\_err} include instrumental noise and background confusion noise. We add a 7\% calibration error to the PACS and SPIRE flux uncertainties \citep{2012A&A...543A.161C, 2014ExA....37..129B}. 

All the data are presented in Tables \ref{tab:data1} and \ref{tab:data2}. We do not apply any correction for Galactic extinction, because it is only nonnegligible for several sources and only affects the UV and optical data, which do not affect $M_{\rm dust}$ (see Appendix \ref{sec:UVopteffect}). To characterize the amount of internal extinction, we have incorporated the \citet{2000ApJ...533..682C} law into our SED fitting (Section \ref{sec:models}).

\section{SED Fitting} \label{sec:fitting}
In this work, we perform UV-FIR SED fitting on our sample using CIGALE (Code Investigating GALaxy Emission; \citealt{2009A&A...507.1793N}). Below we provide a detailed description of our SED fitting procedure in terms of the models and input priors, and present our fitting results.

\subsection{Models}\label{sec:models}
For galaxy SED fitting, in general, CIGALE requires four models in total, which describe the star-formation history (SFH), stellar populations, dust emission, and dust extinction, respectively. We do not use CIGALE's default nebular emission model, as our sources do not exhibit strong nebular emission lines.

For SFH, we use two types of models: one or two exponentially declining recent bursts, with a main stellar population formed earlier \citep{2018ApJ...862....2F}. We refer to these two kinds of models as `single-burst' or `double-burst' models hereafter. The common free parameters in both models are:

(1) \emph{e}-folding time of the main stellar population, $\tau_{\rm main}$;

(2) \emph{e}-folding time of the most recent starburst population, $\tau_{\rm burst}$;

(3) mass fraction of the recent burst(s) relative to the total stellar mass, $f_{\rm burst}$;

(4) age of the main stellar population (the time elapsed since it formed), ${\rm age}_{\rm main}$;

(5) age of the most recent burst (the time elapsed since it started), ${\rm age}_{\rm burst}$.

An additional free parameter, $t_{\rm sep}$, is set in the double-burst model. It describes the time separation between the two recent bursts.

For stellar populations, we incorporate the BC03 \citep{2003MNRAS.344.1000B} model assuming a \citet{2003PASP..115..763C} initial mass function (IMF). For dust emission, we choose the DL07 \citep{2007ApJ...657..810D} model. For dust extinction, we use the \citet{2000ApJ...533..682C} law to model the internal extinction of our sources.

\subsection{Input Priors}\label{sec:priors}
All of the input priors are summarized in Table \ref{table:input}. The prior values given are the allowed discrete values for CIGALE. For SFH, we refer to \citet{2018ApJ...862....2F}  for the number of recent bursts inferred for each galaxy. We give fairly large prior ranges for $f_{\rm burst}$, ${\rm age}_{\rm burst}$, $\tau_{\rm main}$, $\tau_{\rm burst}$, ${\rm age}_{\rm main}$, and $t_{\rm sep}$ to enlarge the parameter space. 

For the BC03 model, we offer a range of three metallicities closest to that inferred from the mass-metallicity relation for each object. The stellar masses come from the SDSS MPA-JHU catalogs (\citealt{2004MNRAS.351.1151B, 2004ApJ...613..898T}). The mass-metallicity relation is from \citet{2005MNRAS.362...41G}. 

For the DL07 model, the mass fraction of PAH, \emph  q$_{\rm PAH}$, is allowed to vary from 0.47 to 4.58. The discrete values in Table \ref{table:input} come directly from \citet{2007ApJ...657..810D}. The prior range of the minimum radiation field (\emph  U$_{\rm min}$) is set to [0.1, 25.0]. The fraction illuminated from \emph  U$_{\rm min}$ to \emph U$_{\rm max}$ ($\gamma$) is allowed to vary between [0.0001, 0.1]. We fix the maximum radiation field (\emph U$_{\rm max}$) to be 10$^6$.

For the dust extinction model, we follow the default setting in CIGALE, which assumes that the stars younger than 10 Myr are subject to more extinction than the stars older than 10 Myr. We allow the color excess of the stellar continuum light for the young population, \emph{E(B--V)}$_{\rm young}$, to vary from 0.01 to 2. As the majority of our sample have \emph{E(B--V)}$_{\rm young}$ $>$ 0.1, we input six evenly distributed prior values from 0.1 to 2, and two values from 0.01 to 0.05. The reduction factor for the color excess of the old population compared to the young one, $f_{\rm att}$, is allowed to vary from 0.3 to 1. We do not add any UV bump or power law to the original \citet{2000ApJ...533..682C} law.

\begin{deluxetable*}{cccc}
\tabletypesize{\footnotesize}
\tablecaption{{\centering}Input Parameters for CIGALE SED Fitting\label{table:input}}
\tablehead{\colhead{Model}          &
           \colhead{Parameter}       &        
           \colhead{Symbol}           &
           \colhead{Prior Values}\\
           \colhead{(1)}              &
           \colhead{(2)}              &
           \colhead{(3)}              &
           \colhead{(4)}}
\startdata 
 SFH&\emph{e}-folding time of the main stellar population&$\tau_{\rm main}$ & 1.0, 2.0, 3.0 Gyr \\
 &\emph{e}-folding time of the recent burst&$\tau_{\rm burst}$ & single-burst: 25, 50, 100, 200, 500 Myr double-burst: 25 Myr\\
 &Mass fraction of the recent burst(s)&$f_{\rm burst}$ & 0.01, 0.05, 0.1, 0.3, 0.5, 0.7, 0.9 \\
 &Age of the main stellar population &${\rm age}_{\rm main}$ & 6.0, 9.0, 11.0 Gyr\\
 &Age of the most recent burst & ${\rm age}_{\rm burst}$ & 30, 100, 200, 500, 1000, 1500, 2000 Myr\\
 &Separation between two recent bursts & $t_{\rm sep}$ & double-burst: 100, 300, 500, 800, 1000 Myr\\
 BC03&Metallicity&[Fe/H]&3 values inferred from \citet{2005MNRAS.362...41G}\\
 DL07&Mass fraction of PAH&$q_{\rm PAH}$&0.47, 1.12, 1.77, 2.50, 3.19, 4.58\\
 &Minimum radiation field&$U_{\rm min}$&0.10, 0.50, 1.00, 2.50, 5.0, 10.0, 25.0\\
 &Maximum radiation field&$U_{\rm max}$&10$^6$\\
 &Dust fraction illuminated from \emph{U}$_{\rm min}$ to \emph{U}$_{\rm max}$&$\gamma$&0.0001, 0.001, 0.01, 0.05, 0.1\\
 &Power-law slope \emph{dU/dM} $\varpropto$ \emph{U}$^{\alpha}$ &$\alpha$&2.0\\
 Dust extinction&Color excess of stellar continuum light for young stars &\emph{E(B -- V)}$_{\rm young}$&0.01, 0.05, 0.1, 0.4, 0.7, 1, 1.5, 2.0\\
 &Reduction factor for \emph{E(B -- V)} of old stars to young stars& $f_{\rm att}$&0.3, 0.5, 0.8, 1.0\\
\enddata
\tablenotetext{}{\textbf{Note.} Configurations of the input parameters used in CIGALE. (1) Model names. (2) Definitions of parameters. (3) Symbols of parameters. (4) Prior values of parameters.}
\end{deluxetable*}

\subsection{Fitting Results}\label{sec:fitting-results}
\begin{figure*}
\centering
\begin{tabular}{c c}
\includegraphics[width=0.42\linewidth, clip]{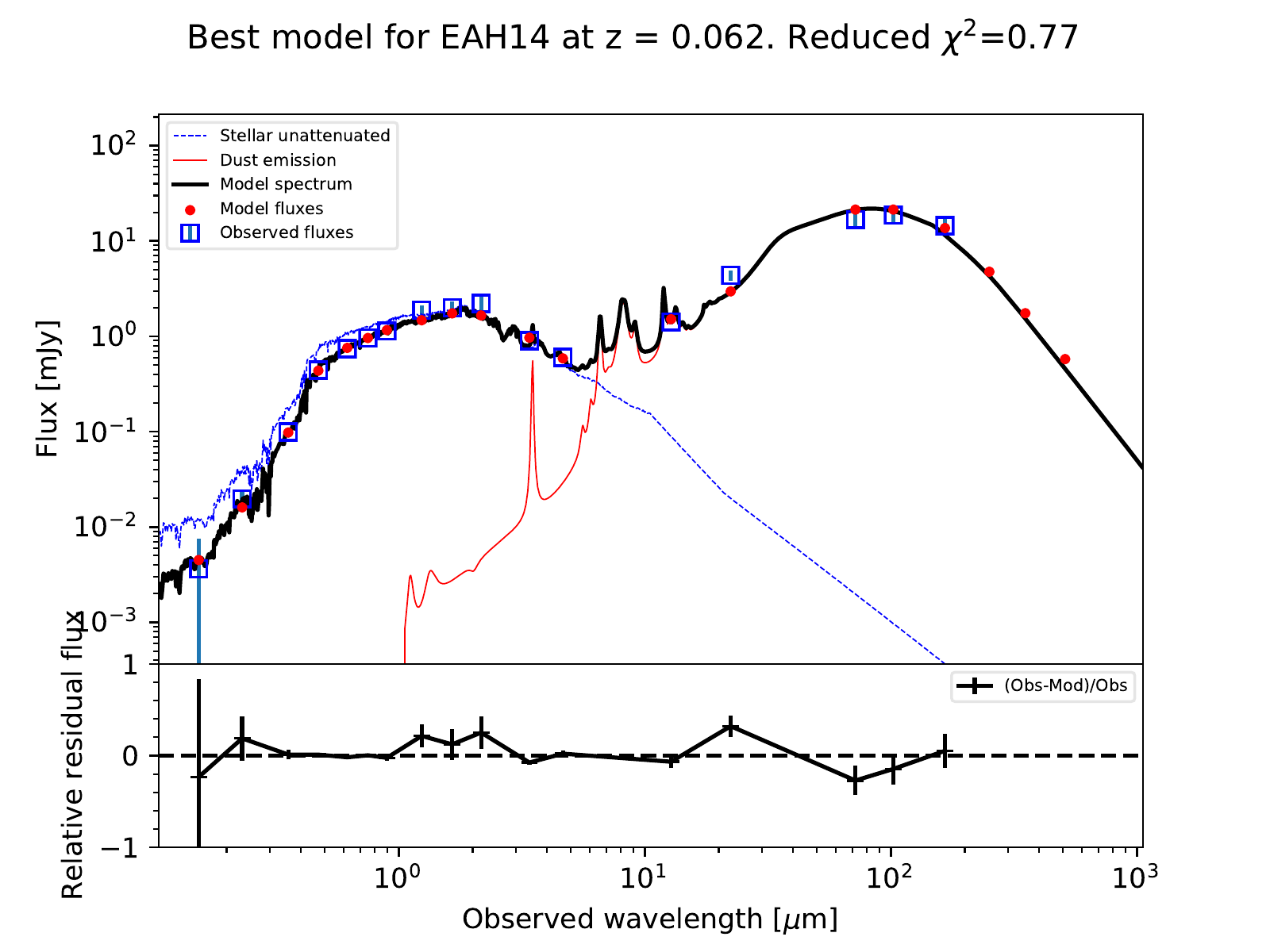} &
\includegraphics[width=0.42\linewidth, clip]{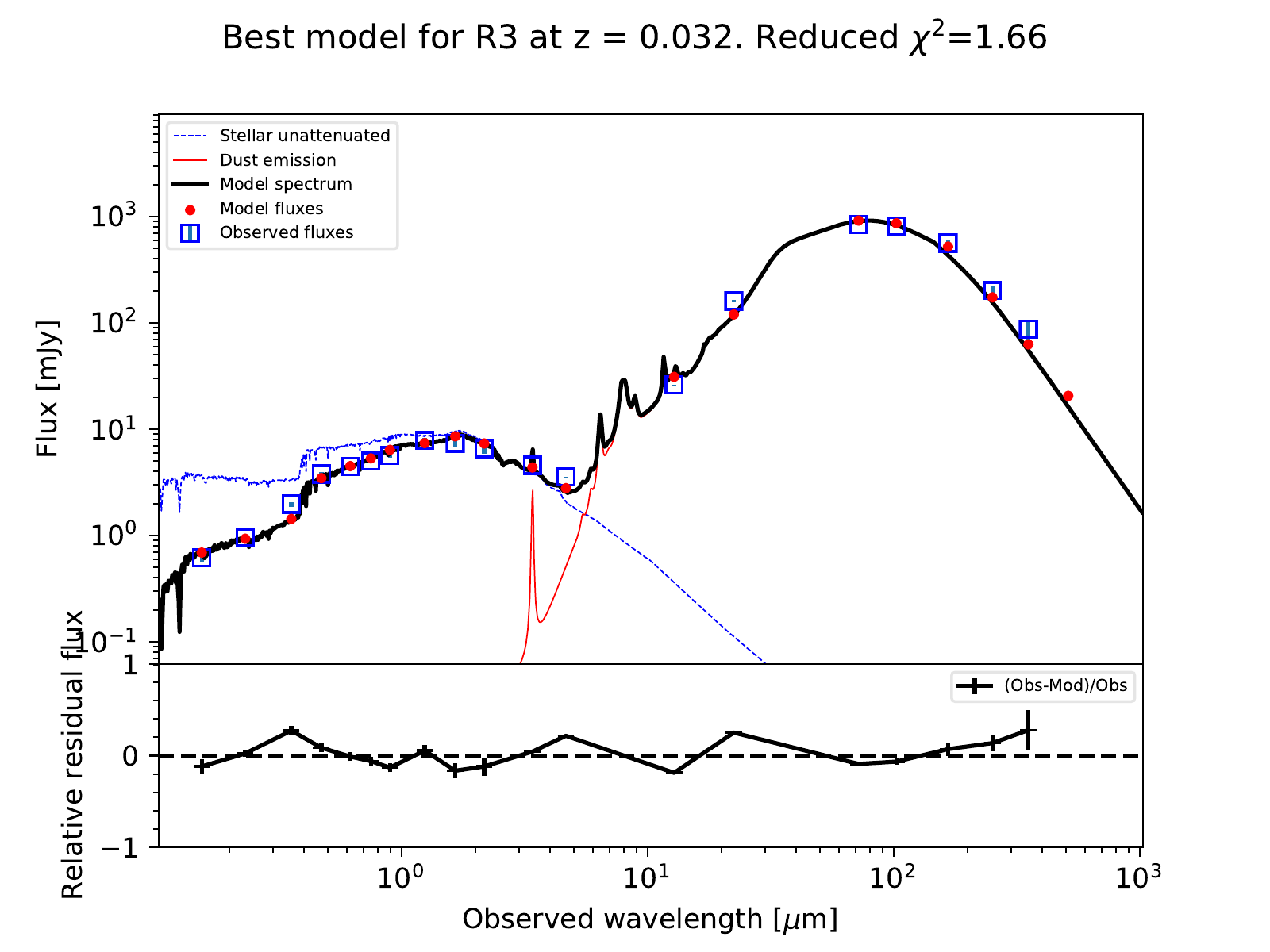} \\
\includegraphics[width=0.42\linewidth, clip]{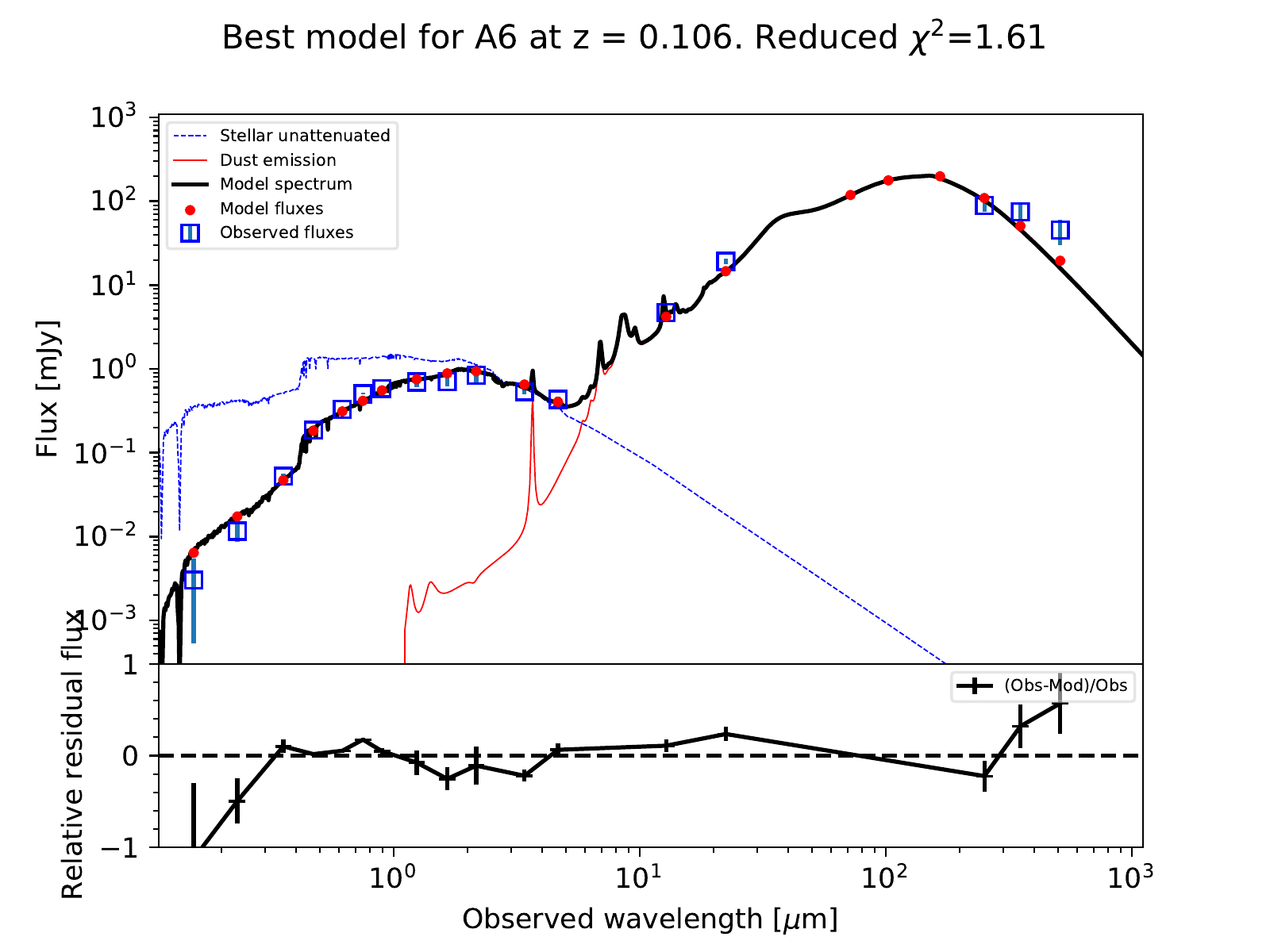} &
\includegraphics[width=0.42\linewidth, clip]{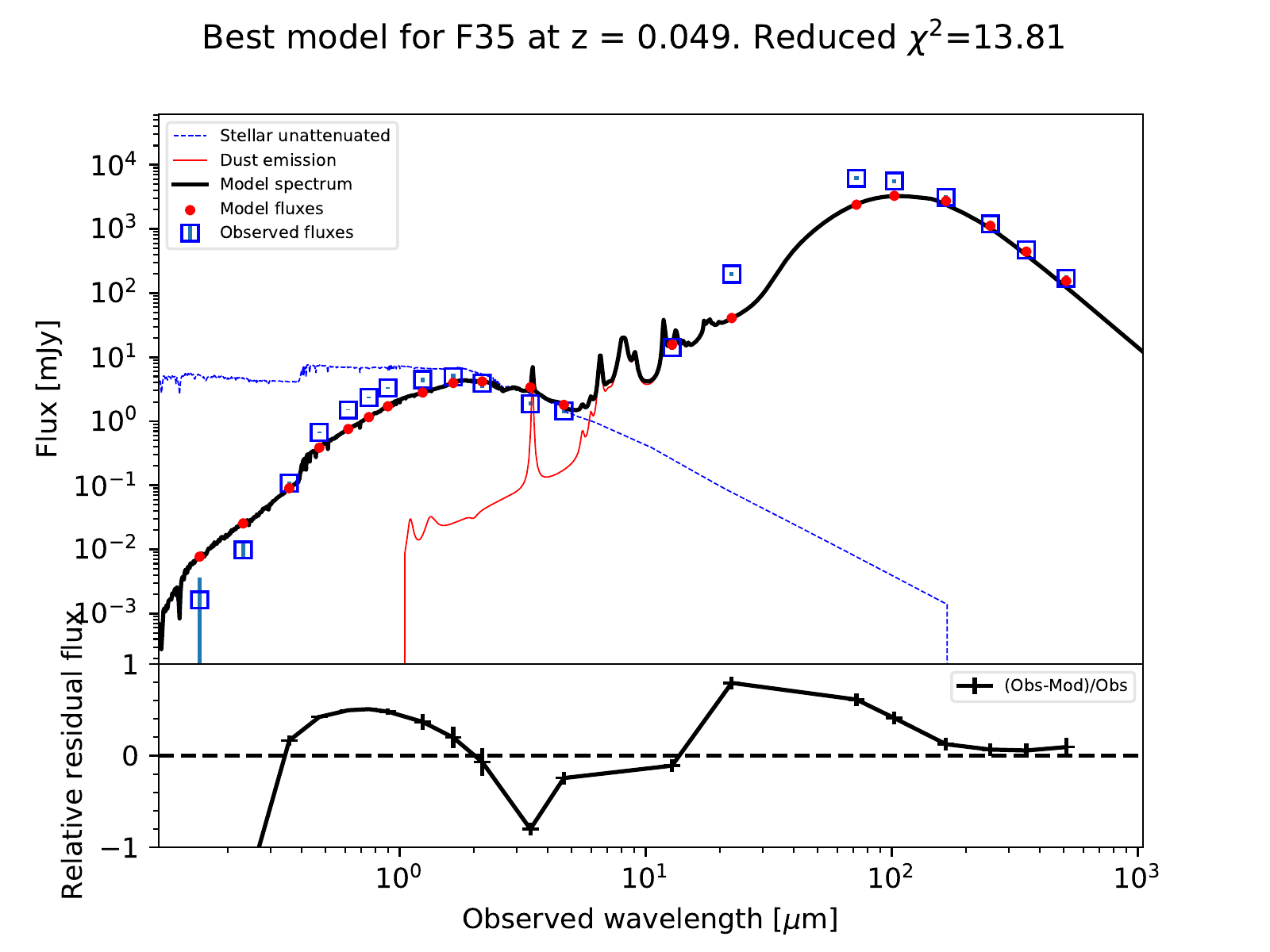} \\
\includegraphics[width=0.42\linewidth, clip]{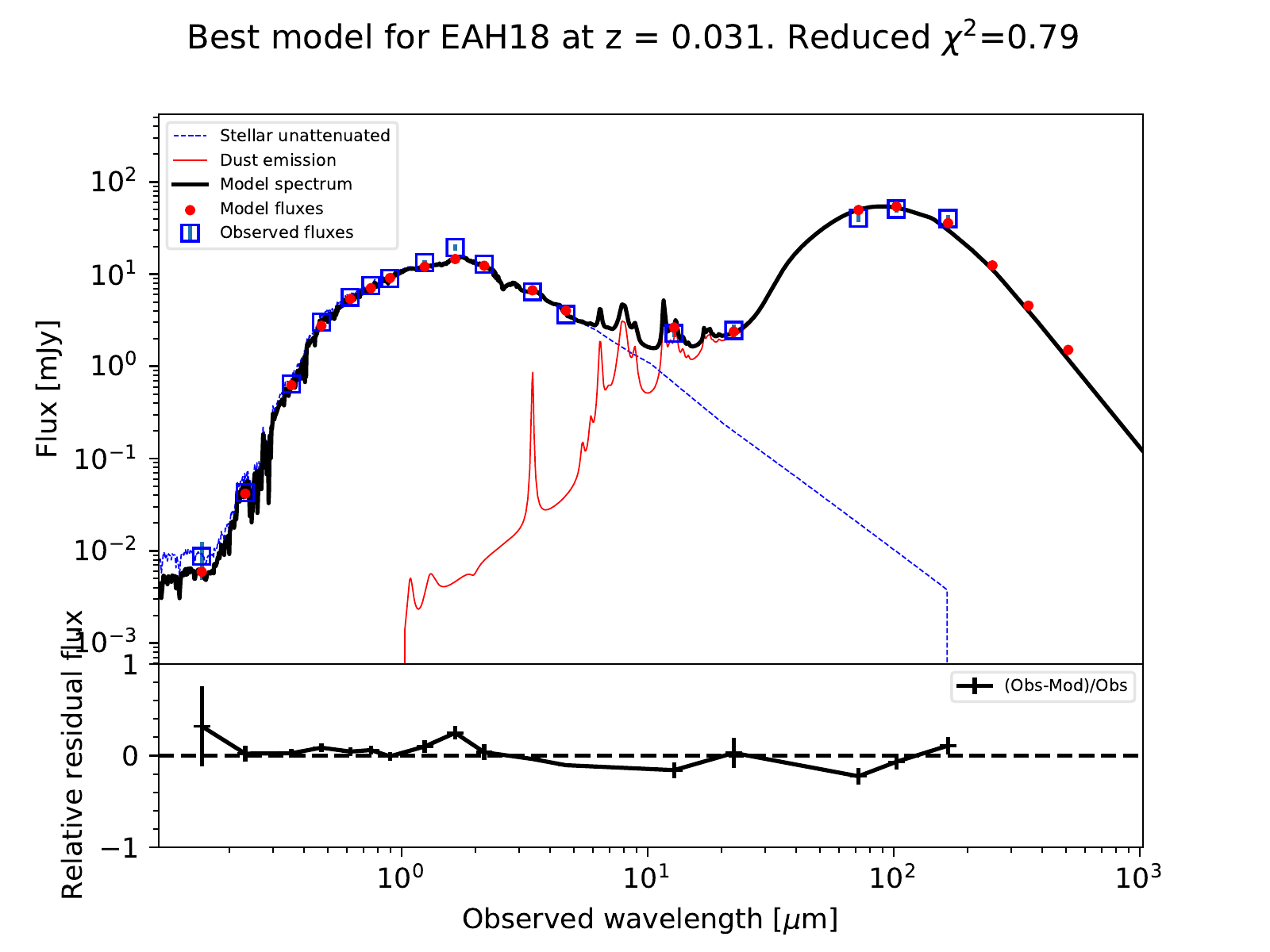} &
\includegraphics[width=0.42\linewidth, clip]{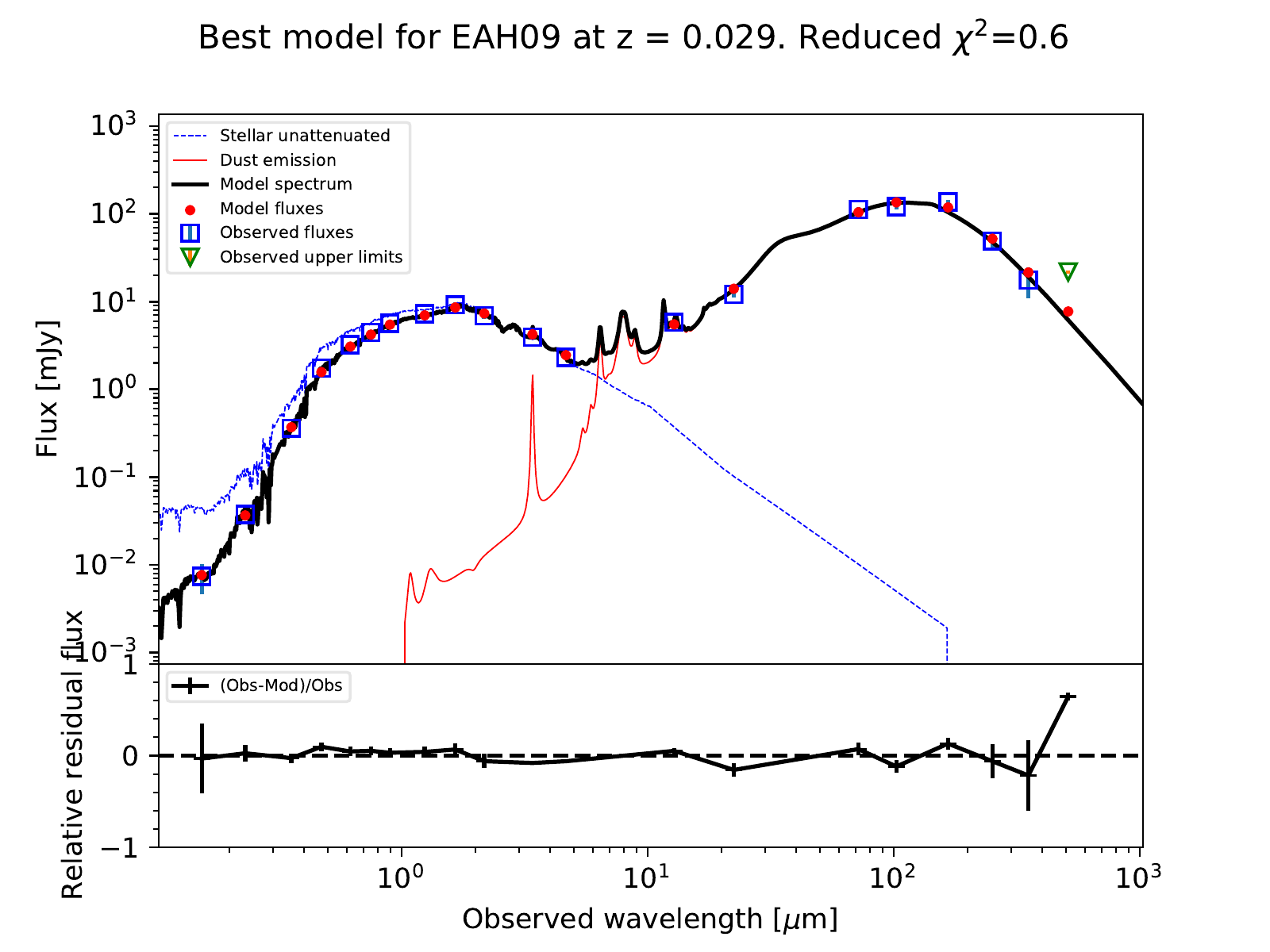} \\
\includegraphics[width=0.42\linewidth, clip]{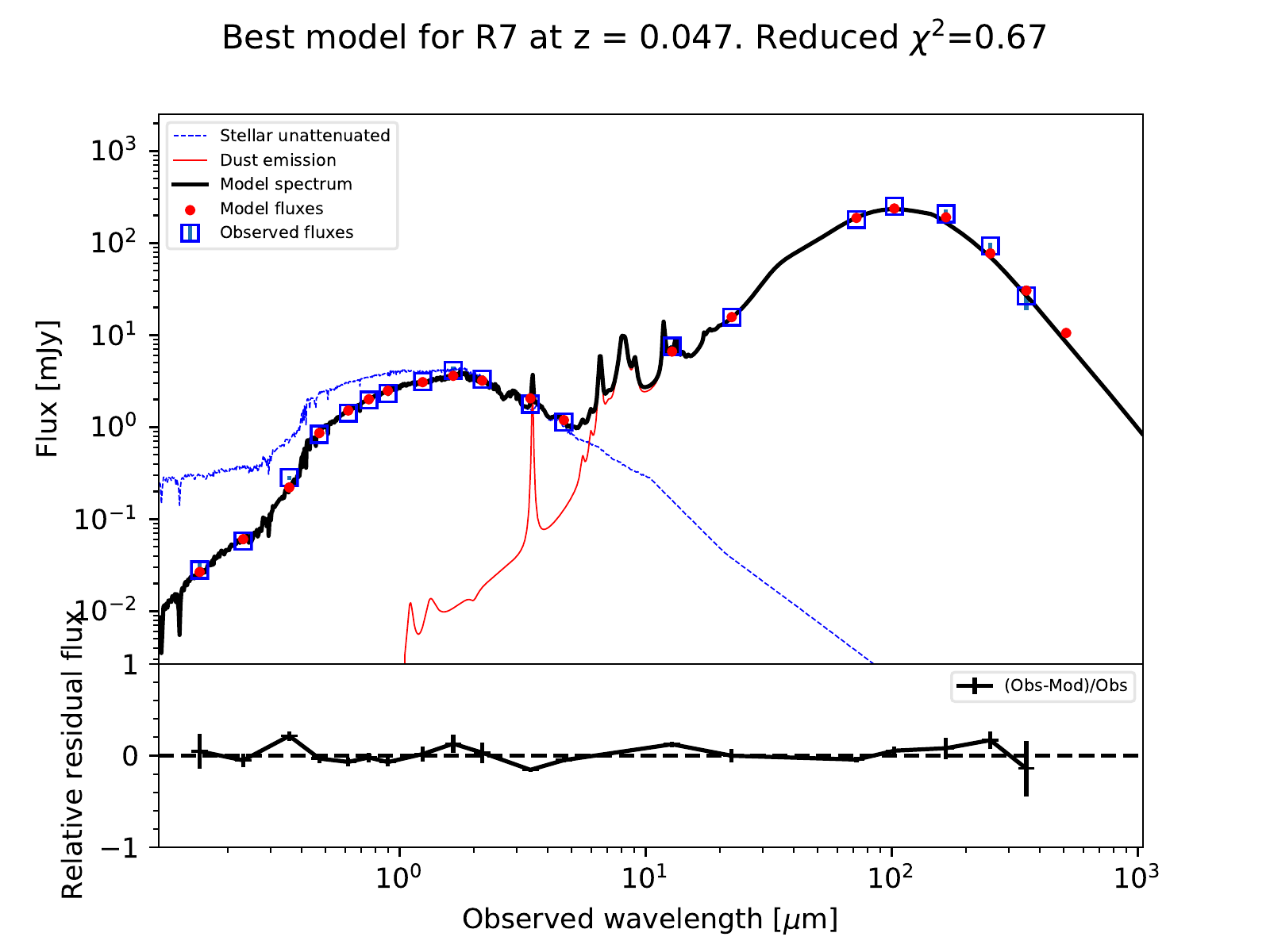} &
\includegraphics[width=0.42\linewidth, clip]{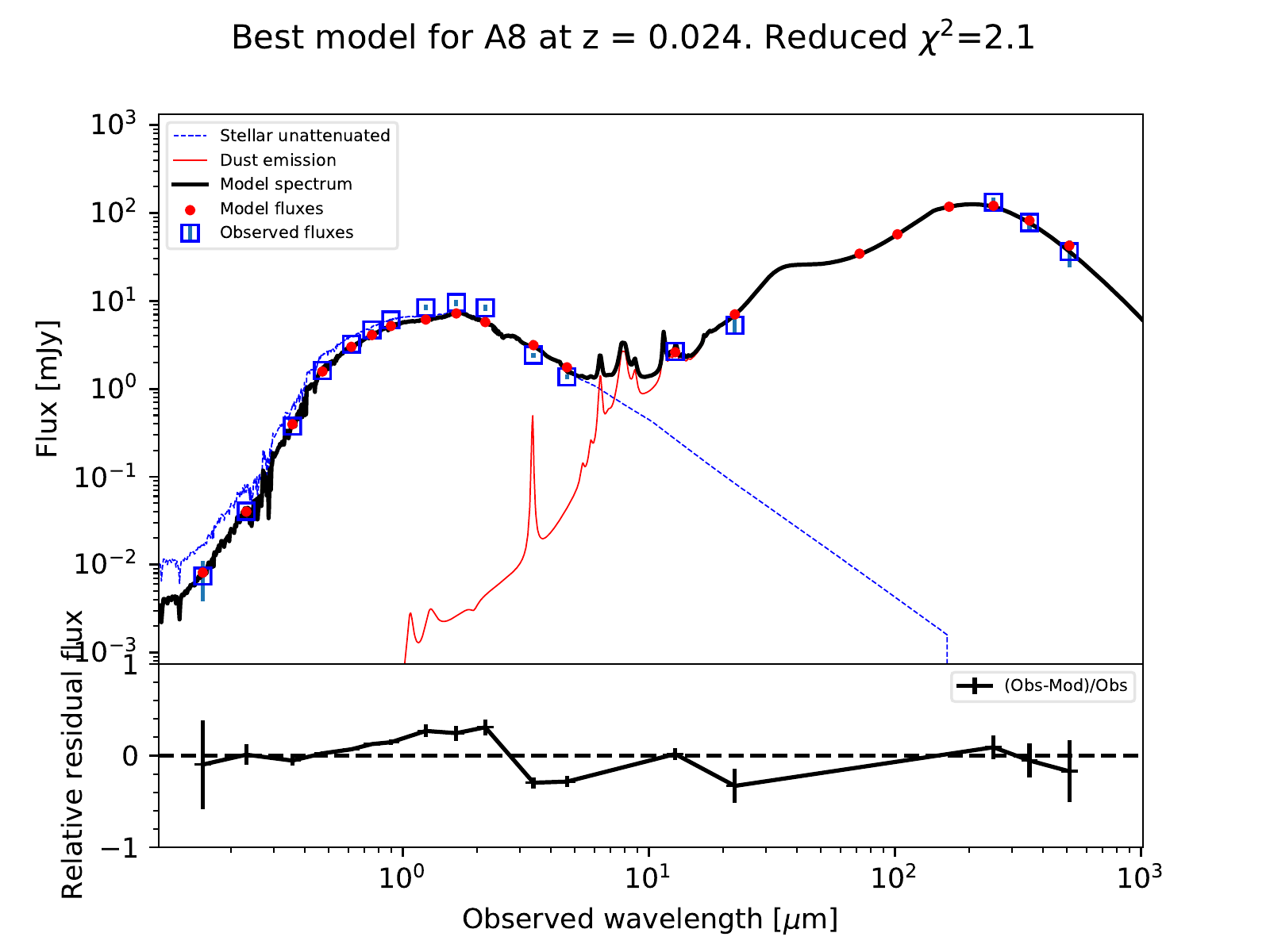}
\end{tabular}
\caption{Eight representative SED fits derived from CIGALE. The top four fits have `single-burst' SFH, whereas the bottom four have `double-burst' SFH. The relative residual flux is defined as $({f_{\rm obs}-f_{\rm model})}/{{f_{\rm obs}}}$. The error bars (plotted in blue) are 1$\sigma$ values. The green triangles represent 5$\sigma$ upper limits. The worst fit here, F35, is particularly extended ($\emph r_{50}$ = 7 arcsec) and dusty.}
\label{fig:fits}
\end{figure*}

In Figure \ref{fig:fits}, we present eight typical SED fits of our sample. Their \emph M$_{\rm dust}$'s range from 10$^{5.78}$ \emph{M$_{\sun}$} to 10$^{8.18}$ \emph{M$_{\sun}$}, while the full range of \emph M$_{\rm dust}$ of our sample is 10$^{5.32}$ \emph{M$_{\sun}$} to 10$^{8.89}$ \emph{M$_{\sun}$}. According to \citet{2018ApJ...862....2F}, the optimal choice of a recent SFH model for EAH14, R3, A6, and F35 is ``single-burst", whereas for EAH18, EAH9, R7, and A8 it is ``double-burst".

\begin{figure}
\includegraphics[width=\linewidth, clip]{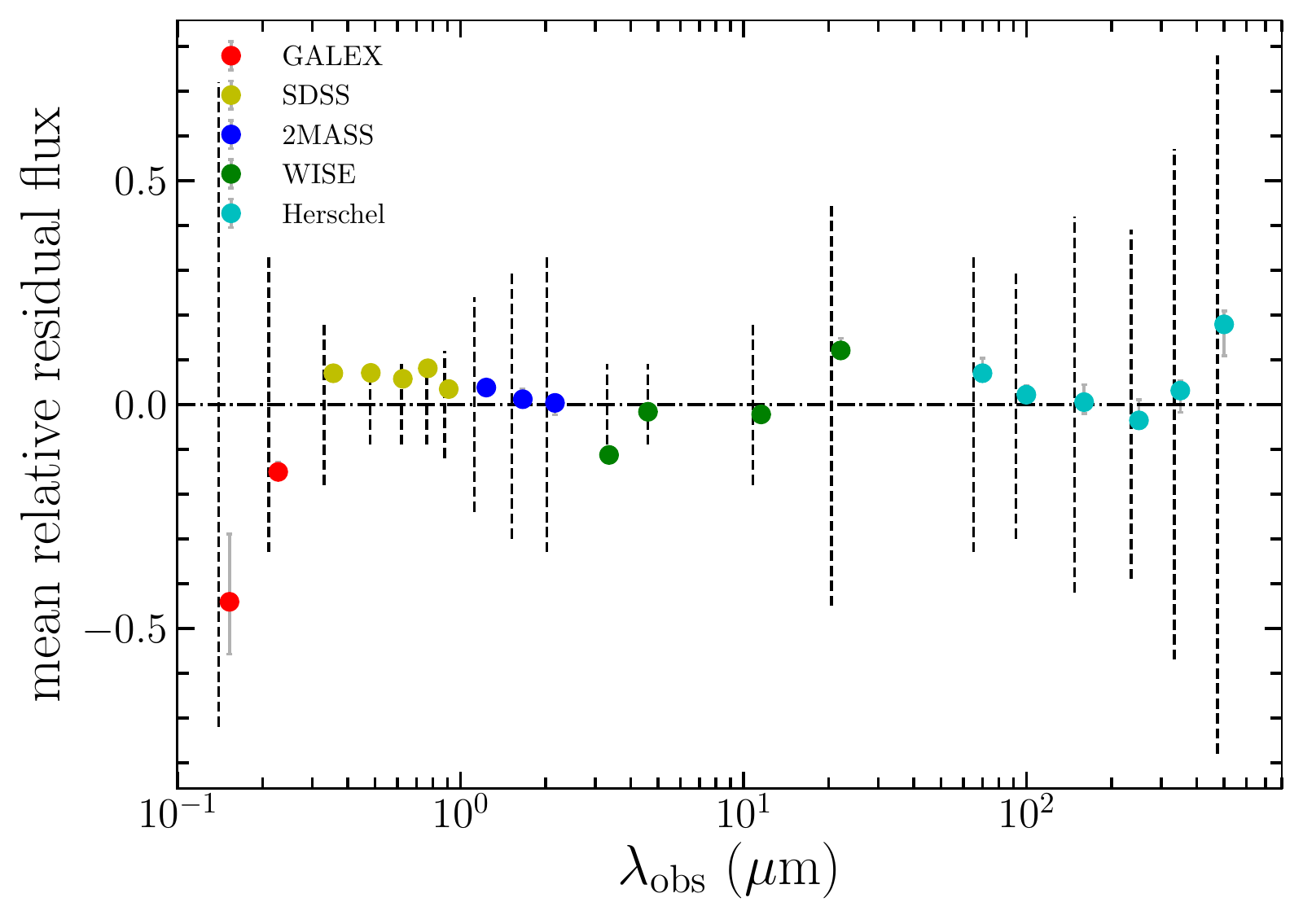}
\caption{Mean relative residual flux (defined as $({f_{\rm obs}-f_{\rm model})}/{{f_{\rm obs}}}$) in different bands (marked in different colors) and its 68\% confidence uncertainty after fitting the SEDs using CIGALE \citep{2009A&A...507.1793N}. For comparison, the vertical dashed lines are the 3$\sigma$ average percentage flux uncertainty, defined as ${\sigma_{\rm obs}}/{{f_{\rm obs}}}$, where $\sigma_{\rm obs}$ is the total flux uncertainty in that band. The horizontal dotted-dashed line represents zero residual flux. In general (19/20 bands), the mean relative residual flux is consistent with the 3$\sigma$ average percentage flux uncertainty, especially for all \emph{Herschel} bands, which are crucial for deriving $M_{\rm dust}$.}
\label{fig:residual}
\end{figure}

To quantify the quality of our SED fits, we calculate the mean relative residual flux for our sample, as shown in Figure \ref{fig:residual}. We define the relative residual flux to be ${(f_{\rm obs}-f_{\rm model})}/{{f_{\rm obs}}}$, in which $f_{\rm obs}$ is the observed flux and $f_{\rm model}$ is the model flux predicted by CIGALE. To determine the uncertainty of the mean relative residual flux, we use the Monte Carlo method to generate realizations of the fluxes and construct probability distributions for each individual flux and the mean relative residual flux. The error bars of the mean relative residual flux in Figure \ref{fig:residual} represent their 68\% confidence level uncertainties. 

From Figure \ref{fig:residual} we conclude that in general (i.e., for 19/20 bands), the mean relative residual flux is consistent with the 3$\sigma$ average percentage flux uncertainty (defined as ${\sigma_{\rm obs}}/{{f_{\rm obs}}}$, where $\sigma_{\rm obs}$ is the total flux uncertainty in that band). For the \emph{W}1 band, the mean relative residual flux is barely consistent with the 3$\sigma$ average percentage flux uncertainty. Such systematics could be due to the uncertainties in the data, the limitations of the stellar population model, or the dust model. 

In terms of reduced $\chi^2$, of all 58 SED fits, only five of the PSBs (EAH02, EAH05, EAH08, A5, and F35) have reduced $\chi^2$ $>$ 5.0. Our tests show that their \emph{M}$_{\rm dust}$ does not change much if we only fit their IR SED (see Appendix \ref{sec:UVopteffect}). All five EAH sources with reduced $\chi^2$ $>$ 3.0 have \emph{M}$_{\rm dust}$ consistent with \citet{2018ApJ...855...51S} (see Appendix \ref{sec:compare}). So we conclude that the relatively large reduced $\chi^2$ of this small fraction of our sample does not affect the robustness of our derived \emph{M}$_{\rm dust}$ or any conclusions in general.

\section{Results and Discussion} \label{sec:results}
\subsection{$M_{dust}$ versus post-burst age}\label{sec:dust-age}
Having derived $M_{\rm dust}$ for our sample using CIGALE, we examine the evolution of specific dust mass (=$M_{\rm dust}/M_{\star}$) versus post-burst age, ${\rm age}_{\rm post-burst}$\footnote{We use the more accurate ${\rm age}_{\rm post-burst}$ derived in \citet{2018ApJ...862....2F} by including optical spectral information (see Appendix \ref{sec:ageMd}).}. Here we define the `post-burst age' to be the time elapsed since the majority (90\%) of the stars formed in the recent burst(s). Thus, for the single-burst model, we have:
\begin{equation}
age_{\rm post-burst}=age_{\rm burst}-2.3\tau_{\rm burst}\ (\rm Myr)
\end{equation}
while for the double-burst model\footnote{Here $\tau_{\rm burst}$ is fixed to 25 Myr.}:
\begin{equation}
age_{\rm post-burst}=age_{\rm burst}-t_{\rm sep}-29\ (\rm Myr)
\end{equation}

\begin{figure}
\includegraphics[width=\linewidth, clip]{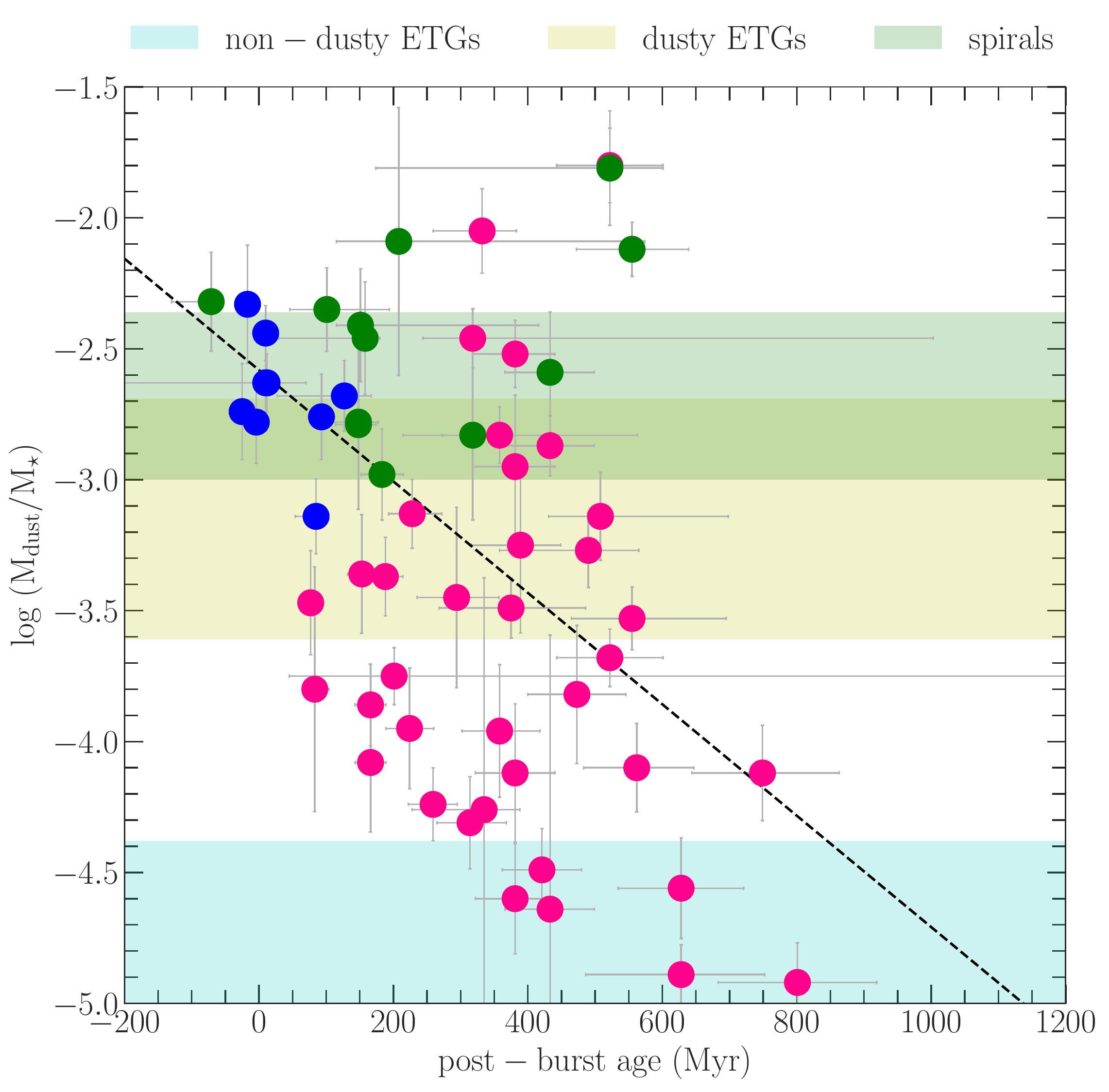}
\caption{Specific dust mass (${M_{\rm dust}}/{{M_{\rm \star}}}$) vs. post-burst age relation. The green points represent the 12 PSBs from \citet{2016ApJS..224...38A}, the red points are the 37 PSBs from \citet{2018ApJ...862....2F}, and the blue points are the nine PSBs from \citet{2015MNRAS.448..258R}. The black dotted line is a linear fit using the method in \citet{2013MNRAS.432.1709C}: log (${M_{\rm dust}}/{{M_{\rm \star}}}$) = \emph{a} $\cdot$ (${\rm age}_{\rm post-burst}$- \emph x$_{0}$) + \emph{b} + $\epsilon$, where \emph{a} = -0.00212 $\pm$ 0.00047, \emph{b} = -3.28 $\pm$ 0.10, \emph x$_{0}$ = 316, and the intrinsic scatter $\epsilon$ = 0.68 $\pm$ 0.09. The Spearman coefficients are \emph{r} = -0.39 $\pm$ 0.13 and log(\emph{p}) = -2.54 $\pm$ 1.24. To compare our results with previous work, we overplot three other samples following \citet{2015MNRAS.448..258R}: the average ${M_{\rm dust}}/{{M_{\rm \star}}}$ for \emph{z} $<$ 0.1 spiral galaxies in \citet{2012MNRAS.419.2545R}, 0.01 $<$ \emph{z} $<$ 0.06 dusty ETGs from \citet{2013MNRAS.431.1929A}, and nondusty ETGs for a range of dust temperatures from \citet{2012MNRAS.419.2545R}. The significant declining trend between ${M_{\rm dust}}/{{M_{\rm \star}}}$ and ${\rm age}_{\rm post-burst}$ implies a dust depletion timescale of 205$^{+58}_{-37}$ Myr, consistent with the CO-traced H$_2$ depletion timescale \citep{2018ApJ...862....2F}.}
\label{fig:Md-psbage}
\end{figure}

As shown in Figure \ref{fig:Md-psbage}, there is a declining trend between specific dust mass and ${\rm age}_{\rm post-burst}$. To quantify the significance of this relation, we perform a Spearman rank test and linear fitting using the method in \citet{2013MNRAS.432.1709C}, which takes errors in both $M_{\rm dust}/M_{\star}$ and ${\rm age}_{\rm post-burst}$ into account. The fitting result is in the form of log(${M_{\rm dust}}/{{M_{\rm \star}}}$) = \emph{a} $\cdot$ (${\rm age}_{\rm post-burst}$- \emph x$_{0}$) + \emph{b} + $\epsilon$, where \emph{a} = -0.00212 $\pm$ 0.00047, \emph{b} = -3.28 $\pm$ 0.10, \emph x$_{0}$ = 316, and intrinsic scatter $\epsilon$ = 0.68 $\pm$ 0.09. We also use the \texttt{ASURV} survival analysis package to calculate the Spearman's rank correlation \citep{1986ApJ...306..490I, 1992ASPC...25..245L}, and perform bootstrap analysis (1000 samples) to derive the confidence intervals of the
Spearman correlation coefficient and the null hypothesis probability \citep{2019ApJ...870...26L}. The null hypothesis is that there is no monotonic
relation between two parameters. We define a significant correlation as one that rejects this hypothesis by having a probability $\leq$ 3 $\times$ 10$^{-3}$ (log(\emph{p}) $\leq$ -2.52), corresponding to approximately 3$\sigma$. The Spearman coefficients are \emph{r} = -0.39 $\pm$ 0.13 and log(\emph{p}) = -2.54 $\pm$ 1.24. To compare our results with previous work, we overplot three other samples following \citet{2015MNRAS.448..258R}. They are the average ${M_{\rm dust}}/{{M_{\rm \star}}}$ for \emph{z} $<$ 0.1 spiral galaxies detected in \citet{2012MNRAS.419.2545R}, for 0.01 $<$ \emph{z} $<$ 0.06 dusty ETGs from \citet{2013MNRAS.431.1929A}, and for nondusty ETGs (representative of red sequence galaxies) for a range of dust temperatures from \citet{2012MNRAS.419.2545R}.

The significant anti-correlation between \emph{M}$_{\rm dust}$/\emph{M}$_{\star}$ and $age_{\rm post-burst}$ suggests that the dust is either destroyed, expelled, or rendered undetectable over the $\sim$1 Gyr after the burst. Assuming the \emph{M}$_{\rm dust}$/\emph{M}$_{\star}$ depletes exponentially after the burst ends, the fitting yields a depletion timescale of 205$^{+58}_{-37}$ Myr. Such a timescale is consistent with the \emph{M}$_{\rm H_2}$/\emph{M}$_{\star}$ depletion timescale (117-230 Myr) derived in \citet{2018ApJ...862....2F}. Considering the typical \emph{M}$_{\rm dust}$/\emph{M}$_{\star}$ of our sample at zero $age_{\rm post-burst}$ ($\sim$ -2.5) and the \emph{M}$_{\rm dust}$/\emph{M}$_{\star}$ range of non-dusty ETGs, it should take $\sim$1-2 Gyr for PSBs reach early-type levels of \emph{M}$_{\rm dust}$/\emph{M}$_{\star}$. This result is consistent with previous claims that PSB stellar populations, color gradients, morphologies, kinematics, and molecular gas (\citealt{2001ApJ...557..150N, 2004ApJ...607..258Y, 2008ApJ...688..945Y, 2013MNRAS.432.3131P, 2016MNRAS.456.3032P, 2018ApJ...862....2F}) will resemble the properties of ETGs in a few Gyr.

The derived \emph{M}$_{\rm dust}$/\emph{M}$_{\star}$ depletion timescale may be, at least partly, due to the low-ionization nuclear emission-line region (LINER) or active galactic nucleus (AGN) feedback, as gas consumption by residual star formation would take much longer time \citep{2018ApJ...862....2F}. The depletion time associated with AGN driven outflows in non-AGN-dominated starburst galaxies could be up to several hundred Myr \citep{2014A&A...562A..21C, 2017MNRAS.470.1687B, 2018MNRAS.480.3993B}, which is consistent with our case here.

We do not find any significant correlation between \emph{M}$_{\rm H_2}$/\emph{M}$_{\rm dust}$ and ${\rm age}_{\rm post-burst}$ (Spearman coefficients \emph{r} = 0.11 $\pm$ 0.16, log(\emph{p}) = -0.33 $\pm$ 0.54). The decrease in \emph{M}$_{\rm dust}$/\emph{M}$_{\star}$ and \emph{M}$_{\rm H_2}$/\emph{M}$_{\star}$ with ${\rm age}_{\rm post-burst}$ and the constancy of the gas-to-dust ratio suggest that the dust and gas decline is driven by the same physical mechanism. Furthermore, the close mutual tracking of the gas and dust indicates that the mechanism removes, consumes, or expels the ISM material, instead of merely altering its state and rendering it undetectable.

\subsection{$M_{\rm dust}$ versus SFR and $M_{\rm H_2}$}\label{sec:SFRMd}Molecular gas, interstellar dust, and star formation are strongly correlated with each other in galaxies. The dust grains produced in supernovae can protect molecular hydrogen from UV radiation and contribute to the formation of molecular clouds, which collapse to form new stars. To quantify these relationships in PSBs, we consider here the relations between $M_{\rm dust}$ and SFR and between $M_{\rm dust}$ and molecular gas mass, $M_{\rm H_2}$. The latter relation is a useful calibration to convert $M_{\rm dust}$ into harder-to-measure $M_{\rm H_2}$.

To consider the SFR-$M_{\rm dust}$ relation for our sample, we convert the H$\alpha$ fluxes from the MPA-JHU catalog \citep{2011ApJS..193...29A} to SFR using the relation from \citet{1994ApJ...435...22K}\footnote{We do not use IR-derived SFRs in this paper, because (1) they are prone to overestimation (\citealt{2014MNRAS.445.1598H, 2018ApJ...855...51S}), and (2) we want to avoid the intrinsic correlation between IR-derived $M_{\rm dust}$ and IR-derived SFRs.}. We estimate the amount of internal dust extinction from the observed Balmer decrement, H$\alpha$/H$\beta$. Assuming the hydrogen nebular emission follows Case B recombination, the intrinsic Balmer flux ratio (H$\alpha$/H$\beta$)$_{0}$ = 2.86 for $T_{e}$ = 10$^{4}$ K. Following 
\citet{2015ApJ...801....1F}, we adopt the reddening curve of \citet{1994ApJ...422..158O}\footnote{Using \citet{2000ApJ...533..682C} would yield a negligible difference: A$_{\rm H\alpha}$/A$_{V}$ = 0.82 instead of 0.84, assuming \emph{R}$_V$ = 4.05.}. When the H$\beta$ line flux is uncertain, we follow \citet{2015ApJ...801....1F} using the mean value of \emph{E(B--V)} of the other PSBs in \citet{2015ApJ...801....1F}. The mean attenuation is \emph{A}$_V$ = 0.92 mag (or \emph{A}$_{\rm H\alpha}$ = 0.77 mag). 

We further correct for potential underlying AGN contribution to H$\alpha$ fluxes following the methodology from \citet{2010MNRAS.405..933W}. We calculate the emission-line ratios  [O III]$\lambda$5007/H$\beta$ and [N II]$\lambda$6583/H$\alpha$ of our PSBs to pinpoint them on the BPT diagram \citep{1981PASP...93....5B, 1987ApJS...63..295V}, and determine the AGN contribution to their H$\alpha$ luminosities. In some cases, a negative SFR correction factor is derived, of which the corresponding 1$\sigma$ upper limit is positive. We designate the SFR in these cases to be 1$\sigma$ upper limit values. In addition, when the H$\beta$ line is not well detected, we use its uncertainty as the 1$\sigma$ flux upper limit to determine the corresponding 1$\sigma$ upper limit for the correction factor. Note that all the SFR 1$\sigma$ upper limits are due to AGN correction, instead of low S/N H$\alpha$ detections. In fact, all of our PSBs have $\geq$ 9$\sigma$ H$\alpha$ detections, except for two 5$\sigma$ detections (EAS07 and EAS08) and two 4$\sigma$ detections (EAS10 and EAH10). All the SFR values are provided in Table \ref{tab:results}.

\begin{figure}
\includegraphics[width=\linewidth, clip]{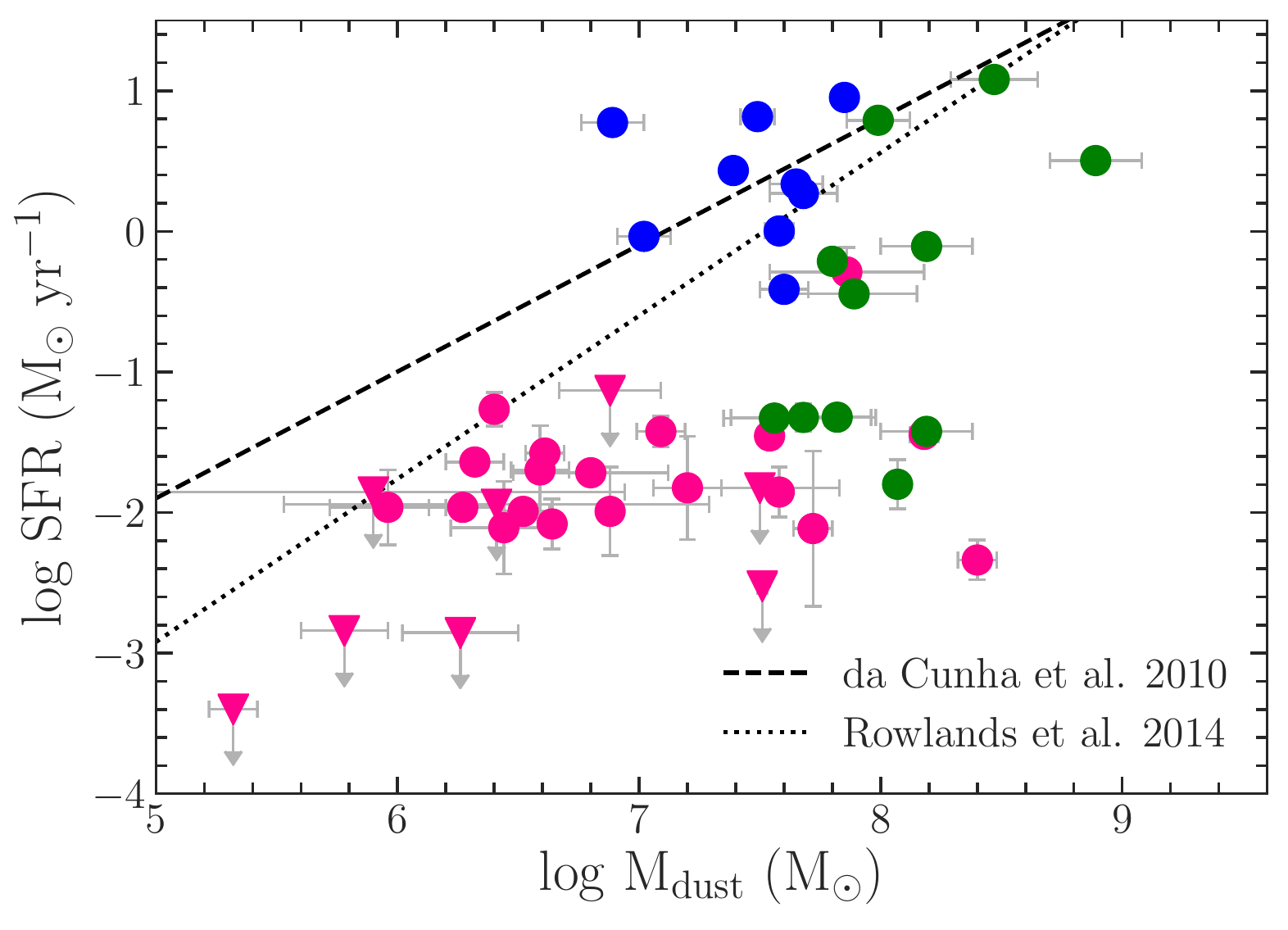}
\caption{Star formation rate (SFR) vs. dust mass ($M_{\rm dust}$) relation. The data points are colored as in Figure 3. The Spearman coefficients are \emph{r} = 0.56 $\pm$ 0.12 and log(\emph{p}) = -3.70 $\pm$ 1.34, indicating a significant correlation. The black dashed line is derived by fitting 1658 local star-forming galaxies from \citet{2010MNRAS.403.1894D}. The black dotted line is a fit to 843 \emph{z} $<$ 0.5 H-ATLAS star-forming galaxies from \citet{2014MNRAS.441.1017R}. The triangles pointing downwards are objects with SFR 1$\sigma$ upper limits. Even only within the \citet{2018ApJ...862....2F} sample (red points), PSBs with the lowest $M_{\rm dust}$ tend to have the lowest SFRs.}
\label{fig:SFR-Mdust}
\end{figure}

Our results are shown in Figure \ref{fig:SFR-Mdust}. For comparison, we plot the lines fit to 1658 local star-forming galaxies from \citet{2010MNRAS.403.1894D} and to 843 \emph{z} $<$ 0.5 H-ATLAS star-forming galaxies from \citet{2014MNRAS.441.1017R}. At fixed \emph{M}$_{\rm dust}$, our PSBs tend to have lower SFR than the local star-forming galaxies. This reflects the nature of PSBs, which have low SFRs by definition. The Spearman coefficients for the PSB log SFR-log $M_{\rm dust}$ relation are \emph{r} = 0.56 $\pm$ 0.12 and log(\emph{p}) = -3.70 $\pm$ 1.34, indicating a significant correlation. Even for the \citet{2018ApJ...862....2F} sample alone, PSBs with the lowest $M_{\rm dust}$ generally have the lowest SFRs.

\begin{figure}
\includegraphics[width=\linewidth, clip]{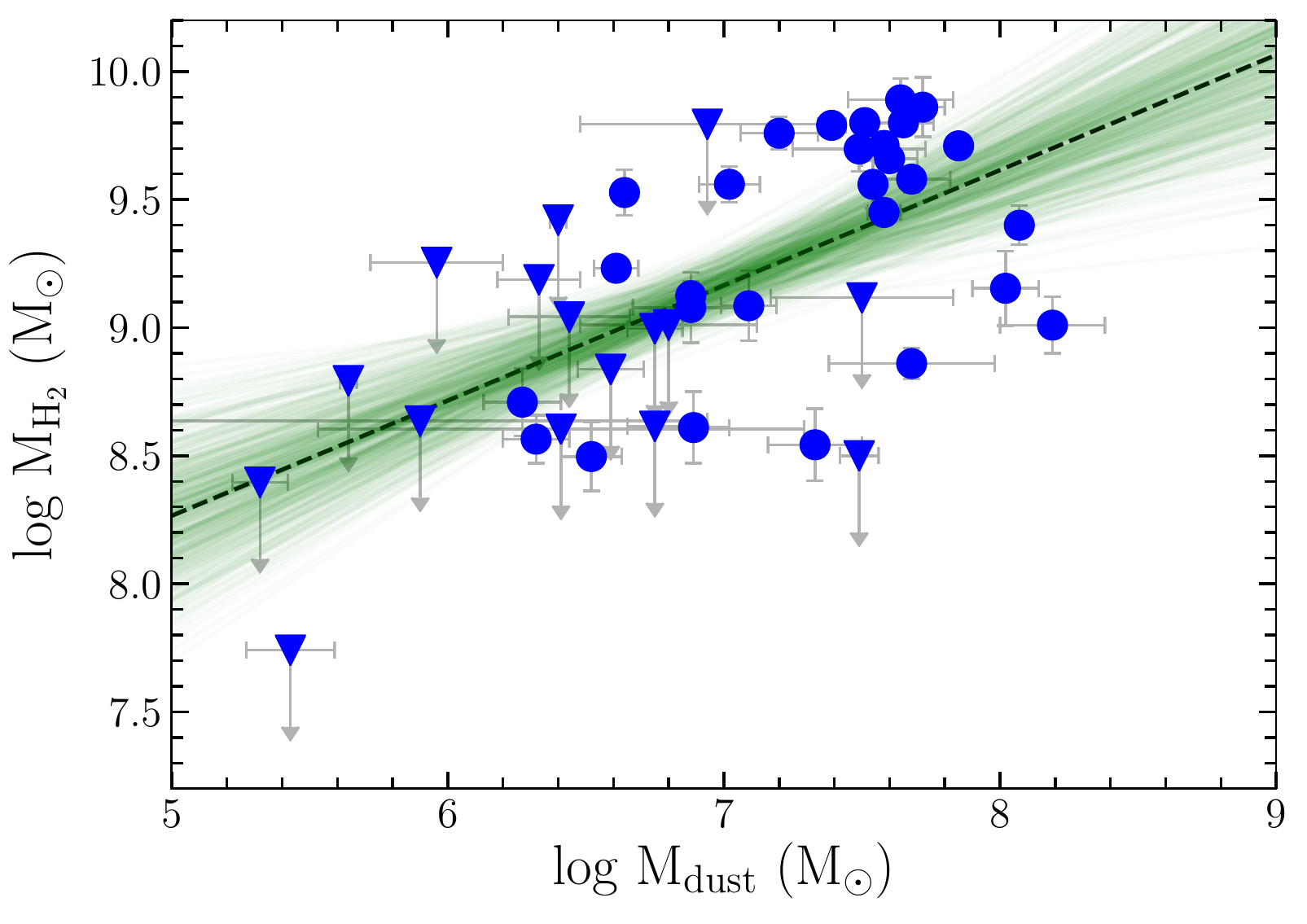}
\caption{The $M_{\rm dust}$-$M_{\rm H_2}$ relation for our 44 PSBs with CO detections (blue solid circles) and 3$\sigma$ upper limits (blue solid triangles). The black dashed line is the best fit to the relation log $M_{\rm H_2}$ = \emph{a} $\cdot$ log $M_{\rm dust}$ + \emph{b} + $\epsilon$, where \emph{a} = 0.45 $\pm$ 0.10, \emph{b} = 6.02 $\pm$ 0.68, and the intrinsic scatter $\epsilon$ = 0.42 $\pm$ 0.05, derived via linear regression \citep{2007ApJ...665.1489K}. The green lines represent 500 evenly spaced samples from the posterior distribution of the model parameters. There is a tight correlation between $M_{\rm dust}$ and $M_{\rm H_2}$, with Spearman coefficients \emph{r} = 0.69 $\pm$ 0.08 and log(\emph{p}) = -5.15 $\pm$ 1.03, which is then useful in estimating $M_{\rm H_2}$ from easier-to-measure $M_{\rm dust}$.}
\label{fig:calibration}
\end{figure}

Next we fit the relationship between $M_{\rm dust}$ and $M_{\rm H_2}$ for those 44 PSBs with both dust and CO measurements (the latter from  \citealt{2015ApJ...801....1F}, \citealt{2015MNRAS.448..258R}, and \citealt{2016ApJ...827..106A}). We adopt the linear regression method from \citet{2007ApJ...665.1489K}, which takes both detections and upper limits into account. The best-fitting result is in the form of log $M_{\rm H_2}$ = \emph{a} $\cdot$ log $M_{\rm dust}$ + \emph{b} + $\epsilon$, where \emph{a} = 0.45 $\pm$ 0.10, \emph{b} = 6.02 $\pm$ 0.68, and intrinsic scatter $\epsilon$ = 0.42 $\pm$ 0.05 (see Figure \ref{fig:calibration}). The Spearman coefficients are \emph{r} = 0.69 $\pm$ 0.08 and log(\emph{p}) = -5.15 $\pm$ 1.03. We use this significant correlation between $M_{\rm dust}$ and $M_{\rm H_2}$ as a calibration, applying it to those 14 PSBs without molecular gas measurements to derive their $M_{\rm H_2}$.

\subsection{Kennicutt-Schmidt relation}\label{sec:KS}
\begin{figure}
\includegraphics[width=\linewidth, clip]{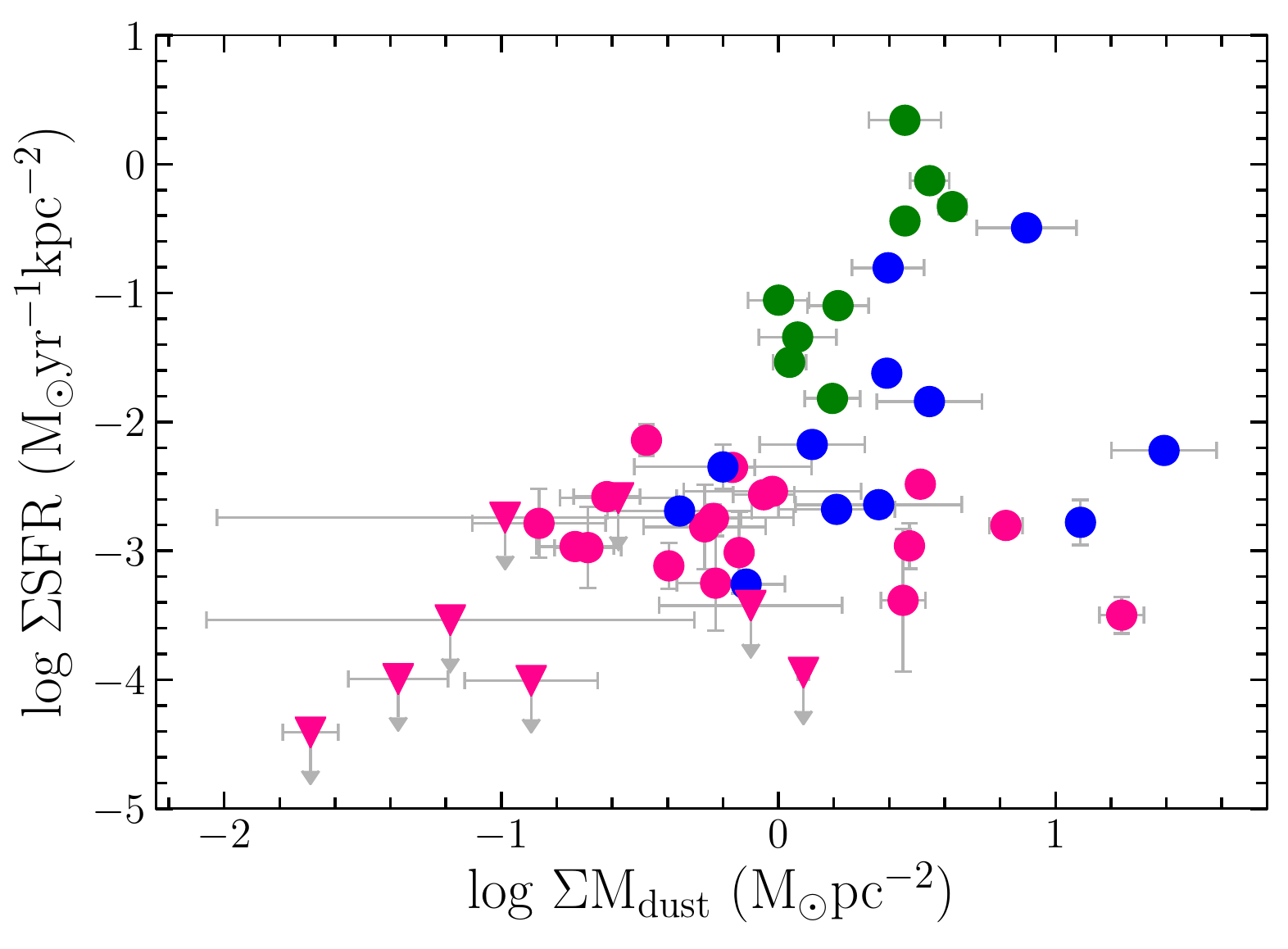}
\includegraphics[width=\linewidth, clip]{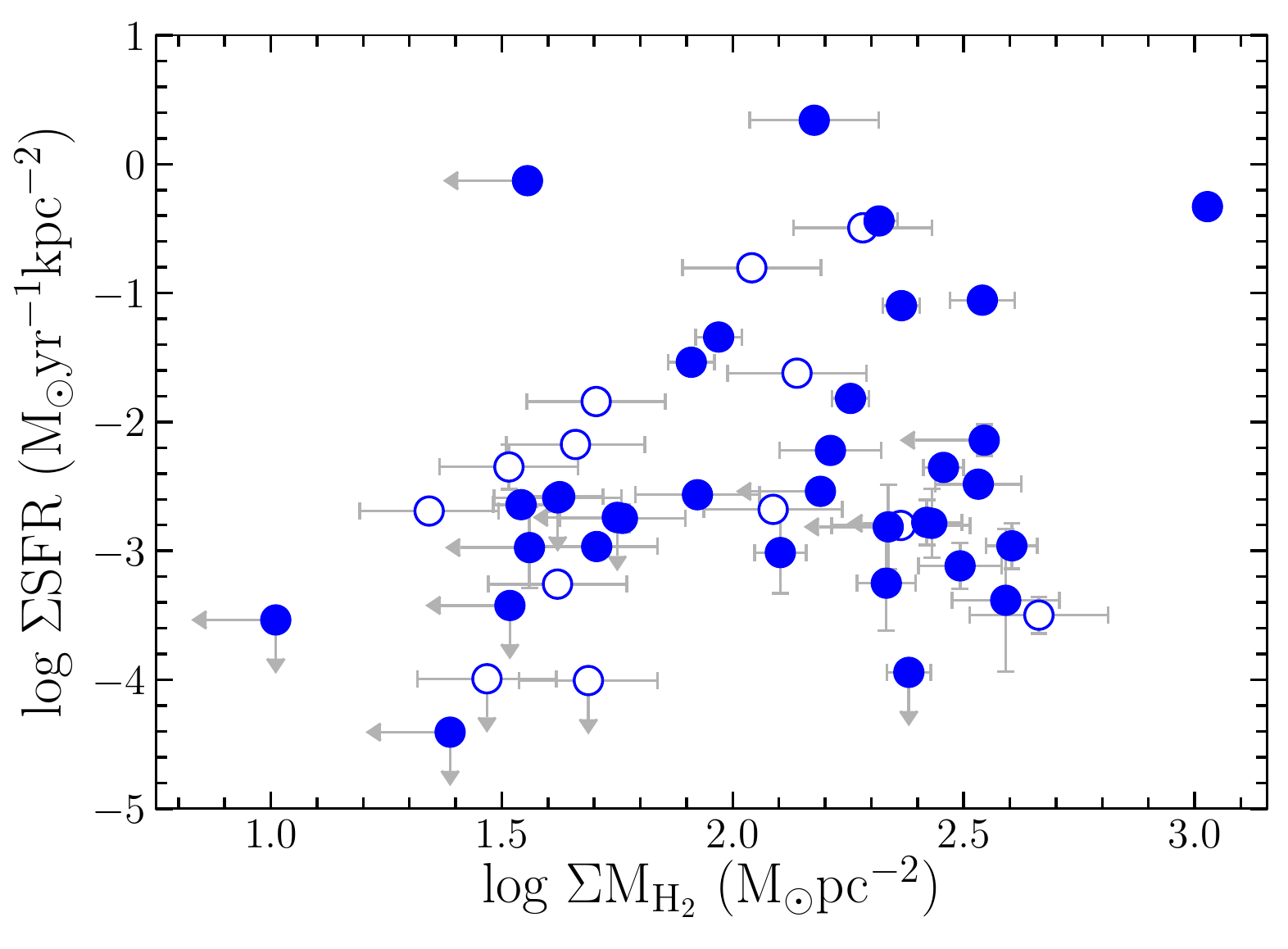}
\caption{The $\Sigma$SFR-$\Sigma$$M_{\rm dust}$ and $\Sigma$SFR-$\Sigma$$M_{\rm H_2}$ planes for our PSBs. Upper: the points are color-coded as in Figure 3. 1$\sigma$ upper limits are marked with \emph{y}-axis arrows. Lower: all the PSBs with CO detections or 3$\sigma$ upper limits (\emph{x}-axis arrows) from Figure \ref{fig:calibration} are marked with filled circles, and the PSBs with $M_{\rm H_2}$ deduced from the $M_{\rm dust}$-$M_{\rm H_2}$ calibration are open circles. Only those PSBs with SFR detections or 1$\sigma$ upper limits (\emph{y}-axis arrows) are shown. The Spearman coefficients are \emph{r} = 0.54 $\pm$ 0.12, log(\emph{p}) = -3.52 $\pm$ 1.26 for the $\Sigma$SFR-$\Sigma M_{\rm dust}$ relation, and \emph{r} = 0.11 $\pm$ 0.17, log(\emph{p}) = -0.34 $\pm$ 0.59 for the $\Sigma$SFR-$\Sigma M_{\rm H_2}$ relation.} 
\label{fig:density}
\end{figure}

The relationships among SFR, $M_{\rm dust}$, and $M_{\rm H_2}$ for PSBs motivate us to explore the KS relation\footnote{The version of the KS relation \citep{1998ApJ...498..541K}
for SFR and $M_{\rm H_2}$ in local normal disk galaxies is $\Sigma$SFR $\propto$ $\Sigma M_{\rm H_2}^{1.0}$ \citep{2013AJ....146...19L}, where the exact slope is sensitive to the tracer (e.g., \citealt{2004ApJ...606..271G}) and the CO-to-H$_2$ conversion factor (e.g., \citealt{2013ARA&A..51..207B}).}. With our $M_{\rm dust}$-$M_{\rm H_2}$ calibration, we deduce $M_{\rm H_2}$ for those PSBs without CO detections, and use the SDSS Petrosian radius\footnote{Without resolved IR, CO, and H$\alpha$ observations for all of our sample, we assume that the dust, molecular gas, and star formation regions are roughly comparable in size and lie within $\emph r_{50}$. We do know that, for a subsample of these galaxies, the dust typically subtends a radius 3-4$\times$ smaller than $\emph r_{50}$ \citep{2018ApJ...855...51S} and that, for four (EAS02, 04, 11, 13), the H$\alpha$ emission does not extend much beyond $\emph r_{50}$ in archival MaNGA \citep{2015ApJ...798....7B} data.
} (\emph r$_{50}$) to convert $M_{\rm dust}$, $M_{\rm H_2}$, and SFR into surface mass densities. The resulting $\Sigma$SFR-$\Sigma M_{\rm dust}$ and $\Sigma$SFR-$\Sigma M_{\rm H_2}$ relations are in Figure \ref{fig:density}.

\begin{figure*}
\includegraphics[width=\linewidth, clip]{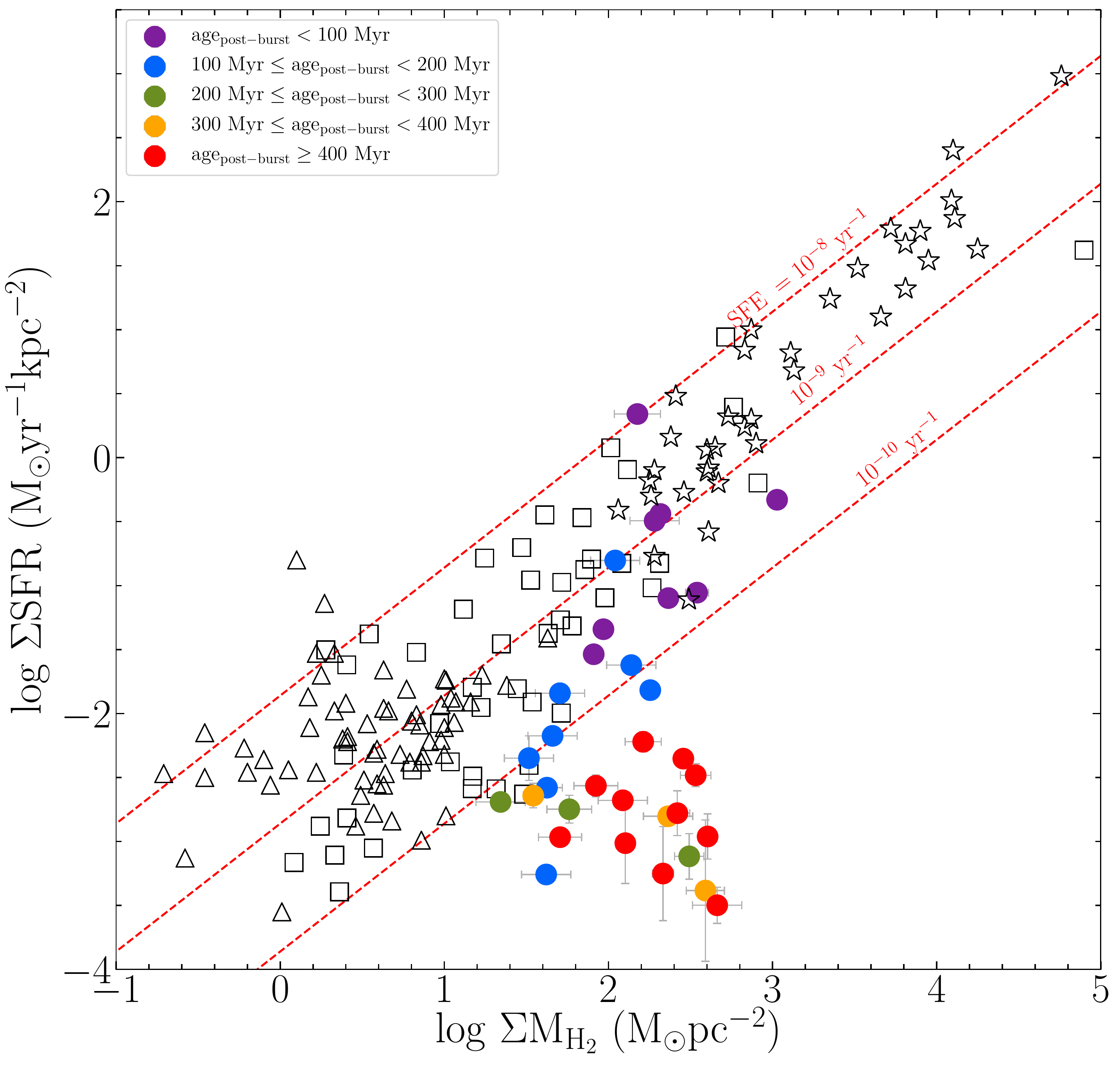}
\caption{Time evolution of PSBs in the Kennicutt-Schmidt plane (the
\emph{x}-axis here is defined as in \citealt{2013AJ....146...19L}). The locus of the PSB sample (filled circles) lies below the KS relation for other galaxies (see \citealt{2015ApJ...801....1F}). Normal star forming (open, black triangles) and starburst (stars) galaxies are from \citet{1998ApJ...498..541K}. ETGs (open squares) are from \citet{2014MNRAS.444.3427D}. Our PSB sample is binned by increasing post-burst ages, from purple to red. To be conservative, we include only detections, not the upper limits from Figure \ref{fig:density}. The PSBs evolve down off the relation over time, due to the faster decrease in $\Sigma$SFR relative to $\Sigma M_{\rm H_2}$, especially during the first 200-300 Myr. This evolution implies a changing SFE. We show three different, constant SFEs (red, dashed lines): the consumption of 1\%, 10\%, and 100\% of the gas reservoir by star formation per 10$^8$ yr, or, equivalently, 10$^{-10}$ -- 10$^{-8}$ $\rm {yr}^{-1}$. Later, $\sim$300 Myr after the burst, the SFE reaches and remains at a low value, $\sim 10^{-11}\ \rm {yr}^{-1}$.}
\label{fig:SFE}
\end{figure*}

Next we compare our 33 PSBs that have SFR detections and $M_{\rm H_2}$ measurements from either CO emission or the $M_{\rm dust}$-$M_{\rm H_2}$ calibration with normal star-forming galaxies and starbursts from \citet{1998ApJ...498..541K} and with ETGs from \citet{2014MNRAS.444.3427D}. For the ETG sample, we recalculate $\Sigma {\rm SFR}$ and $\Sigma M_{\rm H_2}$ using \emph r$_{50}$ to ensure a direct comparison to the PSB sample. For the \citet{1998ApJ...498..541K} sample, we use the original surface densities normalized by the RC2 isophotal radius (where the \emph{B}-band surface brightness drops to 25 mag arcmin$^{-2}$), which is comparable to the H$\alpha$ emitting region for normal spiral galaxies, as the \emph r$_{50}$ is unavailable in the SDSS. \citet{2015ApJ...801....1F} explored the effects of assuming different radii and found consistent results.

Figure \ref{fig:SFE} shows that the locus of our PSBs lies below the KS relation for the other galaxies (as was seen by \citealt{2015ApJ...801....1F}). Scaling the surface densities of these galaxies with a different radius would move them along the KS relation, which does not eliminate the observed offset for our PSBs. Remarkably, when we consider their post-burst ages, the PSBs evolve downward during the first 200-300 Myr, due to the faster decrease in $\Sigma$SFR relative to $\Sigma M_{\rm H_2}$. This evolution also implies a decreasing SFE, defined here as $\Sigma {\rm SFR}/\Sigma M_{\rm H_2}$ (or ${\rm SFR}/M_{\rm H_2})$. The SFE later reaches and remains at a low value, $\sim 10^{-11}\ \rm {yr}^{-1}$, $\geq 300$ Myr after the burst.

\subsection{Star formation efficiency}\label{sec:SFE}
\begin{figure}
\includegraphics[width=\linewidth, clip]{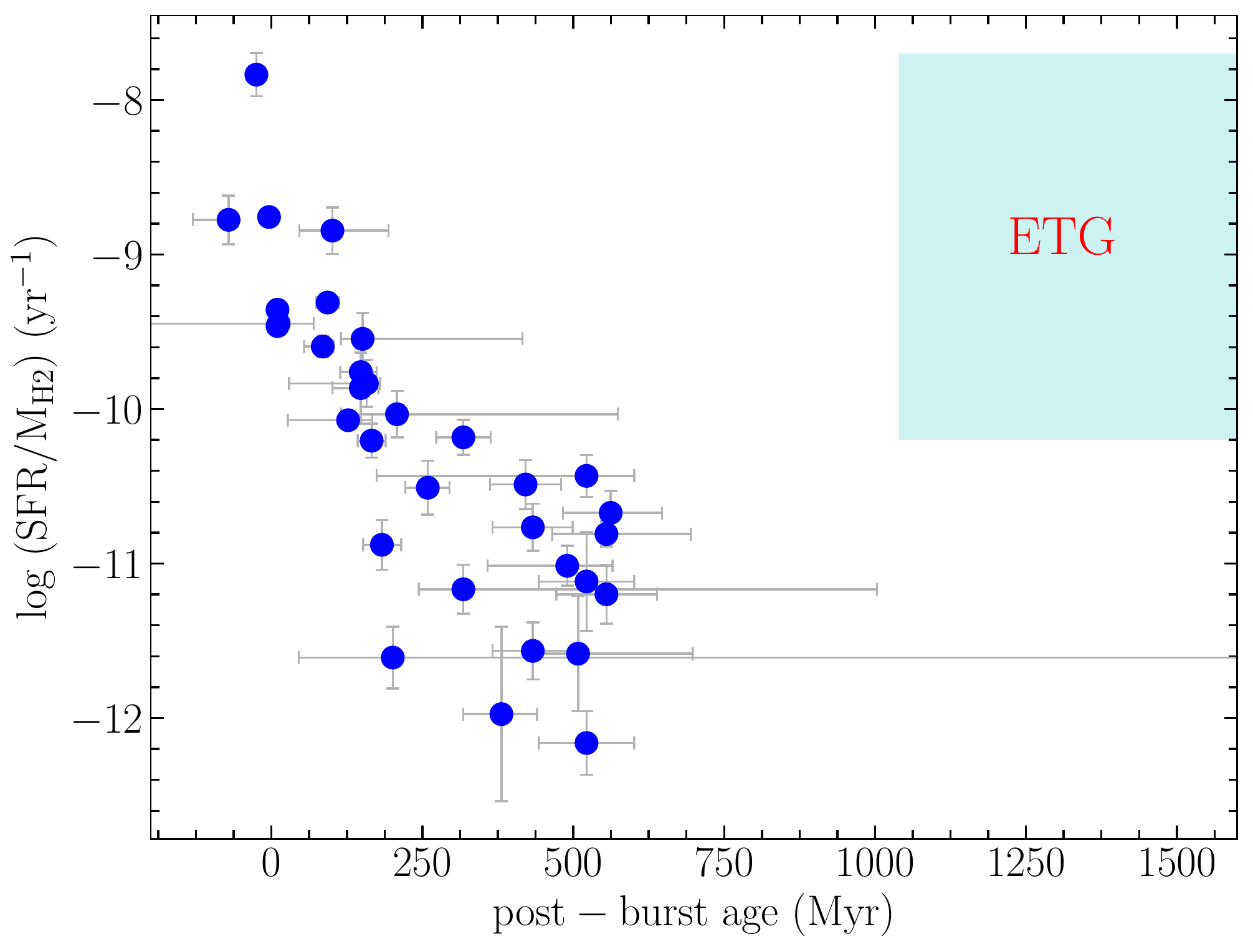} 
\caption{Time evolution of the PSB star formation efficiency. Galaxies are the same as in Figure \ref{fig:SFE}. There is a significant initial decline in SFE with post-burst age; the Spearman coefficients are \emph{r} = -0.88 $\pm$ 0.07, log(\emph{p}) = -3.70 $\pm$ 0.52 for PSBs with ${\rm age}_{\rm post-burst}$ $<$ 300 Myr. After $\sim$200-300 Myr, the SFE decrease slows, reaching an apparent floor of $\sim 10^{-11}\ \rm {yr}^{-1}$. If we assume that the SFR reaches a constant value after $\sim$500 Myr, and that the depletion rate of $M_{\rm H_2}$ does not change, the SFE of PSBs would rise to ETG levels (cyan region; \citealt{2014MNRAS.444.3427D}) in another $\sim$0.5-1.1 Gyr (or $\sim$1-1.6 Gyr after the recent burst ends). This timescale is consistent with the evolution of other PSB properties into ETGs.}
\label{fig:SFE-age}
\end{figure}

The evolution of SFE for our PSB sample is shown directly in Figure \ref{fig:SFE-age}. SFE drops significantly with post-burst age. After the first $\sim$200-300 Myr, the SFE decline slows\footnote{The apparent SFE floor is not due to a limit in our SFR measurement sensitivity: across the ranges of SFEs and post-burst ages plotted here, the S/N of the H$\alpha$ detections is similar and high (see section \ref{sec:SFRMd}).}. To ultimately reach the SFE level of ETGs, there must be an increase in SFE later.

The fast initial SFE decline arises from SFR decreasing more quickly than $M_{\rm H_2}$. The decoupling of SFR from $M_{\rm H_2}$ suggested by this result is consistent with \citet{2018ApJ...862....2F}, who found that the decline of $\emph{M}_{\rm H_2}$/\emph{M}$_{\star}$ is too quick to arise from consumption by star formation and is similar to the observed outflow rates of AGN/LINERs. Such outflows may not only drive the $\emph{M}_{\rm H_2}$/\emph{M}$_{\star}$ decline, but also prevent the large CO-traced molecular gas reservoirs from collapsing and forming denser, star-forming clouds. The rapid drop in SFR is in fact likely due to the absence of denser gas (as traced by HCO$^+$/HCN; \citealt{2018ApJ...861..123F}). Any successful feedback model will need to reproduce these behaviors, namely that $\emph{M}_{\rm H_2}$/\emph{M}$_{\star}$ and $\emph{M}_{\rm dust}$/\emph{M}$_{\star}$ decline similarly over a timescale of several hundred Myr and that denser gas and SFR decline faster than $\emph{M}_{\rm H_2}$ and $\emph{M}_{\rm dust}$.

The slowing of the SFR decline, after the first $\sim$200-300 Myr, was also seen in Figure \ref{fig:SFE}.
If we assume that the SFR's value at $\sim$500 Myr remains constant thereafter,
and that the depletion of $M_{\rm H_2}$ continues \citep{2018ApJ...862....2F}, then the SFE will rise to ETG levels within $\sim$0.5-1.1 Gyr (or equivalently, $\sim$1-1.6 Gyr after the most recent burst ends).

\section{Conclusions}\label{sec:summary}
By performing UV-FIR SED fitting for a sample of 58 PSBs, we have determined \emph{M}$_{\rm dust}$ and quantified the relationship between \emph{M}$_{\rm dust}$ and CO-traced $M_{\rm H_2}$. We have also observed evolution with post-burst age in \emph{M}$_{\rm dust}$/\emph{M}$_{\star}$, SFE, and the KS plane. Our main results are:

(1) There is a significant anticorrelation between the \emph{M}$_{\rm dust}$/\emph{M}$_{\star}$ and the time elapsed since the end of the recent starburst (${\rm age}_{\rm post-burst}$), indicating that the dust is either destroyed, expelled, or rendered undetectable over the $\sim$1 Gyr after the burst. Assuming that the \emph{M}$_{\rm dust}$/\emph{M}$_{\star}$ depletes exponentially after the burst ends yields a depletion timescale of 205$^{+58}_{-37}$ Myr. This timescale is consistent with the CO-traced \emph{M}$_{\rm H_2}$/\emph{M}$_{\star}$ depletion timescale \citep{2018ApJ...862....2F}, suggesting that these dust and molecular gas evolution trends are real and due to the same mechanism. Intriguingly, this observed decline will reduce the dust and CO reservoirs of a PSB to that of an ETG within 1-2 Gyr, when the PSB stellar populations, color gradients, morphologies, and kinematics will likewise resemble those of ETGs.

(2) We determine the $M_{\rm dust}$-$M_{\rm H_2}$ relation from our 44 PSBs with both $M_{\rm dust}$ and CO detections, and apply this calibration to estimate $M_{\rm H_2}$ for the remainder of the sample. We then place the PSBs in the KS plane and find that over time, they move down and away from the KS relation defined by normal star-forming galaxies and starbursts. This evolution is principally due to a rapid drop in the SFR, at least for the first 200-300 Myr after the burst ends.

(3) Direct examination of the evolution of SFE (the ratio of SFR to the CO-traced $M_{\rm H_2}$) reveals a sharp drop during those first 200-300 Myr, i.e., the SFR is decoupled from the $M_{\rm H_2}$ and declines faster. The decrease in SFR in PSBs is likely due to the absence of denser gas \citep{2018ApJ...861..123F}. It is possible that the same mechanism responsible for the decline in $M_{\rm H_2}$/\emph{M}$_{\star}$ and $M_{\rm dust}$/\emph{M}$_{\star}$, whose common short timescale is consistent with AGN/LINER feedback, also prevents the large CO-traced molecular gas reservoirs from collapsing and forming denser, star-forming clouds. After $\sim$200-300 Myr, the $M_{\rm H_2}$ continues to decline, but the SFR levels off, suggesting an SFE floor of 10$^{-11}\ \rm yr^{-1}$. If we assume that the SFR remains constant at this late level, and that the depletion rate of $M_{\rm H_2}$ does not change, then the SFE will rise to ETG levels within $\sim$0.5-1.1 Gyr, a timescale consistent with the evolution of other PSB properties into ETGs.

\acknowledgments
We are grateful to the anonymous referee for carefully reading our manuscript and prviding constructive feedback, which substantially helped improving the quality of this paper. We thank Adam Smercina, Daniel A. Dale, and Dennis Zaritsky for constructive suggestions and helpful discussions. We also thank Yuguang Chen and Michael Zhang for help in installing the \texttt{asurv} package and debugging. Z.L. is grateful for support from the Study Abroad Scholarship for Excellent Students by China Scholarship Council for undergraduate research. K.D.F is supported by program HST-HF2-51391.001-A, provided by NASA through a grant from the Space Telescope Science Institute, which is operated by the Association of Universities for Research in Astronomy, Incorporated, under NASA contract NAS5-26555. A.I.Z. acknowledges funding from NASA grant ADP-NNX10AE88G. L.C.H was supported by the National Key R\&D Program of China (2016YFA0400702) and the National Science Foundation of China (11721303).

This work has made use of the data obtained by \emph{GALEX}, SDSS, 2MASS, \emph{WISE} and \emph{Herschel}. Funding for the Sloan Digital Sky Survey IV has been provided by the Alfred P. Sloan Foundation, the U.S. Department of Energy Office of Science, and the Participating Institutions. SDSS-IV acknowledges
support and resources from the Center for High-Performance Computing at
the University of Utah. The SDSS website is http://www.sdss.org/. SDSS-IV is managed by the Astrophysical Research Consortium for the 
Participating Institutions of the SDSS Collaboration including the 
Brazilian Participation Group, the Carnegie Institution for Science, 
Carnegie Mellon University, the Chilean Participation Group, the French Participation Group, Harvard-Smithsonian Center for Astrophysics, 
Instituto de Astrof\'isica de Canarias, The Johns Hopkins University, 
Kavli Institute for the Physics and Mathematics of the Universe (IPMU) / 
University of Tokyo, Lawrence Berkeley National Laboratory, 
Leibniz Institut f\"ur Astrophysik Potsdam (AIP),  
Max-Planck-Institut f\"ur Astronomie (MPIA Heidelberg), 
Max-Planck-Institut f\"ur Astrophysik (MPA Garching), 
Max-Planck-Institut f\"ur Extraterrestrische Physik (MPE), 
National Astronomical Observatories of China, New Mexico State University, 
New York University, University of Notre Dame, 
Observat\'ario Nacional / MCTI, The Ohio State University, 
Pennsylvania State University, Shanghai Astronomical Observatory, 
United Kingdom Participation Group,
Universidad Nacional Aut\'onoma de M\'exico, University of Arizona, 
University of Colorado Boulder, University of Oxford, University of Portsmouth, 
University of Utah, University of Virginia, University of Washington, University of Wisconsin, 
Vanderbilt University, and Yale University. This
publication makes use of data products from the Two Micron
All-Sky Survey, which is a joint project of the University of
Massachusetts and the Infrared Processing and Analysis
Center/California Institute of Technology, funded by the
National Aeronautics and Space Administration and the
National Science Foundation. 
\emph{WISE} is a joint project of the University
of California, Los Angeles, and the Jet Propulsion Laboratory
of California Institute of Technology, funded by the National
Aeronautics and Space Administration. The \emph{WISE} website
is http://wise.astro.ucla.edu/. \emph{Herschel} is an ESA space observatory with science instruments provided by European-led Principal Investigator consortia and with important participation from NASA.

\clearpage
\startlongtable
\begin{deluxetable*}{ccccccccccccc}
\setlength{\tabcolsep}{2.5pt}
\tabletypesize{\tiny}
\tablecaption{{\centering}Archival UV to NIR Photometry\label{tab:data1}}
\tablehead{
\colhead{ID} & \colhead{R.A.}& \colhead{Decl.} &\multicolumn{2}{c}{\emph{GALEX} (mJy)}&\multicolumn{5}{c}{SDSS (mJy)}&\multicolumn{3}{c}{2MASS (mJy)}\\
\colhead{} & \colhead{(J2000 deg)} & \colhead{(J2000 deg)}& \colhead{FUV}& \colhead{NUV}&\colhead{\emph u}&\colhead{\emph g}&\colhead{\emph r}&\colhead{\emph i}&\colhead{\emph z}&\colhead{\emph J}&\colhead{\emph H}&\colhead{\emph Ks}}
\colnumbers
\startdata 
R1    & 233.13199 & 57.88292 & 5.84E-01  & 7.92E-01  & 1.33E+00 & 2.27E+00 & 2.60E+00 & 3.26E+00 & 3.62E+00 & 3.97E+00 & 5.96E+00 & 3.50E+00 \\
&&&(3.40E-02)       & (2.32E-02)       & (6.69E-02)     & (1.14E-01)     & (1.30E-01)     & (1.63E-01)     & (1.82E-01)     & (3.24E-01)     & (4.78E-01)     & (4.62E-01)      \\
R2    & 228.95127 & 20.02236 & 2.28E-01  & 3.50E-01  & 1.33E+00 & 2.56E+00 & 3.93E+00 & 5.40E+00 & 7.32E+00 & 9.27E+00 & 1.35E+01 & 1.31E+01 \\
&&&(1.67E-02)       & (1.22E-02)       & (6.71E-02)     & (1.28E-01)     & (1.97E-01)     & (2.70E-01)     & (3.67E-01)     & (6.24E-01)     & (8.83E-01)     & (9.76E-01)      \\
R3    & 225.40127 & 16.72968 & 6.18E-01  & 9.56E-01  & 1.97E+00 & 3.79E+00 & 4.45E+00 & 5.01E+00 & 5.65E+00 & 7.83E+00 & 7.37E+00 & 6.56E+00 \\
&&&(4.94E-02)       & (3.53E-02)       & (9.89E-02)     & (1.90E-01)     & (2.23E-01)     & (2.51E-01)     & (2.83E-01)     & (4.82E-01)     & (6.17E-01)     & (6.45E-01)      \\
R4    & 246.45527 & 40.34521 & 6.29E-01  & 9.35E-01  & 1.93E+00 & 4.60E+00 & 6.33E+00 & 8.19E+00 & 9.89E+00 & 1.13E+01 & 1.14E+01 & 1.26E+01 \\
&&&(3.38E-02)       & (2.49E-02)       & (9.77E-02)     & (2.30E-01)     & (3.17E-01)     & (4.10E-01)     & (4.96E-01)     & (7.76E-01)     & (9.43E-01)     & (1.13E+00)      \\
R5    & 244.39756 & 14.05230 & 7.48E-02  & 1.50E-01  & 3.64E-01 & 8.82E-01 & 1.48E+00 & 2.00E+00 & 2.52E+00 & 3.73E+00 & 4.70E+00 & 4.58E+00 \\
&&&(7.20E-03)       & (6.11E-03)       & (1.87E-02)     & (4.42E-02)     & (7.39E-02)     & (1.00E-01)     & (1.27E-01)     & (2.88E-01)     & (3.86E-01)     & (4.32E-01)      \\
R6    & 252.92373 & 41.66838 & 1.83E-01  & 2.78E-01  & 7.59E-01 & 2.39E+00 & 3.96E+00 & 5.28E+00 & 6.36E+00 & 7.66E+00 & 8.22E+00 & 7.54E+00 \\
&&&(1.35E-02)       & (1.00E-02)       & (3.86E-02)     & (1.20E-01)     & (1.98E-01)     & (2.64E-01)     & (3.19E-01)     & (5.60E-01)     & (7.75E-01)     & (7.35E-01)      \\
R7    & 249.49529 & 13.85942 & 2.80E-02  & 5.77E-02  & 2.82E-01 & 8.37E-01 & 1.42E+00 & 1.97E+00 & 2.33E+00 & 3.14E+00 & 4.15E+00 & 3.31E+00 \\
&&&(5.29E-03)       & (4.40E-03)       & (1.48E-02)     & (4.19E-02)     & (7.12E-02)     & (9.87E-02)     & (1.17E-01)     & (2.60E-01)     & (4.02E-01)     & (3.71E-01)      \\
R9    & 239.56848 & 52.48926 & 4.20E-02  & 6.23E-02  & 2.04E-01 & 7.16E-01 & 1.32E+00 & 1.85E+00 & 2.32E+00 & 3.22E+00 & 3.69E+00 & 3.21E+00 \\
&&&(5.36E-03)       & (3.81E-03)       & (1.09E-02)     & (3.59E-02)     & (6.59E-02)     & (9.29E-02)     & (1.17E-01)     & (2.91E-01)     & (4.59E-01)     & (3.89E-01)      \\
R10   & 247.17897 & 22.39712 & 3.50E-02  & 6.66E-02  & 1.87E-01 & 7.41E-01 & 1.43E+00 & 2.09E+00 & 2.84E+00 & 4.78E+00 & 4.82E+00 & 4.63E+00 \\
&&&(8.76E-03)       & (7.68E-03)       & (1.36E-02)     & (3.76E-02)     & (7.19E-02)     & (1.05E-01)     & (1.44E-01)     & (3.22E-01)     & (3.72E-01)     & (4.78E-01)      \\
EAH01 & 128.64046 & 17.34621 & 8.34E-03  & 3.69E-02  & 2.43E-01 & 1.24E+00 & 2.06E+00 & 2.60E+00 & 3.22E+00 & 8.42E+00 & 1.05E+01 & 7.52E+00 \\
&&&(1.48E-03)       & (1.89E-03)       & (1.27E-02)     & (2.51E-02)     & (4.17E-02)     & (5.26E-02)     & (9.83E-02)     & (5.67E-01)     & (7.22E-01)     & (7.63E-01)      \\
EAH02 & 141.58038 & 18.67806 & 3.71E-03  & 3.13E-03  & 2.58E-02 & 1.29E-01 & 3.01E-01 & 4.46E-01 & 6.00E-01 & 6.08E-01 & 8.06E-01 & 5.44E-01 \\
&&&(1.38E-03)       & (1.20E-03)       & (2.59E-03)     & (6.58E-03)     & (1.52E-02)     & (2.25E-02)     & (3.15E-02)     & (9.61E-02)     & (6.93E-02)     & (1.08E-01)      \\
EAH03 & 222.06686 & 17.55165 & 1.63E-03  & 1.23E-02  & 6.33E-02 & 4.51E-01 & 1.01E+00 & 1.49E+00 & 1.96E+00 & 3.08E+00 & 3.71E+00 & 3.28E+00 \\
&&&(3.06E-03)       & (3.28E-03)       & (5.16E-03)     & (9.36E-03)     & (2.08E-02)     & (3.03E-02)     & (6.09E-02)     & (2.79E-01)     & (4.25E-01)     & (4.62E-01)      \\
EAH04 & 318.50226 & 0.53511  & 1.98E-02  & 7.23E-02  & 7.33E-01 & 3.16E+00 & 5.65E+00 & 7.71E+00 & 9.37E+00 & 1.18E+01 & 1.41E+01 & 1.12E+01 \\
&&&(1.65E-03)       & (2.14E-03)       & (3.72E-02)     & (6.35E-02)     & (1.14E-01)     & (1.55E-01)     & (2.83E-01)     & (7.33E-01)     & (1.04E+00)     & (9.08E-01)      \\
EAH05 & 184.26012 & 39.07704 & 1.04E-03  & 1.27E-02  & 6.24E-02 & 2.80E-01 & 5.08E-01 & 6.86E-01 & 8.41E-01 & 1.14E+00 & 1.26E+00 & 9.52E-01 \\
&&&(2.40E-03)       & (3.38E-03)       & (3.57E-03)     & (1.41E-02)     & (2.55E-02)     & (3.44E-02)     & (4.29E-02)     & (1.45E-01)     & (1.78E-01)     & (1.93E-01)      \\
EAH06 & 116.45627 & 31.37838 & 2.59E-03  & 3.68E-03  & 3.04E-01 & 1.57E+00 & 3.03E+00 & 4.18E+00 & 5.29E+00 & 7.29E+00 & 9.03E+00 & 7.65E+00 \\
&&&(4.89E-04)       & (5.16E-04)       & (1.57E-02)     & (7.84E-02)     & (1.52E-01)     & (2.09E-01)     & (2.65E-01)     & (4.34E-01)     & (5.71E-01)     & (5.50E-01)      \\
EAH07 & 167.82484 & 11.55439 & 9.85E-02  & 2.63E-01  & 1.19E+00 & 5.19E+00 & 8.54E+00 & 1.08E+01 & 1.30E+01 & 1.79E+01 & 1.94E+01 & 1.64E+01 \\
&&&(8.80E-03)       & (9.57E-03)       & (6.00E-02)     & (2.60E-01)     & (4.27E-01)     & (5.40E-01)     & (6.54E-01)     & (1.05E+00)     & (1.27E+00)     & (1.28E+00)      \\
EAH08 & 147.07782 & 2.50116  &  ...         & 7.08E-03  & 1.08E-01 & 5.47E-01 & 1.11E+00 & 1.58E+00 & 2.05E+00 & 2.40E+00 & 2.76E+00 & 2.78E+00 \\
&&&  ...     & (2.28E-03)       & (6.36E-03)     & (2.75E-02)     & (5.55E-02)     & (7.94E-02)     & (1.04E-01)     & (2.79E-01)     & (3.62E-01)     & (4.43E-01)      \\
EAH09 & 227.22954 & 37.55827 & 7.42E-03  & 3.76E-02  & 3.61E-01 & 1.74E+00 & 3.22E+00 & 4.44E+00 & 5.64E+00 & 7.27E+00 & 9.19E+00 & 6.90E+00 \\
&&&(2.82E-03)       & (3.36E-03)       & (1.85E-02)     & (8.72E-02)     & (1.61E-01)     & (2.22E-01)     & (2.83E-01)     & (4.25E-01)     & (5.80E-01)     & (5.38E-01)      \\
EAH10 & 158.42798 & 21.12799 & 2.87E-03  & 1.41E-02  & 6.51E-02 & 2.79E-01 & 4.94E-01 & 6.17E-01 & 7.71E-01 & 2.30E+00 & 3.49E+00 & 3.03E+00 \\
&&&(5.14E-04)       & (7.24E-04)       & (1.89E-03)     & (1.51E-03)     & (2.23E-03)     & (3.07E-03)     & (7.09E-03)     & (2.82E-01)     & (4.46E-01)     & (5.21E-01)      \\
EAH11 & 166.41962 & 5.99841  & 3.39E-03  & 1.52E-02  & 1.96E-01 & 1.08E+00 & 2.07E+00 & 2.85E+00 & 3.53E+00 & 4.37E+00 & 4.97E+00 & 4.44E+00 \\
&&&(3.16E-03)       & (3.80E-03)       & (1.03E-02)     & (5.41E-02)     & (1.04E-01)     & (1.42E-01)     & (1.77E-01)     & (2.83E-01)     & (2.84E-01)     & (2.48E-01)      \\
EAH12 & 223.77269 & 13.28101 &  ...         & 7.95E-03  & 6.02E-02 & 3.48E-01 & 7.16E-01 & 1.03E+00 & 1.32E+00 & 1.69E+00 & 1.87E+00 & 1.74E+00 \\
&&&  ...     & (1.03E-03)       & (3.60E-03)     & (1.75E-02)     & (3.59E-02)     & (5.18E-02)     & (6.67E-02)     & (9.50E-02)     & (2.30E-01)     & (1.37E-01)      \\
EAH13 & 155.50328 & 22.16318 & 1.99E-02  & 5.26E-02  & 2.56E-01 & 1.03E+00 & 1.89E+00 & 2.44E+00 & 3.02E+00 & 3.98E+00 & 5.78E+00 & 4.75E+00 \\
&&&(3.48E-03)       & (3.62E-03)       & (1.34E-02)     & (2.09E-02)     & (3.82E-02)     & (4.94E-02)     & (9.19E-02)     & (2.97E-01)     & (4.51E-01)     & (4.52E-01)      \\
EAH14 & 178.27686 & 64.29903 & 3.65E-03  & 1.98E-02  & 9.91E-02 & 4.41E-01 & 7.43E-01 & 9.64E-01 & 1.14E+00 & 1.88E+00 & 1.99E+00 & 2.21E+00 \\
&&&(3.90E-03)       & (4.78E-03)       & (5.44E-03)     & (9.01E-03)     & (1.51E-02)     & (1.96E-02)     & (3.51E-02)     & (2.33E-01)     & (3.33E-01)     & (3.93E-01)      \\
EAH15 & 163.08520 & 5.82822  & 1.46E-02  & 6.10E-02  & 3.78E-01 & 1.71E+00 & 3.07E+00 & 4.14E+00 & 4.97E+00 & 8.31E+00 & 1.13E+01 & 7.83E+00 \\
&&&(1.88E-03)       & (2.48E-03)       & (1.93E-02)     & (3.45E-02)     & (6.19E-02)     & (8.34E-02)     & (1.50E-01)     & (5.70E-01)     & (8.02E-01)     & (7.76E-01)      \\
EAH16 & 141.74037 & 42.52684 & 1.97E-02  & 6.49E-02  & 2.24E-01 & 6.75E-01 & 1.14E+00 & 1.46E+00 & 1.79E+00 & 2.54E+00 & 3.34E+00 & 3.16E+00 \\
&&&(1.25E-03)       & (1.95E-03)       & (1.17E-02)     & (1.37E-02)     & (2.31E-02)     & (2.96E-02)     & (5.48E-02)     & (2.51E-01)     & (4.28E-01)     & (4.26E-01)      \\
EAH17 & 191.21539 & -1.75990 & 4.35E-03  & 8.52E-03  & 7.68E-02 & 3.63E-01 & 7.25E-01 & 1.02E+00 & 1.31E+00 & 1.47E+00 & 1.86E+00 & 1.41E+00 \\
&&&(2.71E-03)       & (2.59E-03)       & (5.08E-03)     & (7.50E-03)     & (1.48E-02)     & (2.09E-02)     & (4.16E-02)     & (1.39E-01)     & (1.65E-01)     & (2.02E-01)      \\
EAH18 & 245.25338 & 21.16836 & 8.73E-03  & 4.26E-02  & 6.43E-01 & 3.02E+00 & 5.62E+00 & 7.51E+00 & 9.01E+00 & 1.34E+01 & 1.94E+01 & 1.28E+01 \\
&&&(3.85E-03)       & (3.63E-03)       & (3.27E-02)     & (1.51E-01)     & (2.81E-01)     & (3.76E-01)     & (4.51E-01)     & (9.00E-01)     & (1.45E+00)     & (1.11E+00)      \\
EAS01 & 11.24684  & -8.88968 & 1.23E-02  & 1.02E-01  & 1.53E+00 & 6.67E+00 & 1.18E+01 & 1.58E+01 & 1.93E+01 & 2.59E+01 & 3.16E+01 & 2.47E+01 \\
&&&(1.74E-03)       & (3.33E-03)       & (7.71E-02)     & (1.34E-01)     & (2.36E-01)     & (3.18E-01)     & (5.80E-01)     & (1.43E+00)     & (1.77E+00)     & (1.67E+00)      \\
EAS02 & 49.22881  & -0.04198 & 6.23E-03  & 3.49E-02  & 4.41E-01 & 2.01E+00 & 3.89E+00 & 5.45E+00 & 6.99E+00 & 7.72E+00 & 9.43E+00 & 8.54E+00 \\
&&&(1.57E-03)       & (1.54E-03)       & (2.28E-02)     & (1.00E-01)     & (1.95E-01)     & (2.73E-01)     & (3.51E-01)     & (5.34E-01)     & (7.82E-01)     & (8.26E-01)      \\
EAS03 & 117.80962 & 34.41820 & 1.09E-02  & 9.09E-03  & 1.36E-01 & 8.93E-01 & 1.91E+00 & 2.79E+00 & 3.58E+00 & 4.65E+00 & 5.31E+00 & 5.09E+00 \\
&&&(3.68E-03)       & (2.12E-03)       & (7.66E-03)     & (4.47E-02)     & (9.57E-02)     & (1.40E-01)     & (1.80E-01)     & (3.16E-01)     & (4.10E-01)     & (4.45E-01)      \\
EAS04 & 126.75582 & 21.70678 & 2.14E-02  & 1.55E-01  & 1.56E+00 & 5.88E+00 & 1.00E+01 & 1.34E+01 & 1.66E+01 & 2.36E+01 & 2.71E+01 & 2.39E+01 \\
&&&(2.14E-03)       & (4.50E-03)       & (7.84E-02)     & (1.18E-01)     & (2.02E-01)     & (2.69E-01)     & (5.01E-01)     & (1.27E+00)     & (1.57E+00)     & (1.40E+00)      \\
EAS05 & 146.11234 & 4.49912  & 1.52E-02  & 4.71E-02  & 2.96E-01 & 1.39E+00 & 2.62E+00 & 3.68E+00 & 4.78E+00 & 6.17E+00 & 7.59E+00 & 5.56E+00 \\
&&&(1.73E-03)       & (2.00E-03)       & (1.54E-02)     & (6.94E-02)     & (1.31E-01)     & (1.84E-01)     & (2.40E-01)     & (4.43E-01)     & (5.55E-01)     & (6.13E-01)      \\
EAS06 & 159.48898 & 46.24451 & 5.55E-04  & 1.72E-02  & 2.18E-01 & 1.18E+00 & 2.56E+00 & 3.81E+00 & 5.03E+00 & 6.74E+00 & 8.13E+00 & 7.34E+00 \\
&&&(2.65E-03)       & (4.05E-03)       & (1.17E-02)     & (2.39E-02)     & (5.15E-02)     & (7.68E-02)     & (1.52E-01)     & (4.22E-01)     & (6.28E-01)     & (5.47E-01)      \\
EAS07 & 169.78174 & 58.05397 & 2.49E-02  & 1.17E-01  & 9.60E-01 & 4.08E+00 & 7.24E+00 & 9.81E+00 & 1.18E+01 & 1.59E+01 & 1.95E+01 & 1.67E+01 \\
&&&(2.01E-03)       & (3.47E-03)       & (6.27E-03)     & (8.26E-03)     & (1.33E-02)     & (1.80E-02)     & (3.25E-02)     & (5.40E-01)     & (1.00E+00)     & (8.14E-01)      \\
EAS08 & 189.90020 & 12.43889 & 2.77E-02  & 7.96E-02  & 7.05E-01 & 2.88E+00 & 5.48E+00 & 7.47E+00 & 9.30E+00 & 1.36E+01 & 1.33E+01 & 1.12E+01 \\
&&&(2.35E-03)       & (2.67E-03)       & (3.61E-02)     & (5.80E-02)     & (1.10E-01)     & (1.50E-01)     & (2.81E-01)     & (7.85E-01)     & (9.38E-01)     & (8.45E-01)      \\
EAS09 & 191.61182 & 50.79206 & 1.88E-02  & 5.76E-02  & 8.32E-01 & 4.18E+00 & 7.86E+00 & 1.10E+01 & 1.38E+01 & 1.95E+01 & 2.58E+01 & 1.96E+01 \\
&&&(2.82E-03)       & (3.53E-03)       & (4.21E-02)     & (8.41E-02)     & (1.58E-01)     & (2.21E-01)     & (4.15E-01)     & (1.08E+00)     & (1.46E+00)     & (1.27E+00)      \\
EAS10 & 196.35760 & 53.59176 & 1.82E-02  & 6.65E-02  & 8.05E-01 & 3.53E+00 & 6.17E+00 & 8.02E+00 & 9.36E+00 & 1.29E+01 & 1.57E+01 & 1.22E+01 \\
&&&(4.35E-03)       & (4.53E-03)       & (5.72E-02)     & (1.90E-01)     & (3.32E-01)     & (4.32E-01)     & (5.46E-01)     & (7.36E-01)     & (9.58E-01)     & (8.55E-01)      \\
EAS11 & 242.58536 & 41.85488 & 6.47E-03  & 9.41E-02  & 1.15E+00 & 4.90E+00 & 8.69E+00 & 1.15E+01 & 1.38E+01 & 1.77E+01 & 2.66E+01 & 2.06E+01 \\
&&&(3.34E-03)       & (4.68E-03)       & (5.81E-02)     & (2.45E-01)     & (4.35E-01)     & (5.74E-01)     & (6.91E-01)     & (1.07E+00)     & (1.62E+00)     & (1.36E+00)      \\
EAS12 & 243.37578 & 51.05988 & 1.39E-01  & 2.46E-01  & 6.27E-01 & 1.72E+00 & 2.19E+00 & 2.54E+00 & 2.79E+00 & 2.46E+00 & 2.30E+00 & 2.02E+00 \\
&&&(7.69E-03)       & (6.60E-03)       & (3.23E-02)     & (8.62E-02)     & (1.10E-01)     & (1.28E-01)     & (1.41E-01)     & (2.66E-01)     & (3.69E-01)     & (4.27E-01)      \\
EAS13 & 246.76067 & 43.47609 & 1.85E-02  & 5.67E-02  & 5.44E-01 & 2.82E+00 & 5.76E+00 & 8.10E+00 & 1.03E+01 & 1.38E+01 & 1.63E+01 & 1.45E+01 \\
&&&(5.61E-03)       & (5.76E-03)       & (3.90E-02)     & (1.52E-01)     & (3.10E-01)     & (4.37E-01)     & (6.02E-01)     & (8.56E-01)     & (1.13E+00)     & (1.17E+00)      \\
EAS14 & 316.28613 & -5.39983 & 4.21E-03  & 2.06E-02  & 2.75E-01 & 1.58E+00 & 3.51E+00 & 5.14E+00 & 6.75E+00 & 8.19E+00 & 9.86E+00 & 8.66E+00 \\
&&&(3.59E-03)       & (3.51E-03)       & (1.54E-02)     & (3.19E-02)     & (7.07E-02)     & (1.04E-01)     & (2.05E-01)     & (5.27E-01)     & (7.62E-01)     & (7.51E-01)      \\
EAS15 & 343.77832 & 0.97776  & 2.76E-02  & 6.67E-02  & 5.01E-01 & 1.97E+00 & 3.91E+00 & 5.55E+00 & 7.10E+00 & 1.52E+01 & 1.57E+01 & 1.35E+01 \\
&&&(6.65E-03)       & (5.97E-03)       & (2.57E-02)     & (3.98E-02)     & (7.86E-02)     & (1.12E-01)     & (2.15E-01)     & (1.04E+00)     & (1.25E+00)     & (1.32E+00)      \\
F34   & 135.03564 & 20.84439 & 3.75E-03  & 1.33E-02  & 1.26E-01 & 5.36E-01 & 9.64E-01 & 1.31E+00 & 1.56E+00 & 1.69E+00 & 2.49E+00 & 2.70E+00 \\
&&&(4.76E-04)       & (6.20E-04)       & (9.44E-03)     & (2.90E-02)     & (5.20E-02)     & (7.08E-02)     & (9.19E-02)     & (2.16E-01)     & (2.89E-01)     & (3.40E-01)      \\
F35   & 155.00089 & 8.22606  & 1.63E-03  & 9.79E-03  & 1.08E-01 & 6.69E-01 & 1.50E+00 & 2.35E+00 & 3.29E+00 & 4.45E+00 & 4.94E+00 & 3.91E+00 \\
&&&(2.05E-03)       & (2.80E-03)       & (6.43E-03)     & (1.37E-02)     & (3.04E-02)     & (4.77E-02)     & (1.01E-01)     & (3.99E-01)     & (5.74E-01)     & (6.05E-01)      \\
F36   & 178.10641 & -1.26746 & 1.30E-03  & 1.11E-02  & 8.27E-02 & 4.20E-01 & 7.80E-01 & 1.28E+00 & 1.38E+00 & 1.89E+00 & 2.35E+00 & 2.33E+00 \\
&&&(4.91E-04)       & (7.01E-04)       & (4.59E-03)     & (8.58E-03)     & (1.58E-02)     & (2.60E-02)     & (4.26E-02)     & (1.80E-01)     & (2.31E-01)     & (1.64E-01)      \\
F37   & 245.25339 & 21.16836 & 6.61E-04  & 4.76E-02  & 6.43E-01 & 3.02E+00 & 5.62E+00 & 7.51E+00 & 9.01E+00 & 1.34E+01 & 1.94E+01 & 1.28E+01 \\
&&&(2.71E-03)       & (3.90E-03)       & (3.27E-02)     & (6.08E-02)     & (1.13E-01)     & (1.51E-01)     & (2.72E-01)     & (9.00E-01)     & (1.45E+00)     & (1.11E+00)      \\
A1    & 183.87215 & 13.73560 & 5.12E-03  & 1.80E-02  & 6.83E-02 & 1.90E-01 & 3.44E-01 & 5.05E-01 & 6.03E-01 & 3.84E-01 & 7.01E-01 & 8.88E-01 \\
&&&(9.14E-04)       & (1.07E-03)       & (5.09E-03)     & (4.19E-03)     & (7.36E-03)     & (1.07E-02)     & (2.11E-02)     & (6.24E-02)     & (1.68E-01)     & (1.28E-01)      \\
A2    & 185.90395 & 8.73016  & 2.09E-02  & 2.81E-02  & 1.07E-01 & 3.02E-01 & 5.09E-01 & 7.39E-01 & 8.85E-01 & 7.28E-01 & 7.51E-01 & 9.84E-01 \\
&&&(2.37E-03)       & (2.01E-03)       & (8.22E-03)     & (6.67E-03)     & (1.10E-02)     & (1.61E-02)     & (3.37E-02)     & (9.37E-02)     & (1.36E-01)     & (1.89E-01)      \\
A3    & 188.27840 & 62.26692 & 8.29E-03  & 1.76E-02  & 7.37E-02 & 2.82E-01 & 6.79E-01 & 1.03E+00 & 1.37E+00 & 1.73E+00 & 1.74E+00 & 2.21E+00 \\
&&&(1.37E-03)       & (1.09E-03)       & (5.09E-03)     & (5.99E-03)     & (1.40E-02)     & (2.14E-02)     & (4.56E-02)     & (2.45E-01)     & (3.71E-01)     & (3.73E-01)      \\
A4    & 197.36469 & 30.17020 & 5.79E-02  & 1.22E-01  & 2.26E-01 & 5.65E-01 & 8.82E-01 & 1.11E+00 & 1.25E+00 & 2.02E+00 & 2.32E+00 & 2.54E+00 \\
&&&(8.83E-03)       & (7.65E-03)       & (1.17E-02)     & (1.15E-02)     & (1.79E-02)     & (2.25E-02)     & (3.85E-02)     & (2.22E-01)     & (3.07E-01)     & (3.49E-01)      \\
A5    & 198.76463 & 24.61884 & 1.39E-02  & 1.61E-01  & 2.00E+00 & 8.59E+00 & 1.62E+01 & 2.25E+01 & 2.76E+01 & 3.78E+01 & 4.86E+01 & 3.86E+01 \\
&&&(3.31E-03)       & (6.55E-03)       & (1.02E-01)     & (1.73E-01)     & (3.27E-01)     & (4.51E-01)     & (8.30E-01)     & (1.99E+00)     & (2.59E+00)     & (2.17E+00)      \\
A6    & 200.09401 & 32.78479 & 3.02E-03  & 1.16E-02  & 5.25E-02 & 1.86E-01 & 3.29E-01 & 5.06E-01 & 5.85E-01 & 7.00E-01 & 7.07E-01 & 8.45E-01 \\
&&&(2.48E-03)       & (2.86E-03)       & (3.87E-03)     & (4.00E-03)     & (6.94E-03)     & (1.06E-02)     & (2.00E-02)     & (9.44E-02)     & (8.81E-02)     & (1.78E-01)      \\
A7    & 200.74947 & 27.11643 & 1.81E-01  & 3.34E-01  & 1.44E+00 & 6.68E+00 & 1.22E+01 & 1.66E+01 & 2.02E+01 & 2.45E+01 & 2.89E+01 & 2.41E+01 \\
&&&(1.01E-02)       & (9.12E-03)       & (7.25E-02)     & (1.34E-01)     & (2.44E-01)     & (3.33E-01)     & (6.07E-01)     & (1.39E+00)     & (1.82E+00)     & (1.55E+00)      \\
A8    & 203.56174 & 34.19415 & 7.41E-03  & 4.05E-02  & 3.75E-01 & 1.61E+00 & 3.23E+00 & 4.68E+00 & 6.07E+00 & 8.40E+00 & 9.55E+00 & 8.32E+00 \\
&&&(3.58E-03)       & (4.53E-03)       & (2.00E-02)     & (3.27E-02)     & (6.51E-02)     & (9.46E-02)     & (1.85E-01)     & (5.82E-01)     & (7.85E-01)     & (7.09E-01)      \\
A9    & 212.76057 & 25.51935 & 1.79E-02  & 7.28E-02  & 6.89E-01 & 1.96E+00 & 4.36E+00 & 6.32E+00 & 9.69E+00 & 1.41E+01 & 1.89E+01 & 1.57E+01 \\
&&&(7.59E-03)       & (8.04E-03)       & (3.49E-02)     & (3.96E-02)     & (8.79E-02)     & (1.27E-01)     & (2.92E-01)     & (8.27E-01)     & (1.16E+00)     & (1.03E+00)      \\
A10   & 213.29910 & -0.39937 & 9.76E-03  & 1.58E-02  & 3.65E-02 & 1.44E-01 & 3.02E-01 & 4.63E-01 & 6.19E-01 & 6.07E-01 & 4.44E-01 & 6.67E-01 \\
&&&(8.46E-04)       & (1.09E-03)       & (3.11E-03)     & (3.20E-03)     & (6.44E-03)     & (9.84E-03)     & (2.23E-02)     & (1.36E-01)     & (1.01E-01)     & (2.38E-01)      \\
A11   & 232.70238 & 55.32884 & 2.25E-03  & 1.20E-02  & 9.41E-02 & 4.26E-01 & 8.03E-01 & 1.12E+00 & 1.38E+00 & 1.99E+00 & 2.14E+00 & 2.13E+00 \\
&&&(2.54E-03)       & (3.15E-03)       & (5.71E-03)     & (8.75E-03)     & (1.64E-02)     & (2.27E-02)     & (4.31E-02)     & (2.13E-01)     & (2.77E-01)     & (3.37E-01)      \\
A12   & 237.80305 & 14.69640 & 2.43E-02  & 5.43E-02  & 3.92E-01 & 1.73E+00 & 3.05E+00 & 4.12E+00 & 5.09E+00 & 5.78E+00 & 6.98E+00 & 6.20E+00 \\
&&&(4.88E-03)       & (4.83E-03)       & (2.02E-02)     & (3.48E-02)     & (6.15E-02)     & (8.30E-02)     & (1.54E-01)     & (4.51E-01)     & (6.32E-01)     & (6.25E-01) 
\enddata
\tablenotetext{}{\scriptsize \textbf{Note.} (1) Object ID. R1-R11 are from \citet{2015MNRAS.448..258R}; EAH01-EAH18, EAS01-EAS15 and F34-F37 are from \citet{2018ApJ...862....2F}, and A1-A12 are from \citet{2016ApJS..224...38A}. (2)-(3) Right ascension and declination. (4)-(5) \emph{GALEX} fluxes. (6)-(10) SDSS fluxes. (11)-(13) 2MASS fluxes. All fluxes are given in mJy. The total flux uncertainties, which correspond to 68\% confidence levels, are given in parentheses when available. No correction for extinction has been applied.}
\end{deluxetable*}

\startlongtable
\begin{deluxetable*}{ccccccccccc}
\tabletypesize{\scriptsize}
\tablewidth{0pt} 
\tablecaption{{\centering}Archival MIR-FIR Photometry\label{tab:data2}}
\tablehead{
\colhead{ID}&\multicolumn{4}{c}{\emph{WISE} (mJy)}&\multicolumn{3}{c}{\emph{Herschel}-PACS (mJy)}&\multicolumn{3}{c}{\emph{Herschel}-SPIRE (mJy)}\\
\colhead{}&\colhead{\emph W1(3.6$\mu$m)}&\colhead{\emph W2(4.5$\mu$m)}&\colhead{\emph W3(12$\mu$m)}&\colhead{\emph W4(22$\mu$m)} &\colhead{70$\mu$m}&\colhead{100$\mu$m}&\colhead{160$\mu$m}&\colhead{250$\mu$m}&\colhead{350$\mu$m}&\colhead{500$\mu$m}}
\colnumbers
\startdata 
R1  & 3.71E+00 & 2.93E+00 & 3.00E+01 & 1.62E+02 & 1.15E+03 & 1.24E+03 & 9.70E+02 & 3.90E+02 & 1.68E+02 & 6.09E+01 \\
&(2.39E-02)     & (2.70E-02)   & (1.66E-01)   & (1.79E+00)   & (2.33E+01) & (3.35E+01)  & (3.16E+01)  & (1.49E+01)  & (1.96E+01)  & (1.72E+01)  \\
R2  & 7.79E+00 & 6.23E+00 & 5.35E+01 & 2.30E+02 & 3.58E+03 & 3.76E+03 & 2.58E+03 & 9.10E+02 & 3.21E+02 & 1.05E+02 \\
&(4.30E-02)     & (4.01E-02)   & (2.95E-01)   & (1.48E+00)   & (5.64E+01) & (8.14E+01)  & (6.16E+01)  & (1.94E+01)  & (1.65E+01)  & (1.88E+01)  \\
R3  & 4.57E+00 & 3.56E+00 & 2.62E+01 & 1.60E+02 & 8.42E+02 & 8.13E+02 & 5.62E+02 & 2.02E+02 & 8.71E+01 &... \\
&(2.94E-02)     & (2.94E-02)   & (1.93E-01)   & (3.24E+00)   & (2.09E+01) & (2.63E+01)  & (4.23E+01)  & (1.78E+01)  & (1.88E+01)  &...   \\
R4  & 6.61E+00 & 4.66E+00 & 3.54E+01 & 8.19E+01 & 1.52E+03 & 1.96E+03 & 2.01E+03 & 8.16E+02 & 3.48E+02 & 1.31E+02 \\
&(3.65E-02)     & (3.00E-02)   & (1.95E-01)   & (9.79E-01)   & (3.41E+01) & (4.79E+01)  & (5.62E+01)  & (1.59E+01)  & (1.86E+01)  & (1.93E+01)  \\
R5  & 3.38E+00 & 2.99E+00 & 3.78E+01 & 1.81E+02 & 1.15E+03 & 1.28E+03 & 1.01E+03 & 4.13E+02 & 1.67E+02 & 5.25E+01 \\
&(2.18E-02)     & (2.48E-02)   & (2.43E-01)   & (2.16E+00)   & (1.85E+01) & (3.15E+01)  & (3.98E+01)  & (1.40E+01)  & (1.31E+01)  & (1.20E+01)  \\
R6  & 3.40E+00 & 2.18E+00 & 1.05E+01 & 2.76E+01 & 4.11E+02 & 5.46E+02 & 6.01E+02 &... &...&...\\
&(1.88E-02)     & (1.60E-02)   & (8.72E-02)   & (9.41E-01)   & (1.22E+01) & (1.79E+01)  & (2.90E+01)  &...   &...   &...   \\
R7  & 1.78E+00 & 1.14E+00 & 7.57E+00 & 1.57E+01 & 1.80E+02 & 2.51E+02 & 2.07E+02 & 9.29E+01 & 2.67E+01 &... \\
&(3.93E-02) & (2.93E-02)   & (2.30E-01)   & (1.18E+00)   & (5.28E+00) & (1.20E+01)  & (2.41E+01)  & (8.90E+00)  & (8.10E+00)  &...   \\
R9  & 1.39E+00 & 8.31E-01 & 3.29E+00 & 5.25E+00 & 1.49E+02 & 2.58E+02 & 3.10E+02 & ... &...&...\\
&(8.96E-03)     & (9.94E-03)   & (3.63E-02)   & (4.49E-01)   & (7.44E+00) & (1.28E+01)  & (1.80E+01)  &...&...&...\\
R10 & 2.74E+00 & 1.92E+00 & 1.09E+01 & 2.33E+01 & 7.44E+02 & 1.01E+03 & 9.57E+02 & 4.06E+02 & 1.57E+02 & 6.07E+01 \\
&(1.51E-02)     & (2.12E-02)   & (9.04E-02)   & (9.87E-01)   & (2.02E+01) & (2.96E+01)  & (4.46E+01)  & (1.29E+01)  & (1.38E+01)  & (1.31E+01)  \\
F34 & 1.03E+00 & 6.20E-01 & 1.30E+00 & 4.05E+00 &...& 3.33E+01 &...&...& 5.93E+01 & 4.21E+01 \\
&(6.64E-02) & (4.20E-02)   & (1.72E-01)   & (1.22E+00)   & ... & (1.11E+01)  &...&...& (1.06E+01)  & (9.85E+00)  \\
F35 & 1.88E+00 & 1.45E+00 & 1.41E+01 & 1.98E+02 & 6.17E+03 & 5.59E+03 & 3.12E+03 & 1.20E+03 & 4.70E+02 & 1.70E+02 \\
&(1.30E-01) & (9.98E-02)   & (9.55E-01)   & (1.42E+01)   & (4.40E+02) & (4.03E+02)  & (2.24E+02)  & (8.61E+01)  & (3.54E+01)  & (1.70E+01)  \\
F36 & 1.34E+00 & 8.88E-01 & 3.57E+00 & 1.34E+01 &...&...& 9.37E+01 & 6.09E+01 & 5.03E+01 &...          \\
&(8.61E-02)     & (5.90E-02)   & (2.68E-01)   & (1.52E+00)   &...  &...   & (6.01E+01)  & (1.56E+01)  & (1.52E+01)  &...   \\
F37 & 4.71E+00 & 2.61E+00 & 1.95E+00 & 4.39E+00 & 5.90E+01 & 7.77E+01 & 6.48E+01 &...&...&...\\
&(2.84E-01)     & (1.58E-01)   & (1.44E-01)   & (8.73E-01)   & (6.37E+00) & (1.07E+01)  & (1.27E+01)  &...&...&...\\
A1  & 4.02E-01 & 3.34E-01 & 1.95E+00 & 7.02E+00 &...  & 7.48E+01 & 8.46E+01 & 4.84E+01 & ... &...          \\
&(2.66E-02)     & (2.50E-02)   & (1.83E-01)   & (1.19E+00)   & ... & (1.33E+01)  & (2.77E+01)  & (1.52E+01)  &...   &...   \\
A2  & 4.48E-01 & 3.07E-01 & 1.91E+00 & 3.20E+00 &... & 7.60E+01 & 1.47E+02 & 7.39E+01 & ... &...\\
&(2.98E-02)     & (2.37E-02)   & (1.89E-01)   & (1.10E+00)   & ... & (1.52E+01)  & (2.88E+01)  & (1.27E+01)  &  ...&...   \\
A3  & 9.19E-01 & 7.44E-01 & 2.18E+00 & 6.41E+00 &...         & ...& ... & 1.31E+02 & 5.86E+01 & 3.41E+01 \\
&(5.91E-02) & (4.83E-02)   & (1.82E-01)   & (1.07E+00)   &...  &...   &...   & (1.35E+01)  & (1.15E+01)  & (7.68E+00)  \\
A4  & 1.23E+00 & 9.71E-01 & 6.23E+00 & 1.74E+01 &... &...    &...& 1.04E+02 & 8.53E+01 & 6.35E+01 \\
&(7.87E-02) & (6.27E-02)   & (4.07E-01)   & (1.37E+00)   &...  & ...  &...   & (1.38E+01)  & (1.67E+01)  & (1.48E+01)  \\
A5  & 1.70E+01 & 1.03E+01 & 4.16E+01 & 5.78E+02 & 1.97E+04 & 1.81E+04 & 1.05E+04 & 4.15E+03 & 1.69E+03 & 5.90E+02 \\
&(1.02E+00)     & (6.23E-01)   & (2.51E+00)   & (3.48E+01)   & (1.41E+03) & (1.31E+03)  & (7.55E+02)  & (2.94E+02)  & (1.21E+02)  & (4.77E+01)  \\
A6  & 5.35E-01 & 4.33E-01 & 4.72E+00 & 1.92E+01 &    ...      &     ...     &    ...      & 8.94E+01 & 7.48E+01 & 4.55E+01 \\
&(3.47E-02)     & (2.97E-02)   & (3.27E-01)   & (1.50E+00)   & ... &  ... & ...  & (1.52E+01)  & (1.79E+01)  & (1.52E+01)  \\
A7  & 1.04E+01 & 5.90E+00 & 1.28E+01 & 1.24E+01 &     ...     &     ...     &    ...      & 2.81E+02 & 1.31E+02 & 5.22E+01 \\
&(6.27E-01)     & (3.56E-01)   & (7.90E-01)   & (1.36E+00)   & ... & ...  & ... & (2.35E+01)  & (1.58E+01)  & (1.24E+01)  \\
A8  & 2.42E+00 & 1.37E+00 & 2.65E+00 & 5.29E+00 &     ...     &    ...      &      ...    & 1.33E+02 & 7.80E+01 & 3.63E+01 \\
&(1.46E-01)     & (8.31E-02)   & (1.77E-01)   & (9.91E-01)   & ... & ...  & ...  & (1.72E+01)  & (1.47E+01)  & (1.23E+01)  \\
A9  & 6.61E+00 & 4.28E+00 & 1.37E+01 & 4.19E+01 & 2.34E+03 &    ...      & 2.24E+03 & 8.95E+02 & 3.39E+02 & 1.08E+02 \\
&(3.98E-01)     & (2.59E-01)   & (8.28E-01)   & (2.71E+00)   & (1.67E+02) & ...  & (1.64E+02)  & (6.33E+01)  & (2.53E+01)  & (1.06E+01)  \\
A10 & 5.48E-01 & 5.28E-01 & 2.43E+00 & 3.32E+00 &   ...       & 8.40E+01 & 9.99E+01 & 7.03E+01 & 3.44E+01 & 3.65E+01 \\
&(3.58E-02)     & (3.67E-02)   & (1.95E-01)   & (9.26E-01)   & ... & (1.29E+01)  & (3.55E+01)  & (1.11E+01)  & (1.12E+01)  & (1.18E+01)  \\
A11 & 1.01E+00 & 8.11E-01 & 2.23E+00 & 6.72E+00 & 4.24E+01 & 4.15E+01 &  ...        &    ...      &    ...      & 3.40E+01 \\
&(6.49E-02)     & (5.21E-02)   & (1.58E-01)   & (7.53E-01)   & (7.20E+00) & (1.27E+01)  & ...  &  ... & ...  & (1.02E+01)  \\
A12 & 3.08E+00 & 1.80E+00 & 2.81E+00 & 7.86E+00 & 3.03E+01 & 5.37E+01 & 8.29E+01 & 6.19E+01 & 4.45E+01 & 3.13E+01 \\
&(1.86E-01)     & (1.14E-01)   & (2.29E-01)   & (3.10E+00)   & (6.83E+00) & (1.01E+01)  & (1.74E+01)  & (9.81E+00)  & (1.04E+01)  & (9.17E+00)          
\enddata
\tablenotetext{}{\textbf{Note.} (1) Object ID. R1-R11 are from \citet{2015MNRAS.448..258R}; EAH01-EAH18, EAS01-EAS15 and F34-F37 are from \citet{2018ApJ...862....2F}, and A1-A12 are from \citet{2016ApJS..224...38A}. (2)-(5) \emph{WISE} fluxes. (6)-(11) \emph{Herschel} fluxes. All fluxes are given in mJy. The total flux uncertainties, which correspond to 68\% confidence levels, are given in parentheses when available. The data of EAS/EAH sources are presented in \citet{2018ApJ...855...51S}. No correction for extinction has been applied.}
\end{deluxetable*}

\newcommand{\msun}{$M_{\odot}$}

\startlongtable
\begin{deluxetable*}{lccccccc}
\tabletypesize{\scriptsize}
\tablecaption{{\centering}Galaxy Properties \label{tab:results}}
\tablehead{
\colhead{ID}&\colhead{\emph r$_{50}$}&\colhead{log \emph M$_{\star}$}&\colhead{log \emph M$_{\rm dust}$}&\colhead{log \emph M$_{\rm dust}$/\emph M$_{\star}$}&\colhead{log \emph M$_{\rm gas}$}&\colhead{SFR$_{\rm uncor}$}&\colhead{SFR$_{\rm cor}$}\\
\colhead{}&\colhead{(arcsec)}&\colhead{(\msun)}&\colhead{(\msun)}&\colhead{}&\colhead{(\msun)} &\colhead{\textbf{(\msun\ yr$^{-1}$)}}&\colhead{\textbf{(\msun\ yr$^{-1}$)}}}
\colnumbers
\startdata 
R1    & 2.08       & 9.82$^{+0.24}_{-0.19}$         & 7.49$\pm${0.07}    & -2.33$\pm${0.23}        & $<$8.50    &       6.74$\pm${0.05}          & 6.55$\pm${0.06}            \\
R2    & 3.77       & 10.63$^{+0.06}_{-0.25}$      & 7.85$\pm${0.02}    & -2.78$\pm${0.15}        & 9.71$\pm${0.04}         & 9.25$\pm${0.05}          & 8.97$\pm${0.06}               \\
R3    & 1.41       & 9.63$^{+0.14}_{-0.12}$      & 6.89$\pm${0.13}    & -2.74$\pm${0.18}        & 8.61$\pm${0.14}         & 6.42$\pm${0.04}          & 5.95$\pm${0.06}               \\
R4    & 4.90       & 10.28$^{+0.17}_{-0.13}$      & 7.65$\pm${0.11}    & -2.63$\pm${0.19}        & 9.80$\pm${0.04}         & 1.94$\pm${0.02}          & 2.17$\pm${0.02}               \\
R5    & 1.95       & 9.83$^{+0.10}_{-0.08}$      & 7.39$\pm${0.05}    & -2.44$\pm${0.10}        & 9.79$\pm${0.02}         & 3.15$\pm${0.02}          & 2.71$\pm${0.03}               \\
R6    & 4.15       & 10.44$^{+0.09}_{-0.08}$      & 7.68$\pm${0.14}    & -2.76$\pm${0.16}        & 9.58$\pm${0.05}         & 2.47$\pm${0.02}          & 1.86$\pm${0.02}               \\
R7    & 1.92       & 10.16$^{+0.10}_{-0.08}$      & 7.02$\pm${0.11}    & -3.14$\pm${0.14}        & 9.56$\pm${0.07}         & 1.05$\pm${0.01}          & 0.92$\pm${0.01}               \\
R9    & 2.90       & 10.28$^{+0.09}_{-0.09}$      & 7.60$\pm${0.10}    & -2.68$\pm${0.13}        & 9.66$\pm${0.04}         & 0.42$\pm${0.01}          & 0.39$\pm${0.01}               \\
R10   & 4.72       & 10.21$^{+0.10}_{-0.09}$      & 7.58$\pm${0.06}    & -2.63$\pm${0.11}        & 9.45$\pm${0.05}         & 0.98$\pm${0.01}          & 1.01$\pm${0.01}               \\
EAH01 & 2.09       & 10.45$^{+0.12}_{-0.11}$      & 7.58$\pm${0.02}    & -2.87$\pm${0.12}        & 9.71$\pm${0.06}         & 0.06$\pm${0.01}          & 0.01$\pm${0.01}               \\
EAH02 & 1.71       & 10.39$^{+0.11}_{-0.10}$      & 6.64$\pm${0.02}    & -3.75$\pm${0.11}        & 9.53$\pm${0.09}         & 0.03$\pm${0.01}          & 0.01$\pm${0.01}               \\
EAH03 & 3.17       & 10.34$^{+0.11}_{-0.10}$      & 7.51$\pm${0.02}    & -2.83$\pm${0.11}        & 9.80$\pm${0.05}         & 0.02$\pm${0.01}          &     $<0.003$              \\
EAH04 & 2.98       & 10.18$^{+0.11}_{-0.09}$      & 6.32$\pm${0.12}    & -3.86$\pm${0.16}        & 8.56$\pm${0.09}         & 0.07$\pm${0.01}          & 0.02$\pm${0.01}               \\
EAH05 & 1.42       & 10.81$^{+0.15}_{-0.13}$      & 7.54$\pm${0.02}    & -3.27$\pm${0.14}        & 9.56$\pm${0.09}         & 0.06$\pm${0.01}          & 0.04$\pm${0.01}               \\
EAH06 & 1.93       & 10.12$^{+0.13}_{-0.10}$      & 6.75$\pm${0.10}    & -3.37$\pm${0.15}        & $<$9.00        & 0.37$\pm${0.04}          &     ...               \\
EAH07 & 3.69       & 9.88 $^{+0.10}_{-0.07}$      & 6.75$\pm${0.10}    & -3.13$\pm${0.13}        & $<$8.62        & 0.22$\pm${0.02}          &     ...               \\
EAH08 & 1.67       & 10.07$^{+0.11}_{-0.11}$      & 8.02$\pm${0.12}    & -2.05$\pm${0.16}        & 9.15$\pm${0.15}         & 0.04$\pm${0.01}          &     ...               \\
EAH09 & 1.81       & 11.01$^{+0.11}_{-0.11}$      & 6.52$\pm${0.11}    & -4.49$\pm${0.16}        & 8.50$\pm${0.13}         & 0.06$\pm${0.01}          & 0.01$\pm${0.01}               \\
EAH10 & 1.22       & 10.24$^{+0.12}_{-0.08}$      & 7.72$\pm${0.08}    & -2.52$\pm${0.13}        & 9.86$\pm${0.12}         & 0.04$\pm${0.03}          & 0.01$\pm${0.01}               \\
EAH11 & 1.09       & 10.64$^{+0.10}_{-0.09}$      & 6.33$\pm${0.15}    & -4.31$\pm${0.18}        & $<$9.19         & 0.17$\pm${0.02}          &     ...               \\
EAH12 & 0.96       & 9.89$^{+0.12}_{-0.10}$      & 6.40$\pm${0.03}    & -3.49$\pm${0.11}        & $<$9.42         & 0.18$\pm${0.03}          & 0.05$\pm${0.02}               \\
EAH13 & 1.50       & 11.00$^{+0.13}_{-0.11}$      & 7.64$\pm${0.19}    & -3.36$\pm${0.22}        & 9.89$\pm${0.08}         & 0.62$\pm${0.14}          &     ...               \\
EAH14 & 1.18       & 10.04$^{+0.13}_{-0.10}$      & 5.96$\pm${0.24}    & -4.08$\pm${0.27}        & $<$9.25         & 0.10$\pm${0.01}          & 0.01$\pm${0.01}               \\
EAH15 & 1.52       & 10.40$^{+0.13}_{-0.12}$      & 6.44$\pm${0.22}    & -3.96$\pm${0.25}        & $<$9.04         & 0.06$\pm${0.01}          & 0.01$\pm${0.01}               \\
EAH16 & 1.37       & 10.74$^{+0.09}_{-0.07}$      & 6.94$\pm${0.46}    & -3.80$\pm${0.47}        & $<$9.79         & 0.51$\pm${0.09}          &     ...               \\
EAH17 & 1.49       & 10.05$^{+0.10}_{-0.10}$      & 6.80$\pm${0.32}    & -3.25$\pm${0.34}        & $<$9.01         & 0.05$\pm${0.01}          & 0.02$\pm${0.01}               \\
EAH18 & 3.33       & 10.38$^{+0.12}_{-0.10}$      & 5.78$\pm${0.18}    & -4.60$\pm${0.21}        & 8.68$\pm${0.15}         & 0.03$\pm${0.01}          &     $<0.001$               \\
EAS01 & 4.39       & 10.24$^{+0.12}_{-0.11}$      & 5.32$\pm${0.10}    & -4.92$\pm${0.15}        & $<$8.40         & 0.01$\pm${0.01}          &     $<0.0004$               \\
EAS02 & 3.72       & 10.37$^{+0.10}_{-0.10}$      & 6.27$\pm${0.14}    & -4.10$\pm${0.17}        & 8.71$\pm${0.13}         & 0.03$\pm${0.01}          & 0.01$\pm${0.01}               \\
EAS03 & 2.34       & 10.34$^{+0.10}_{-0.09}$      & 7.20$\pm${0.14}    & -3.14$\pm${0.17}        & 9.76$\pm${0.06}         & 0.17$\pm${0.02}          & 0.02$\pm${0.01}               \\
EAS04 & 3.12       & 9.99$^{+0.11}_{-0.10}$      & 5.43$\pm${0.16}    & -4.56$\pm${0.19}        & $<$7.74         & 0.01$\pm${0.01}          &     ...               \\
EAS05 & 2.74       & 11.33$^{+0.10}_{-0.10}$      & 7.09$\pm${0.10}    & -4.24$\pm${0.14}        & 9.09$\pm${0.14}         & 0.11$\pm${0.02}          & 0.04$\pm${0.01}               \\
EAS06 & 2.91       & 10.14$^{+0.09}_{-0.09}$      & 6.61$\pm${0.08}    & -3.53$\pm${0.12}        & 9.23$\pm${0.04}         & 0.06$\pm${0.01}          & 0.03$\pm${0.01}               \\
EAS07 & 2.34       & 10.54$^{+0.12}_{-0.11}$      & 5.90$\pm${1.04}    & -4.64$\pm${1.05}        & 8.64         & 0.04$\pm${0.02}          &     $<0.01$              \\
EAS08 & 4.25       & 10.67$^{+0.10}_{-0.10}$      & 6.41$\pm${0.88}    & -4.26$\pm${0.89}        & $<$8.60         & 0.05$\pm${0.03}          &     $<0.01$               \\
EAS09 & 3.27       & 10.56$^{+0.11}_{-0.10}$      & 6.88$\pm${0.03}    & -3.68$\pm${0.11}        & 9.12$\pm${0.06}         & 0.04$\pm${0.01}          & 0.01$\pm${0.01}               \\
EAS10 & 2.03       & 10.53$^{+0.12}_{-0.10}$      & 5.64$\pm${0.03}    & -4.89$\pm${0.11}        & $<$8.79         & 0.02$\pm${0.01}          &     ...               \\
EAS11 & 3.04       & 10.71$^{+0.14}_{-0.13}$      & 6.59$\pm${0.12}    & -4.12$\pm${0.18}        & $<$8.84         & 0.08$\pm${0.02}          & 0.02$\pm${0.01}               \\
EAS12 & 6.39       & 10.80$^{+0.10}_{-0.11}$      & 7.33$\pm${0.17}    & -3.47$\pm${0.20}        & 8.54$\pm${0.14}         & 0.03$\pm${0.01}          &     ...               \\
EAS13 & 3.80       & 10.95$^{+0.10}_{-0.10}$      & 7.50$\pm${0.33}    & -3.45$\pm${0.34}        & $<$9.12         & 0.09$\pm${0.02}          &     $<0.02$               \\
EAS14 & 3.72       & 11.31$^{+0.11}_{-0.11}$      & 7.49$\pm${0.24}    & -3.82$\pm${0.26}        & 9.70$\pm${0.09}         & 0.41$\pm${0.07}          &     ...               \\
EAS15 & 2.82       & 10.83$^{+0.10}_{-0.09}$      & 6.88$\pm${0.21}    & -3.95$\pm${0.23}        & 9.08$\pm${0.14}         & 0.17$\pm${0.05}          &     $<0.07$               \\
F34   & 1.77       & 10.20$^{+0.12}_{-0.11}$      & 8.40$\pm${0.08}    & -1.80$\pm${0.14}        & 9.83$\pm${0.15}         & 0.03$\pm${0.01}          & 0.01$\pm${0.01}               \\
F35   & 2.73       & 10.64$^{+0.10}_{-0.09}$      & 8.18$\pm${0.06}    & -2.46$\pm${0.11}        & 9.80$\pm${0.15}         & 0.08$\pm${0.01}          & 0.04$\pm${0.01}               \\
F36   & 1.11       & 10.28$^{+0.13}_{-0.13}$      & 7.33$\pm${0.24}    & -2.95$\pm${0.27}        & 9.32$\pm${0.15}         & 0.61$\pm${0.13}          &     ...               \\
F37   & 3.33       & 10.38$^{+0.12}_{-0.10}$      & 6.26$\pm${0.24}    & -4.12$\pm${0.26}        & 8.82$\pm${0.15}         & 0.03$\pm${0.01}          &     $<0.001$               \\
A1    & 2.18       & 10.64$^{+0.10}_{-0.09}$      & 7.86$\pm${0.32}    & -2.78$\pm${0.33}        & 9.57$\pm${0.15}         & 3.89$\pm${0.09}          & 0.51$\pm${0.21}               \\
A2    & 5.38       & 9.98$^{+0.35}_{-0.53}$      & 7.89$\pm${0.26}    & -2.09$\pm${0.51}        & 9.59$\pm${0.15}         & 0.35$\pm${0.01}          & 0.36$\pm${0.01}               \\
A3    & 2.70       & 11.30$^{+0.10}_{-0.10}$      & 8.89$\pm${0.19}    & -2.41$\pm${0.21}       & 10.06$\pm${0.15}        & 14.25$\pm${0.44}          & 3.19$\pm${0.54}               \\
A4    & 1.11       & 10.79$^{+0.06}_{-0.05}$      & 8.47$\pm${0.18}    & -2.32$\pm${0.19}        & 9.86$\pm${0.15}        & 27.71$\pm${0.90}          & 12.04$\pm${1.29}               \\
A5    & 6.41       & 10.19$^{+0.09}_{-0.10}$      & 8.07$\pm${0.04}    & -2.12$\pm${0.10}        & 9.40$\pm${0.08}         & 0.13$\pm${0.01}          & 0.02$\pm${0.01}               \\
A6    & 1.77       & 10.34$^{+0.10}_{-0.09}$      & 7.99$\pm${0.13}    & -2.35$\pm${0.16}        & 9.63$\pm${0.15}         & 7.04$\pm${0.08}          & 6.17$\pm${0.12}               \\
A7    & 7.41       & 10.80$^{+0.10}_{-0.11}$      & 7.82$\pm${0.14}    & -2.98$\pm${0.18}        & 9.55$\pm${0.15}         & 0.19$\pm${0.01}          & 0.05$\pm${0.01}               \\
A8    & 5.43       & 10.15$^{+0.10}_{-0.09}$      & 7.56$\pm${0.21}    & -2.59$\pm${0.23}        & 9.43$\pm${0.15}         & 0.14$\pm${0.01}          & 0.05$\pm${0.01}               \\
A9    & 4.40       & 10.59$^{+0.10}_{-0.09}$      & 7.80$\pm${0.02}    & -2.79$\pm${0.10}        & 9.54$\pm${0.15}         & 1.13$\pm${0.02}          & 0.61$\pm${0.03}               \\
A10   & 2.62       & 10.65$^{+0.10}_{-0.10}$      & 8.19$\pm${0.19}    & -2.46$\pm${0.21}        & 9.73$\pm${0.15}         & 1.04$\pm${0.03}          & 0.78$\pm${0.05}               \\
A11   & 1.51       & 10.00$^{+0.11}_{-0.11}$      & 8.19$\pm${0.19}    & -1.81$\pm${0.22}        & 9.01$\pm${0.11}         & 0.19$\pm${0.01}          & 0.04$\pm${0.01}               \\
A12   & 2.66       & 10.51$^{+0.13}_{-0.11}$      & 7.68$\pm${0.30}    & -2.83$\pm${0.32}        & 8.86$\pm${0.06}         & 0.22$\pm${0.01}          & \ 0.05$\pm${0.01} 
\enddata
\tablenotetext{}{\textbf{Note.} (1) Object ID. (2) Petrosian radius from SDSS 14th Data Release (DR14, \citealt{2018ApJS..235...42A}; containing 50\% light). (3) Stellar masses from the SDSS MPA-JHU (\citealt{2004MNRAS.351.1151B, 2004ApJ...613..898T}) catalog. (4) CIGALE-derived dust masses. (5) Specific dust masses. (6) Molecular gas masses from \citet{2015ApJ...801....1F},  \citet{2016ApJ...827..106A}, or converted from dust masses if no CO measurement is available (with an estimated uncertainty of 0.15 dex). (7) Star formation rates derived from H$\alpha$ fluxes from the MPA-JHU catalog (corrected for internal dust extinction). (8) Star formation rates corrected for AGN contribution.}
\end{deluxetable*}

\bibliography{dust-psb}
\appendix

\section{Minimal \emph{Herschel} data required for reliable $M_{\MakeLowercase{\rm dust}}$}\label{sec:Herschelbands}
\begin{figure}
\centering
\begin{tabular}{c c c}
\includegraphics[width=0.49\linewidth, clip]{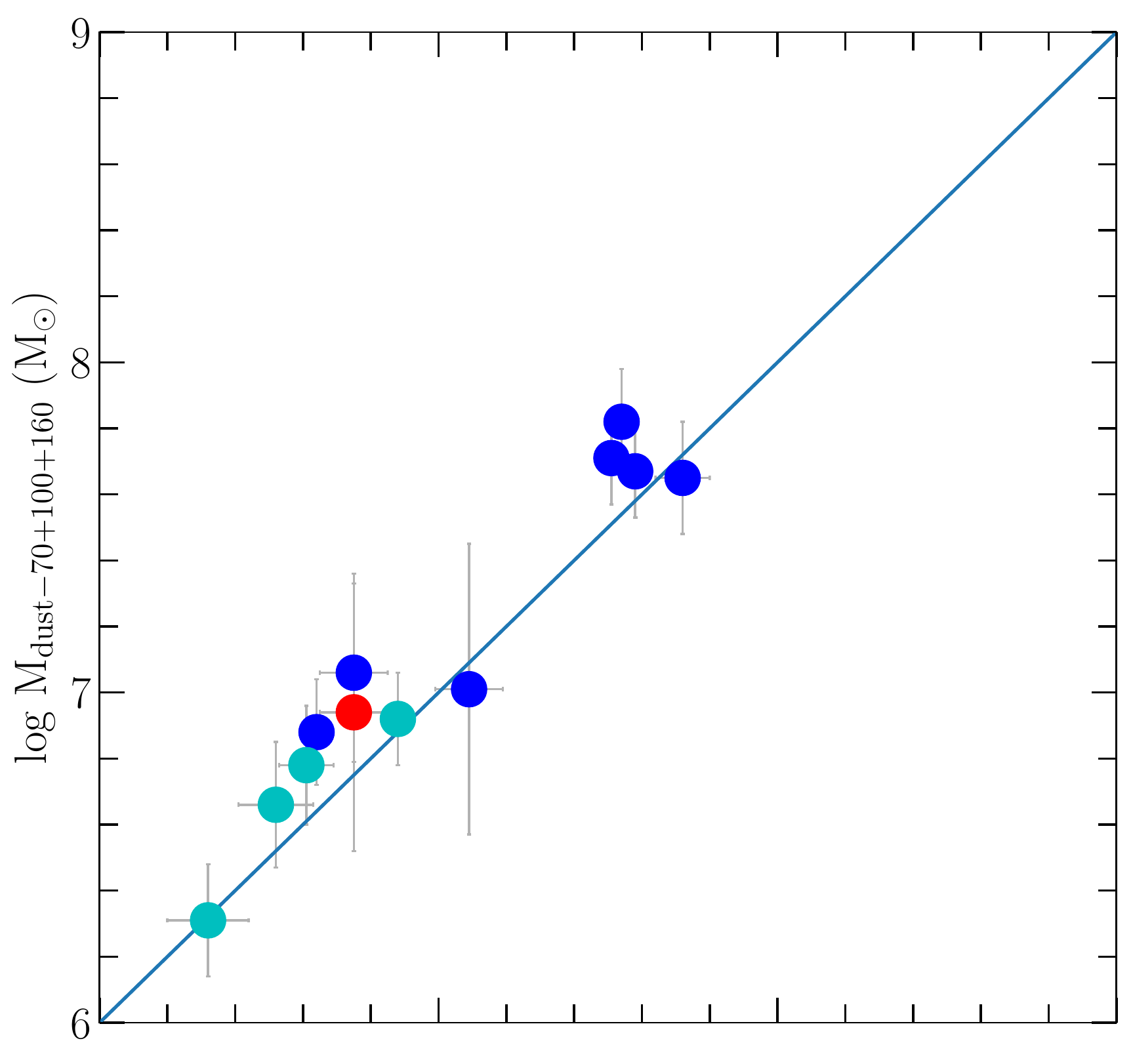} &
\includegraphics[width=0.49\linewidth, clip]{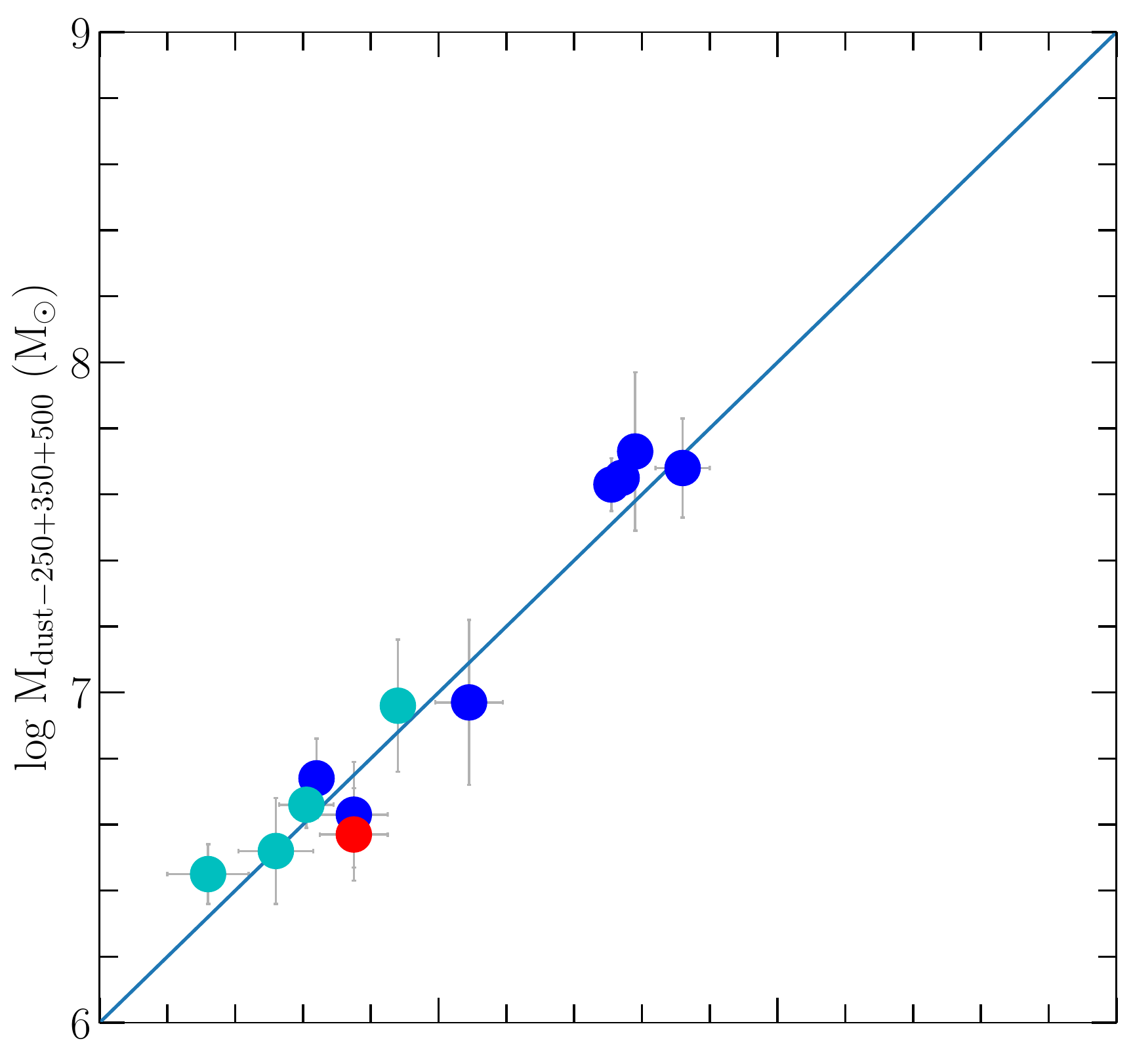}&\\
\includegraphics[width=0.49\linewidth, clip]{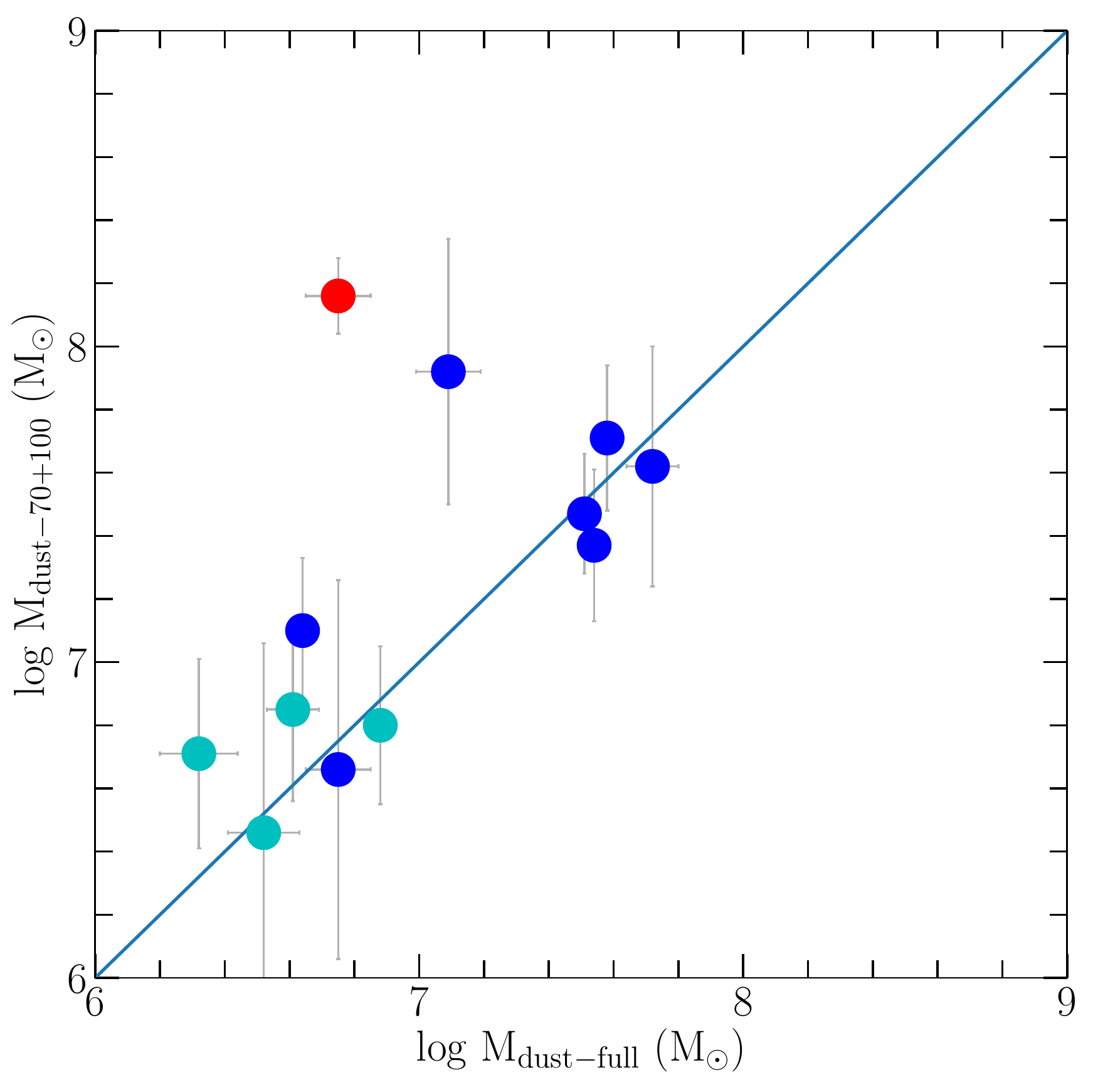}&
\includegraphics[width=0.49\linewidth, clip]{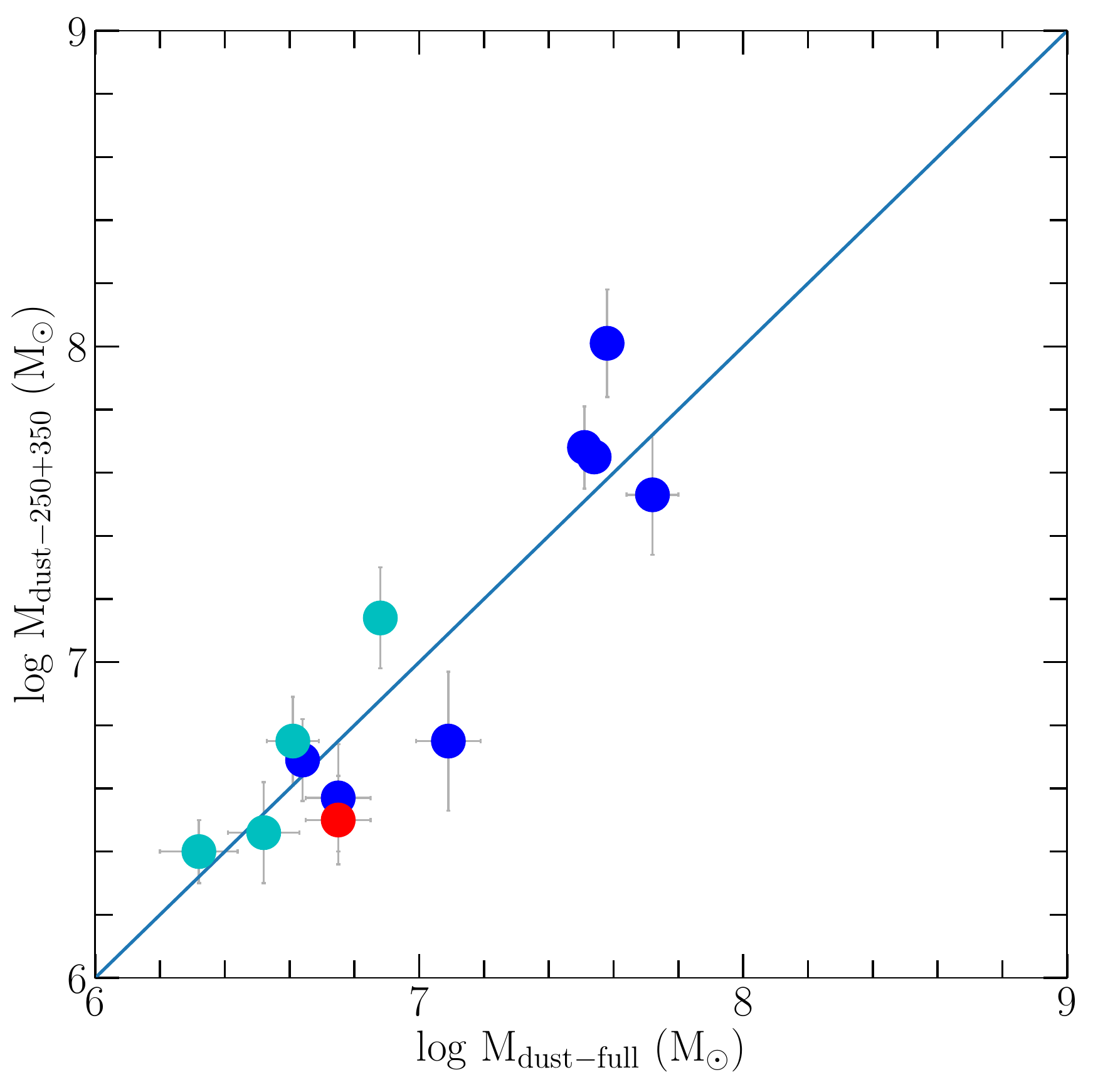}
\end{tabular}
\caption{Comparison of $M_{\rm dust}$ derived by CIGALE using different \emph{Herschel} bands. The x-axis includes all six \emph{Herschel} bands and the y-axis includes different subsets of \emph{Herschel} bands. Different colors represent different redshift ranges (cyan: \emph{z} $\leq$ 0.03; red: 0.03 $<$ \emph{z} $\leq$ 0.04; blue: \emph{z} $>$ 0.04). All error bars (shown in gray) correspond to 68\% confidence levels. The one-to-one line is plotted in blue for comparison. It can be seen that 70 $\mu$m+100 $\mu$m+160 $\mu$m or 250 $\mu$m+350 $\mu$m+500 $\mu$m bands are enough to derive reliable $M_{\rm dust}$, while outliers exist where only 70 $\mu$m+100 $\mu$m or 250 $\mu$m+350 $\mu$m bands are used.}
\label{fig:Herscheldata}
\end{figure}

According to \citet{2007ApJ...657..810D}, determining the position of the FIR SED peak is crucial for obtaining a reliable $M_{\rm dust}$. Here we determine the fewest and best bands for fitting the SEDs of our sample. We perform SED fitting tests using 12 EAH/EAS PSBs (i.e., EAH01-07, EAH09-10, EAS05-06, and EAS09) that have archival fluxes in all six \emph{Herschel} bands and that are best fit by the model at 70 and 100$\mu$m. Then, we remove different sets of bands, refit the SEDs of these galaxies, and compare the derived $M_{\rm dust}$ with those derived from fitting all six \emph{Herschel} bands. As shown in Figure \ref{fig:Herscheldata}, 70$\mu$m+100$\mu$m+160$\mu$m or 250$\mu$m+350$\mu$m+500$\mu$m bands are enough to derive reliable $M_{\rm dust}$, while outliers exist where only 70$\mu$m+100$\mu$m or 250$\mu$m+350$\mu$m bands are used\footnote{If we loosen our criterion to $\leq$2 \emph{Herschel} bands, 18 additional PSBs could be included, the majority of which have 70$\mu$m+100$\mu$m or 250$\mu$m+350$\mu$m bands. As our tests show that 70$\mu$m+100$\mu$m or 250$\mu$m+350$\mu$m bands are not enough to yield a reliable $M_{\rm dust}$, we decide not to include them in our sample.}. To be conservative, we remove all the PSBs with $<$3 \emph{Herschel} bands and include only those 58 PSBs with at least three \emph{Herschel} measurements in our analysis in this paper. 

\section{Effect of including UV and optical data on $M_{\MakeLowercase{\rm dust}}$}\label{sec:UVopteffect}
\begin{figure}
\centering
\includegraphics[width=0.5\linewidth, clip]{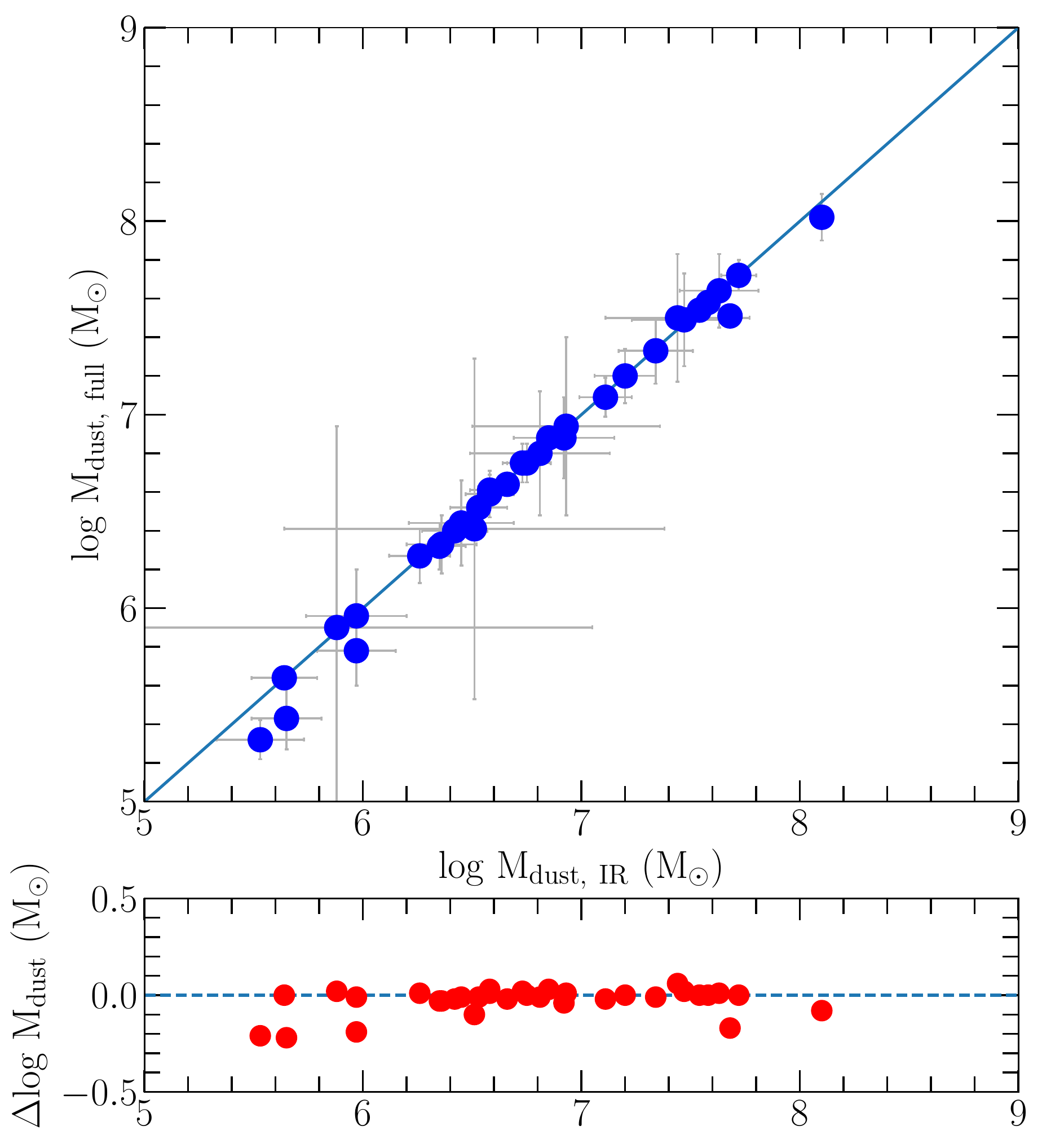}
\caption{Comparison of $M_{\rm dust}$ derived by CIGALE by fitting only the MIR-FIR SED (\emph{x}-axis) versus the full UV-FIR SED (\emph{y}-axis), with the difference in $M_{\rm dust}$ \emph{(y-x)} plotted in the bottom panel. All error bars (shown in gray) correspond to 68\% confidence levels. The one-to-one solid line (upper) and the zero fiducial dashed line (lower) are plotted in blue for comparison. The $M_{\rm dust}$ is very robust whether or not we use UV and optical data in our fitting.}
\label{fig:IRcompare}
\end{figure}

We also test the effects of fitting the SEDs for 33 PSBs (i.e., EAH01-18 and EAS01-15) from the UV to FIR using CIGALE, as we do in this paper, compared to only fitting with MIR-FIR data on $M_{\rm dust}$. As shown in Figure \ref{fig:IRcompare}, $M_{\rm dust}$ is very robust whether or not we use UV and optical data in our fitting. This is expected, as $M_{\rm dust}$ is basically determined by the position and height of the IR peak.

\section{$M_{\MakeLowercase{\rm dust}}$ comparison with Smercina et al. (2018)}\label{sec:compare}
\begin{figure}
\centering
\includegraphics[width=0.5\linewidth, clip]{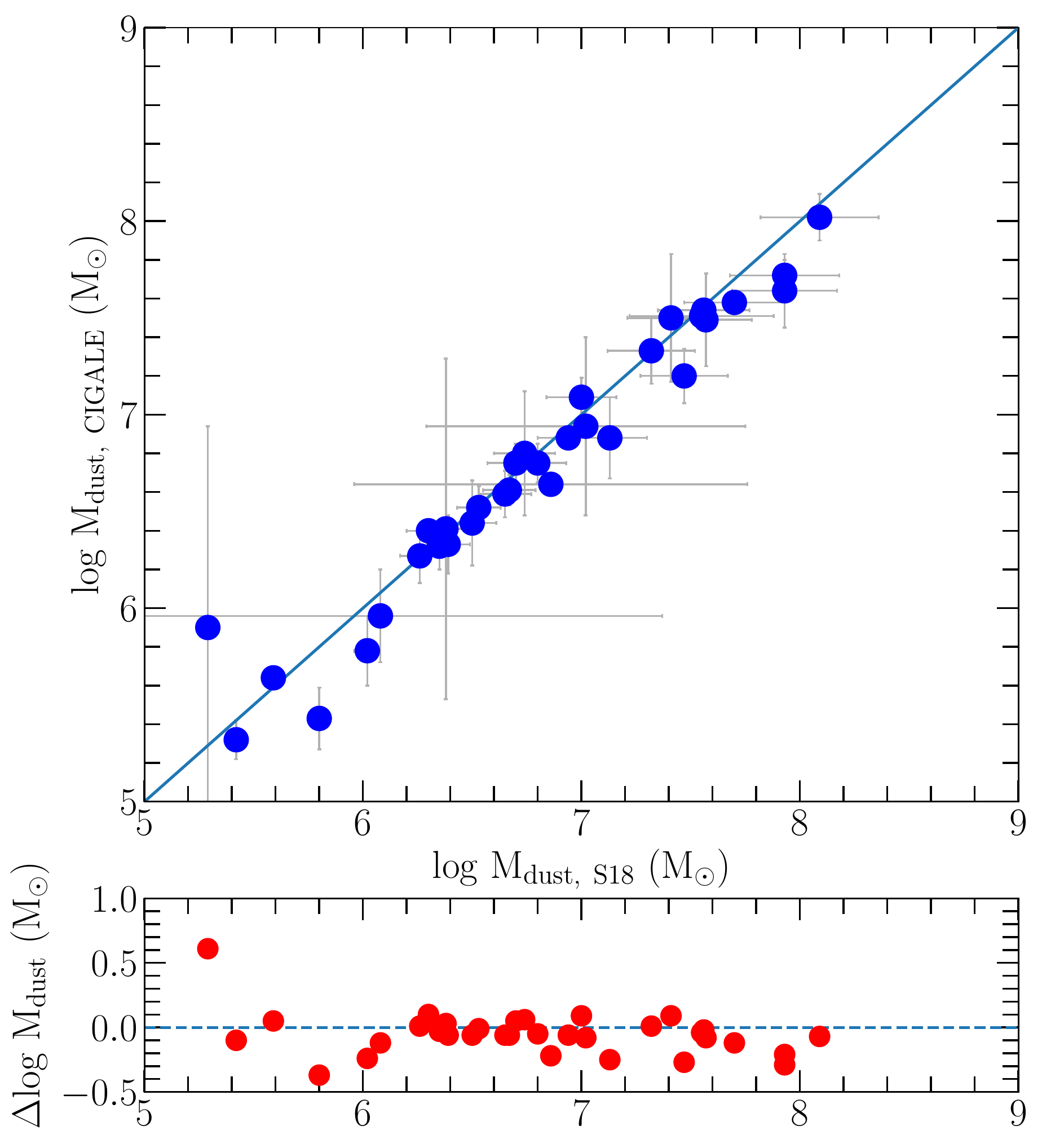}
\caption{Comparison of $M_{\rm dust}$ derived by us using CIGALE (\emph{y}-axis) and by \citet{2018ApJ...855...51S} using an alternate stellar model (\emph{x}-axis), with the difference in $M_{\rm dust}$ \emph{(y-x)} plotted at the bottom. All error bars (shown in gray) correspond to 68\% confidence levels. The one-to-one solid line (upper) and the zero fiducial dashed line (lower) are plotted in blue for comparison. When the error bars are taken into account, the $M_{\rm dust}$'s are consistent in almost all cases.}
\label{fig:compare}
\end{figure}

\citet{2018ApJ...855...51S} derive $M_{\rm dust}$ for 33 PSBs (i.e., EAH01-18 and EAS01-15) by performing SED fitting using the DL07 model and a \emph{T} = 5000 K blackbody stellar model. To compare our results with theirs, we combine the \emph{WISE} and \emph{Herschel} fluxes from \citet{2018ApJ...855...51S} with our UV-optical fluxes and perform SED fitting using CIGALE. As shown in Figure \ref{fig:compare}, when the error bars are taken into account, the $M_{\rm dust}$'s are consistent\footnote{The log\ $M_{\rm dust}$ for EAH03 in \citet{2018ApJ...855...51S} should be 7.55 $M_{\odot}$, instead of 8.70 $M_{\odot}$. We thank A. Smercina and D. A. Dale for their help in identifying this problem.} in almost all cases: 32/33 PSBs have -0.5 $M_{\odot}$ $<$ $\Delta$log\ $M_{\rm dust}$ $<$ 0.25 $M_{\odot}$, with on average $\Delta$log\ $M_{\rm dust}$ = -0.04$^{+0.05}_{-0.04}$ $M_{\odot}$. The small differences in $M_{\rm dust}$ and its error bars arise mostly from differences in our model assumptions and in our definitions of $\chi^2$.

\section{Effect of burst ages on $M_{\MakeLowercase{\rm dust}}$}\label{sec:ageMd}
\begin{figure}
\begin{tabular}{cc}
\includegraphics[width=0.5\linewidth, clip]{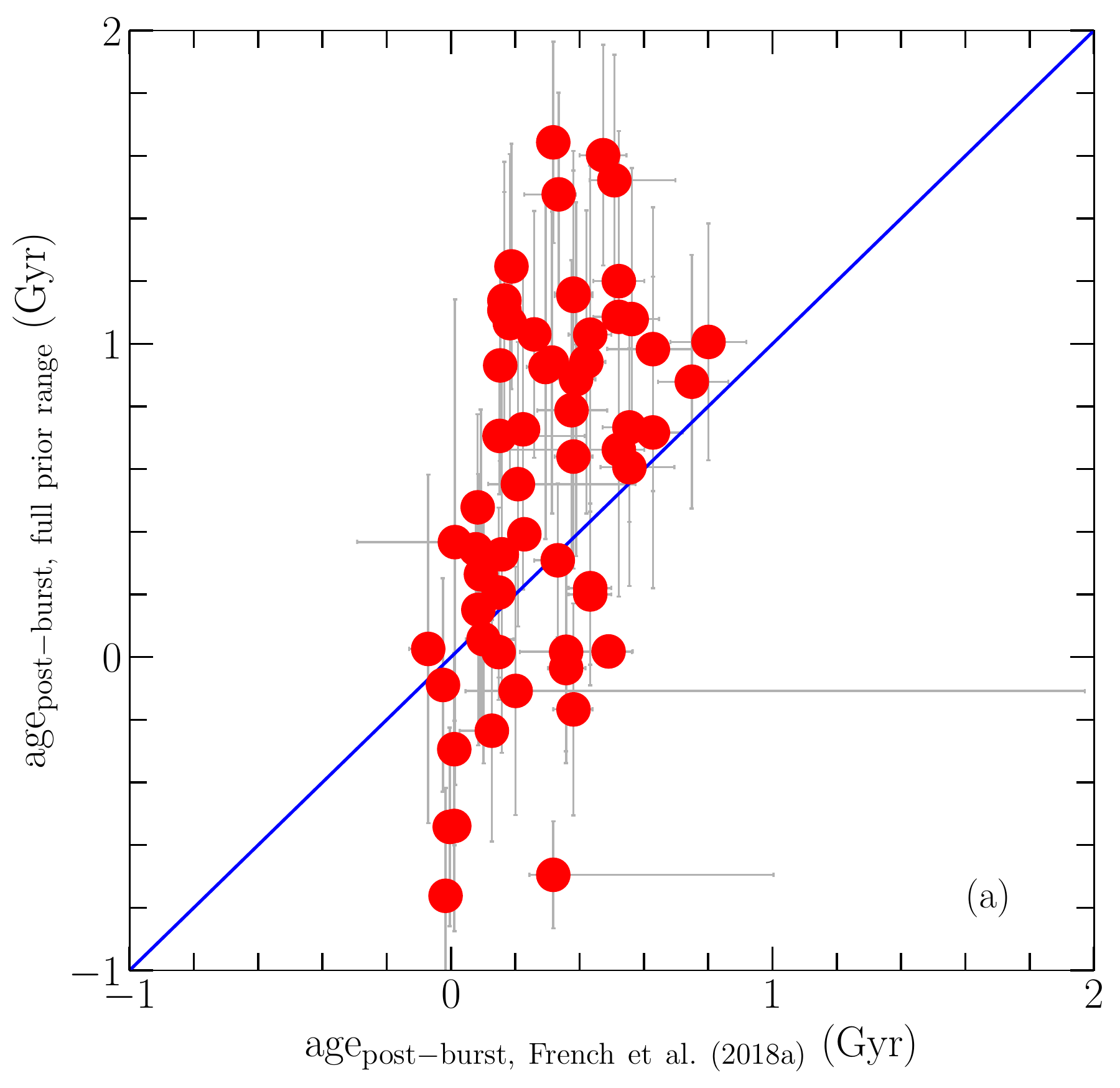}
\includegraphics[width=0.485\linewidth, clip]{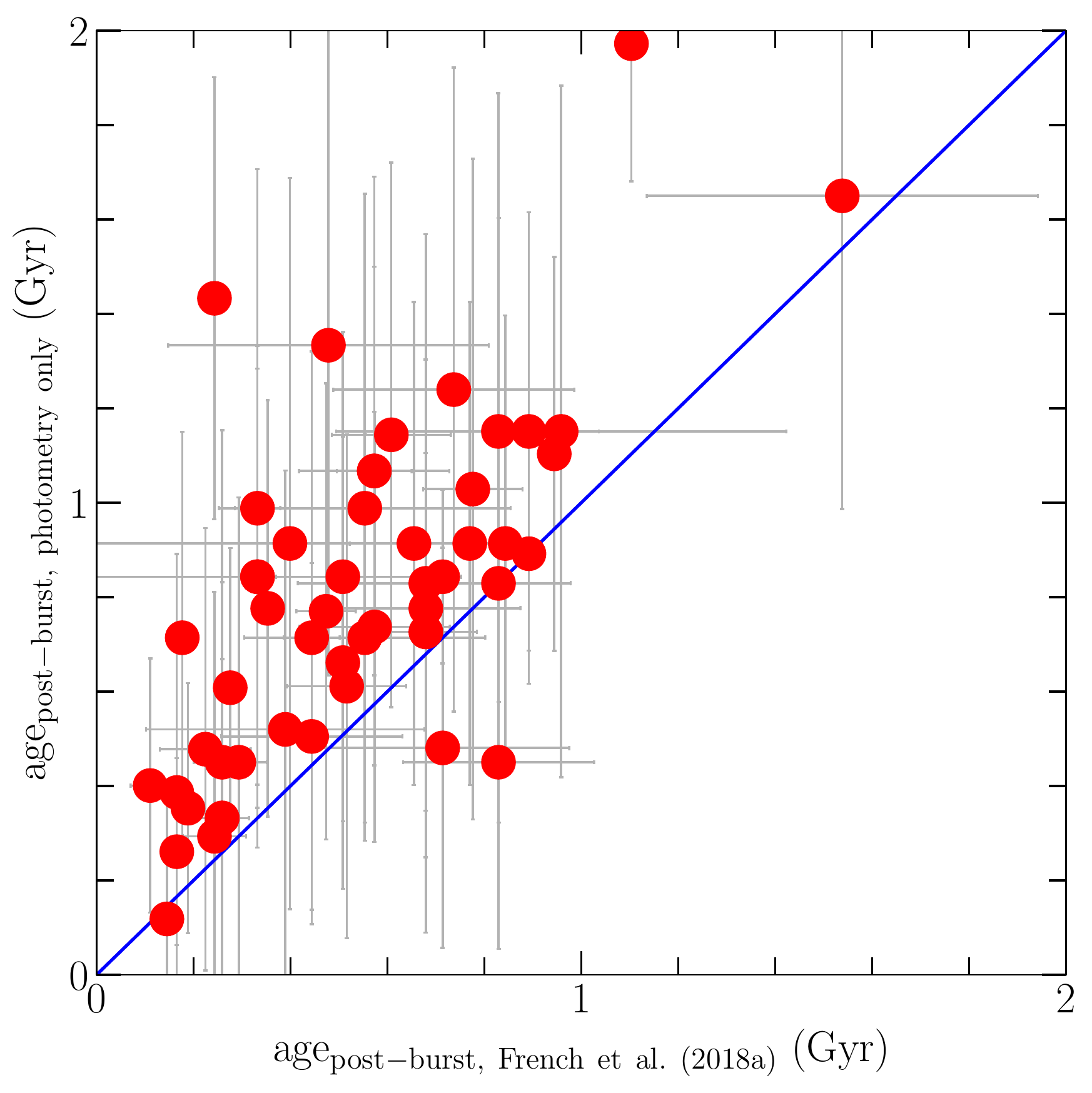}\\
\includegraphics[width=0.5\linewidth, clip]{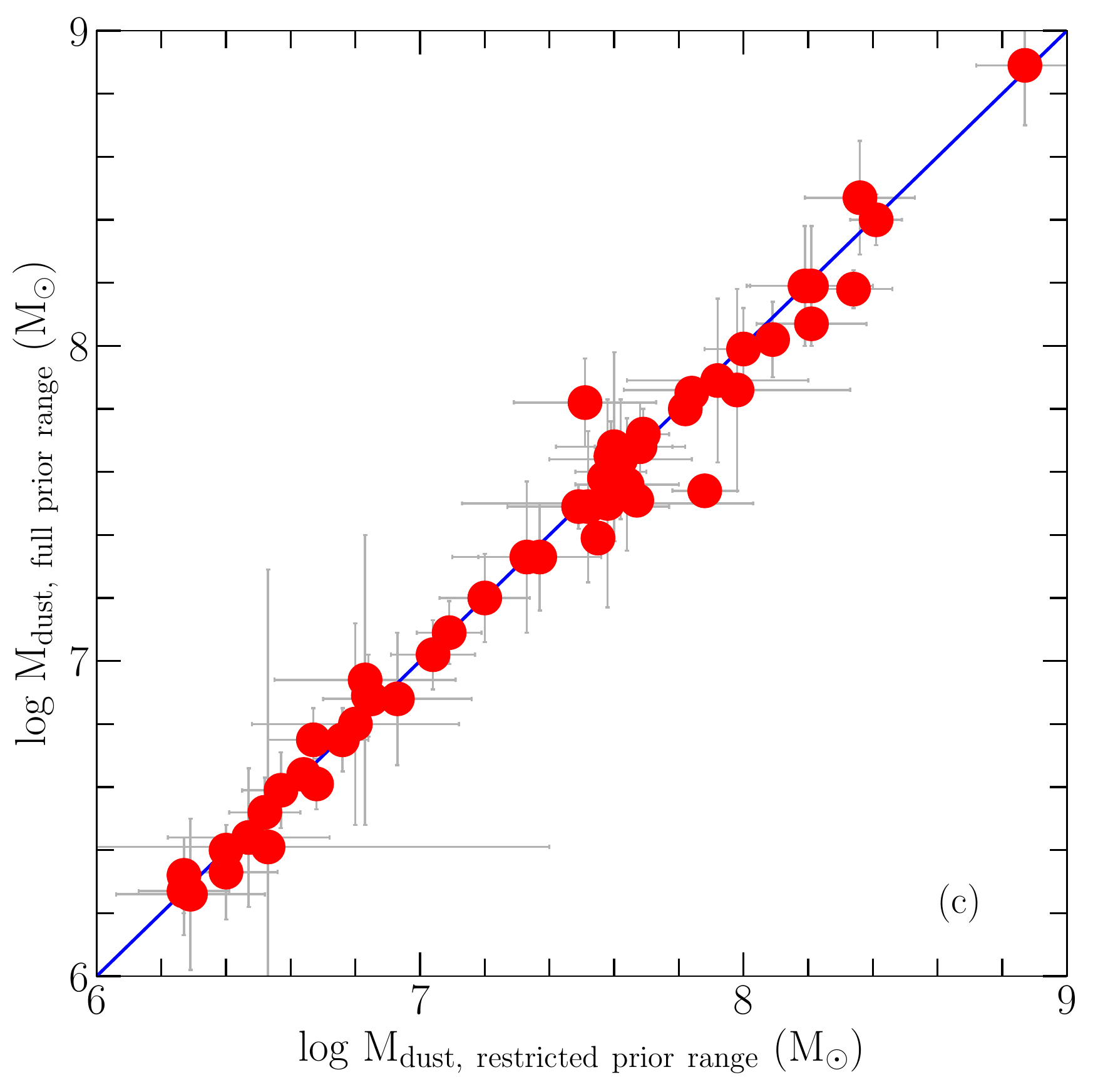}
\end{tabular}
\caption{Quantifying the effects of priors on post-burst age and $M_{\rm dust}$. We see from (a) that the post-burst ages derived from our CIGALE fits without restricted prior ranges have more scatter and are on average higher than the ages from \citet{2018ApJ...862....2F}, where optical spectral lines are included in the fits. From (b), using the age-dating method in \citet{2018ApJ...862....2F}, we see this effect of adding those optical spectral lines into age-dating. Nevertheless, from (c) we see that $M_{\rm dust}$ is robust to the choice of priors.}
\label{fig:gd}
\end{figure}

We find a discrepancy between the (post-)burst ages derived from CIGALE and \citet{2018ApJ...862....2F}, who perform age-dating by combining \emph{GALEX} photometry, SDSS photometry, and SDSS spectra. We have tested two types of ${\rm age}_{\rm burst}$ priors for CIGALE: one is the full range described in Table \ref{table:input}, the other is restricted to the 2-$\sigma$ value ranges provided in \citet{2018ApJ...862....2F} for our total sample of 58 PSBs. As shown in Figure \ref{fig:gd}a, even provided with a full range of priors, the average of the (post-)burst ages from CIGALE is still higher than that from \citet{2018ApJ...862....2F} (0.55 versus 0.30 Gyr), with a larger standard deviation (0.59 versus 0.20 Gyr). Having excluded all other possible factors that may contribute to such discrepancy (models, priors, and photometric data), we conclude that it is the inclusion of optical spectral information that causes the difference. As illustrated in Figure \ref{fig:gd}b, for a subsample of PSBs in \citet{2018ApJ...862....2F}, the (post-)burst ages derived by fitting with only photometry ({\emph{y}}-axis) are systematically higher and have larger scatter than those derived by fitting photometry and spectral lines together ({\emph{x}}-axis) using the age-dating method in \citet{2018ApJ...862....2F}. This age discrepancy almost has no impact on $M_{\rm dust}$, however, as shown in Figure \ref{fig:gd}c. The decreasing trend of $M_{\rm dust}$ versus $age_{\rm (post-)burst}$ is also not notably affected. The age-dating from \citet{2018ApJ...862....2F} is likely more accurate, so we adopt the \citet{2018ApJ...862....2F} PSB ages throughout this paper when considering the evolution of PSB properties.

\section{SED fits of all 58 PSBs in our sample}\label{sec:SEDfits}
Here we present all 58 SED fits of our sample, eight of which are already presented in Figure \ref{fig:fits}.

\begin{figure}
\figurenum{13a}
\centering
\begin{tabular}{c}
\includegraphics[width=\linewidth, clip]{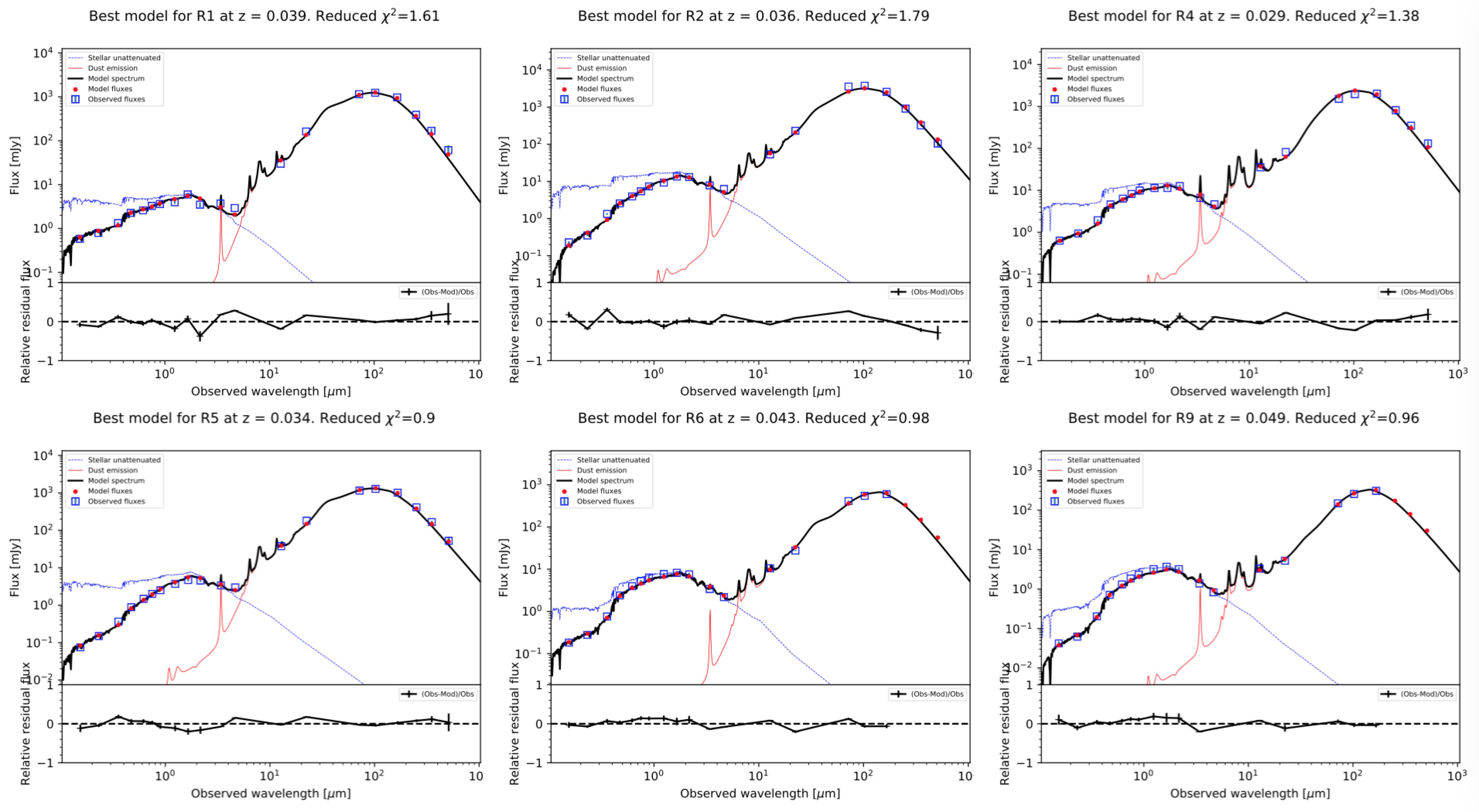}\\
\includegraphics[width=\linewidth, clip]{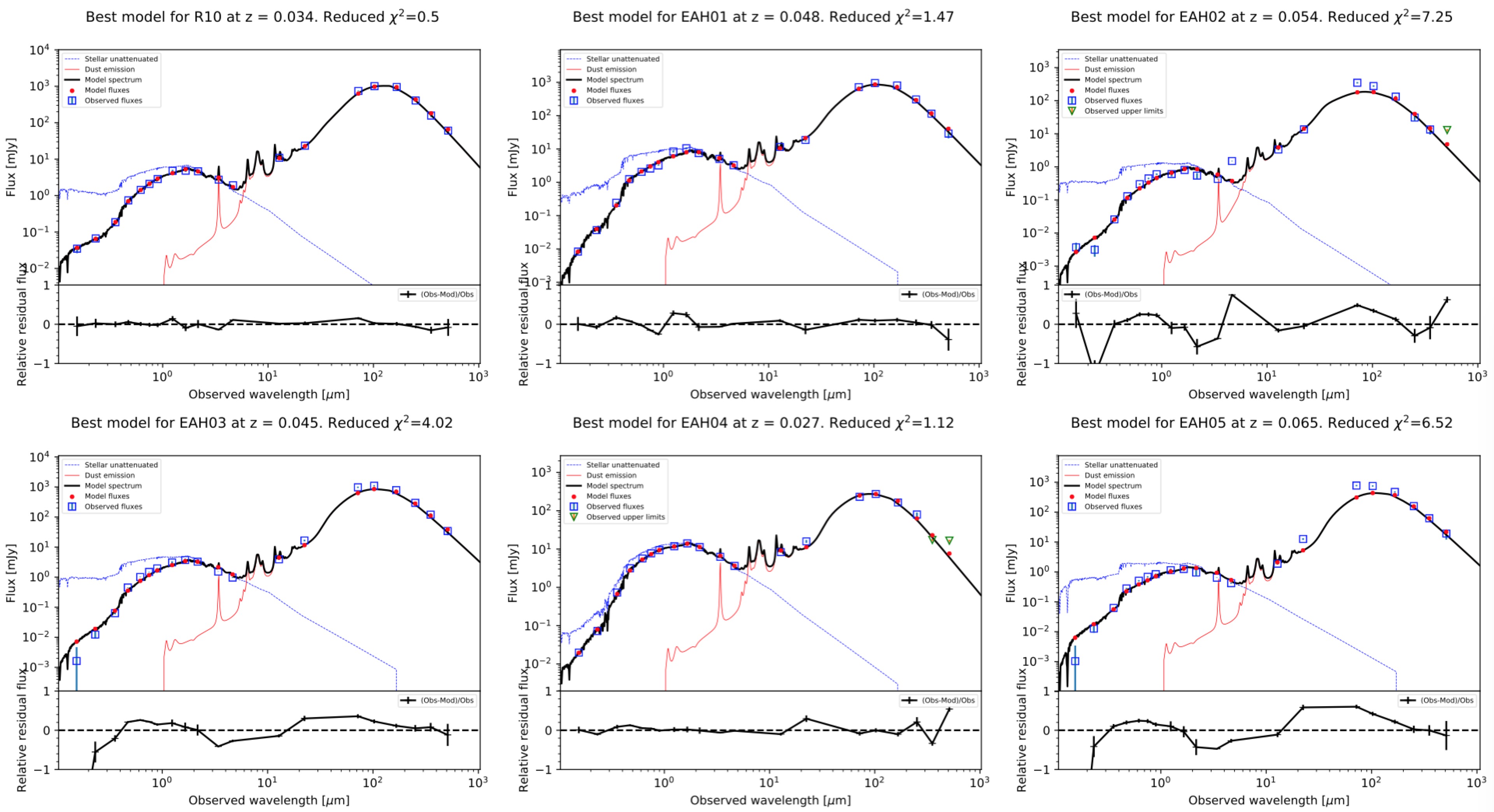}\\
\end{tabular}
\caption{
All 58 SED fits of our sample derived from CIGALE (eight of them are already presented in Figure \ref{fig:fits}).}
\label{fig:fitsa}
\end{figure}

\begin{figure}
\figurenum{13b}
\centering
\begin{tabular}{c}
\includegraphics[width=\linewidth, clip]{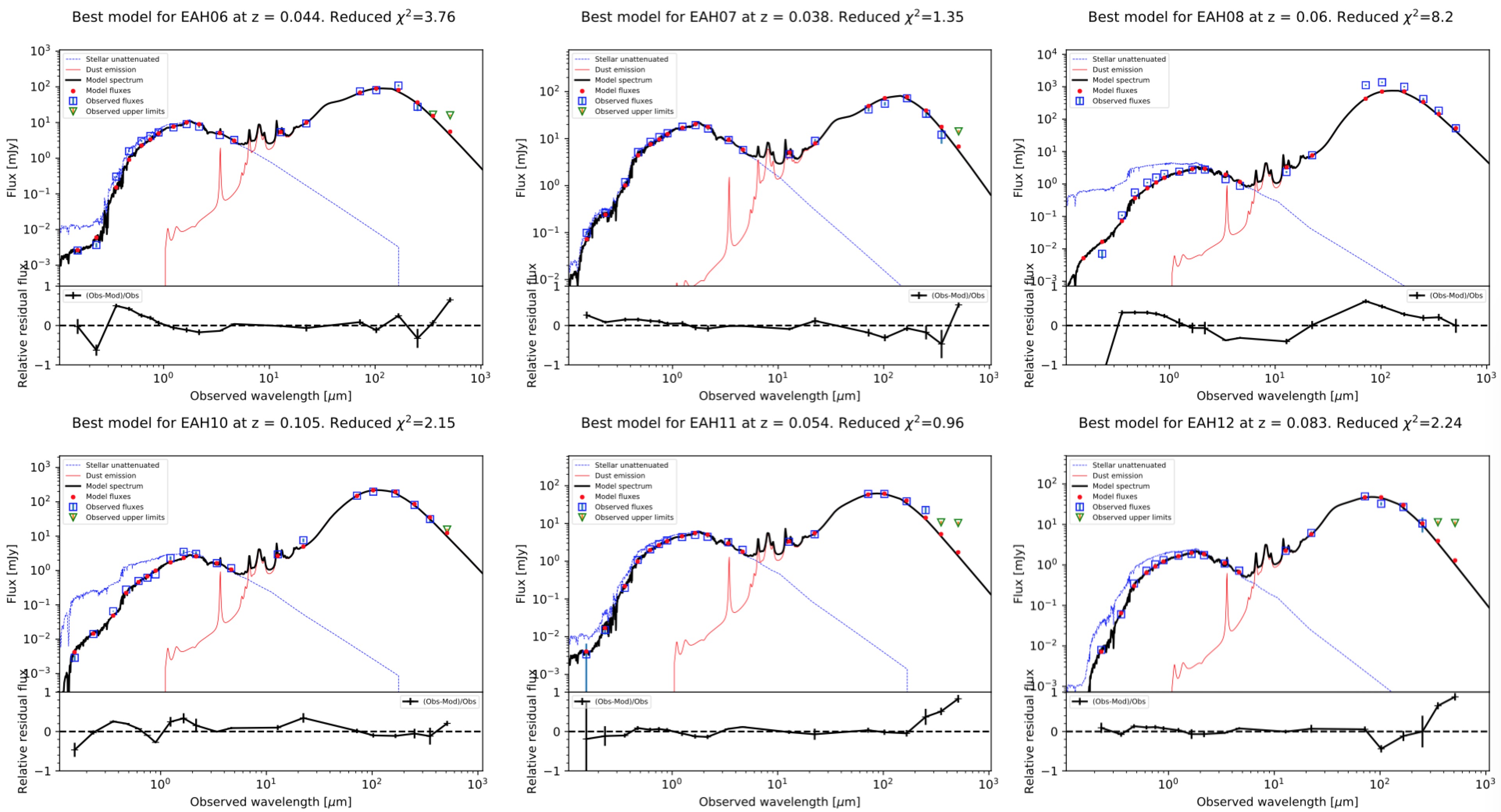}\\
\includegraphics[width=\linewidth, clip]{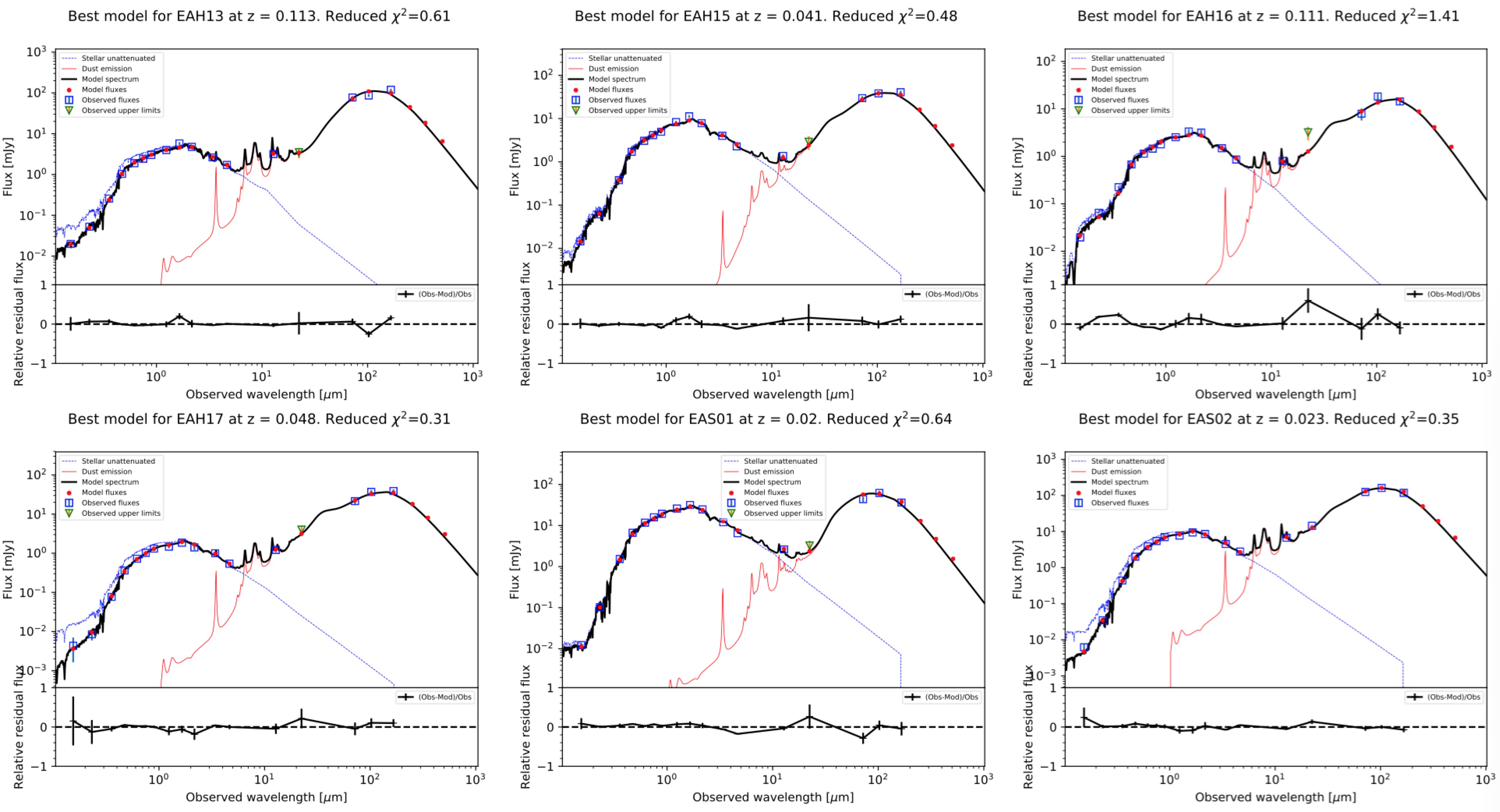}\\
\end{tabular}
\caption{Continued.}
\label{fig:fitsb}
\end{figure}

\begin{figure}
\figurenum{13c}
\centering
\begin{tabular}{c}
\includegraphics[width=\linewidth, clip]{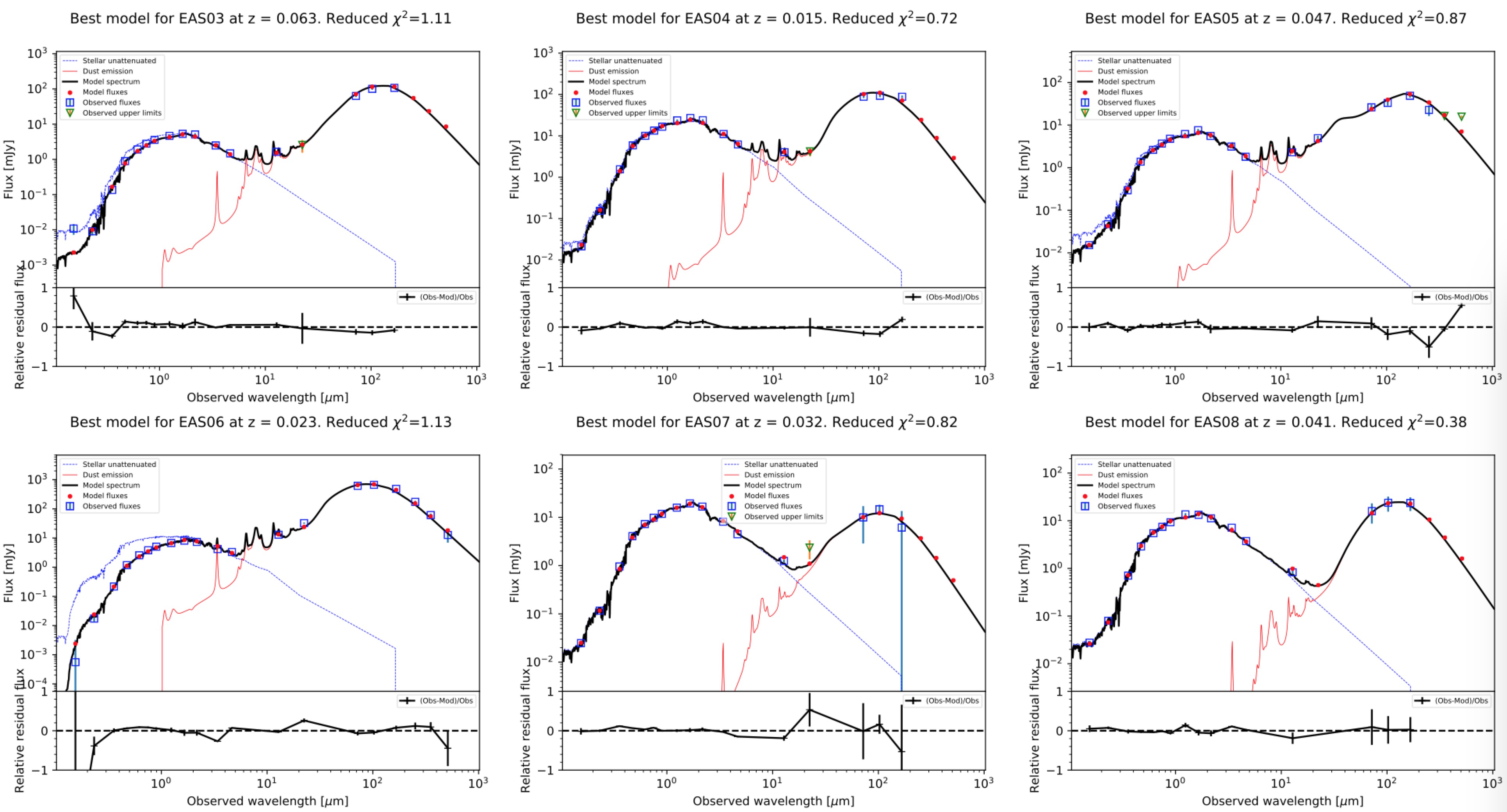}\\
\includegraphics[width=\linewidth, clip]{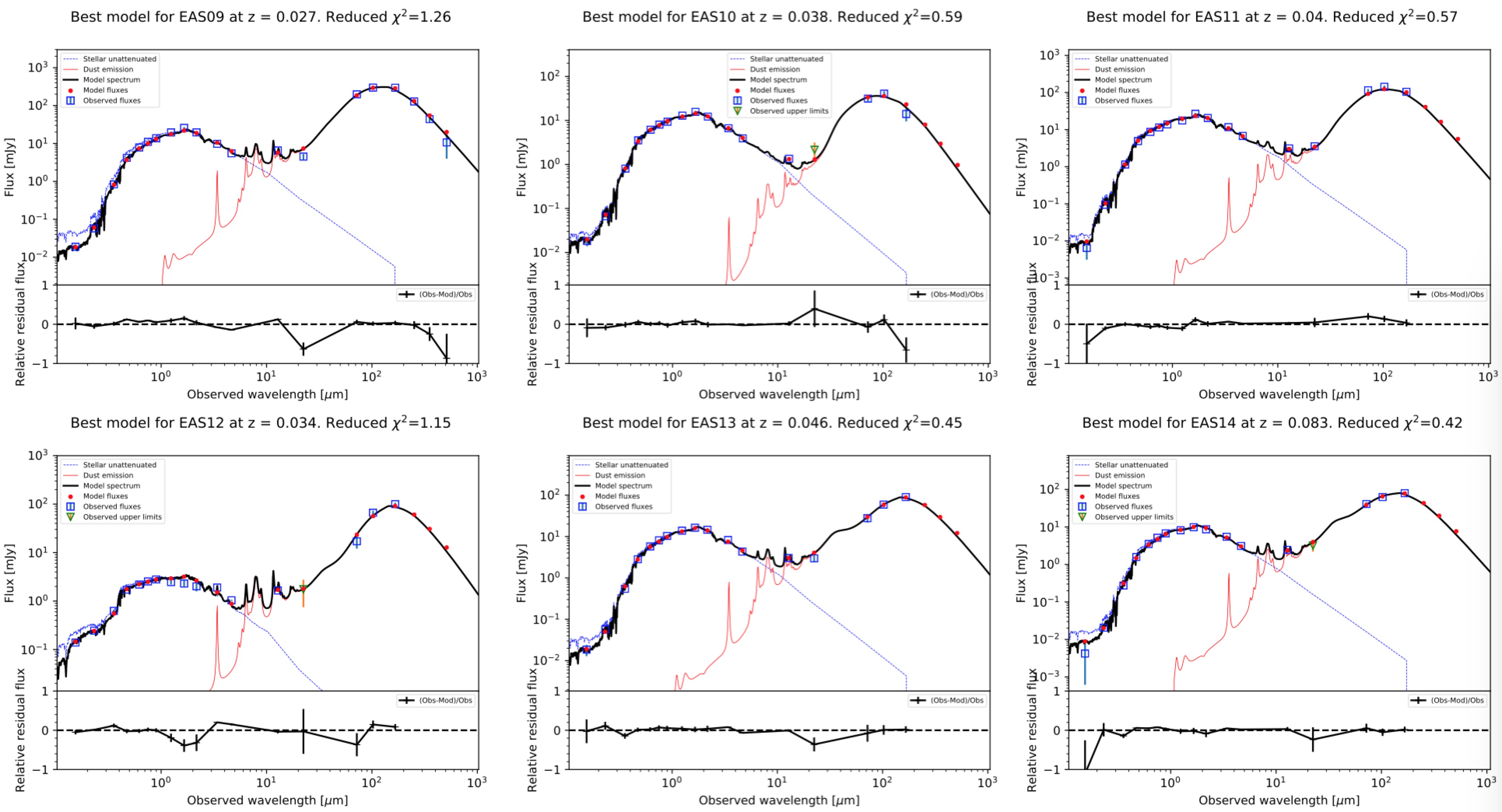}\\
\end{tabular}
\caption{Continued.}
\label{fig:fitsc}
\end{figure}

\begin{figure}
\figurenum{13d}
\centering
\begin{tabular}{c}
\includegraphics[width=\linewidth, clip]{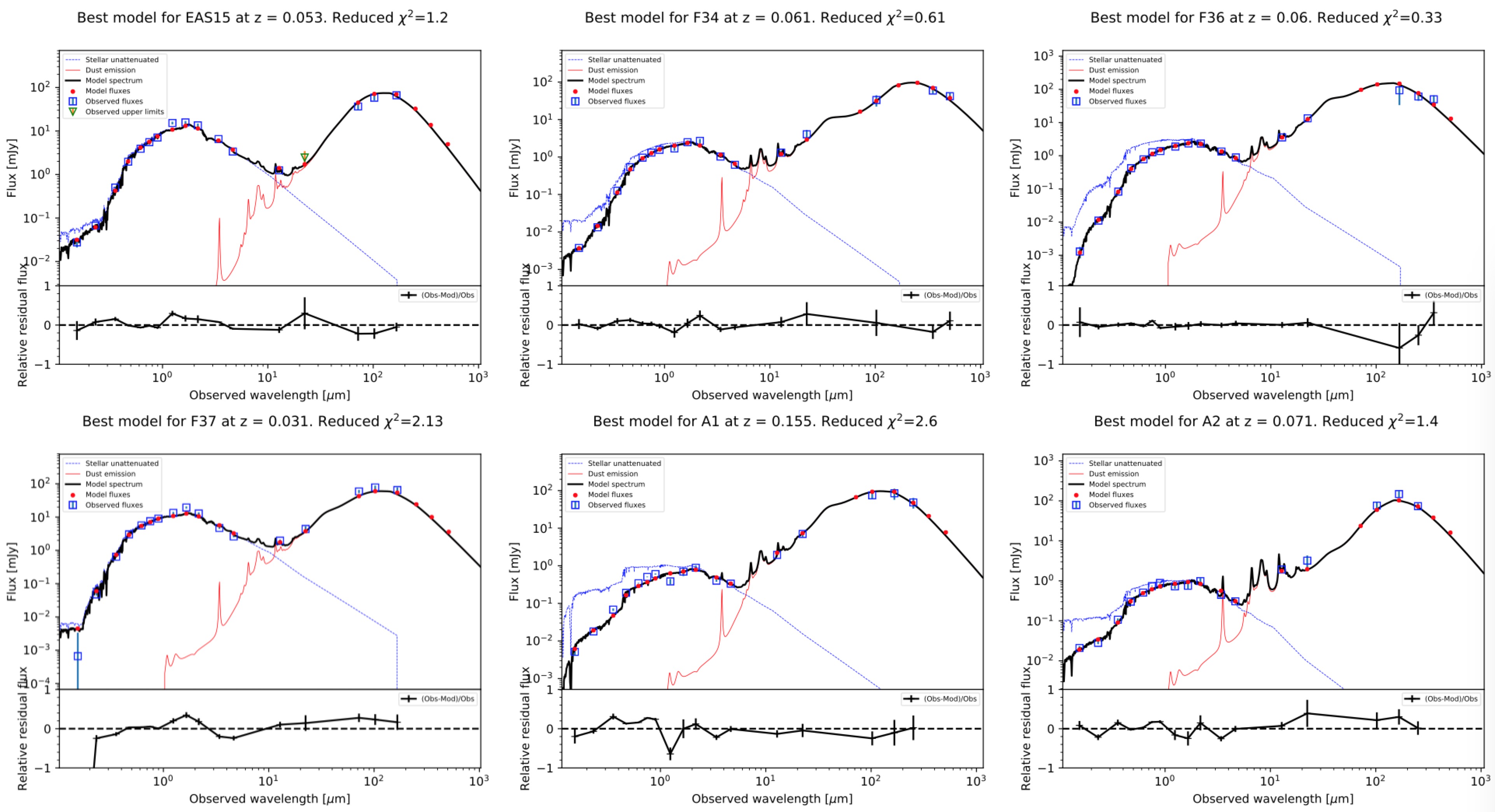}\\
\includegraphics[width=\linewidth, clip]{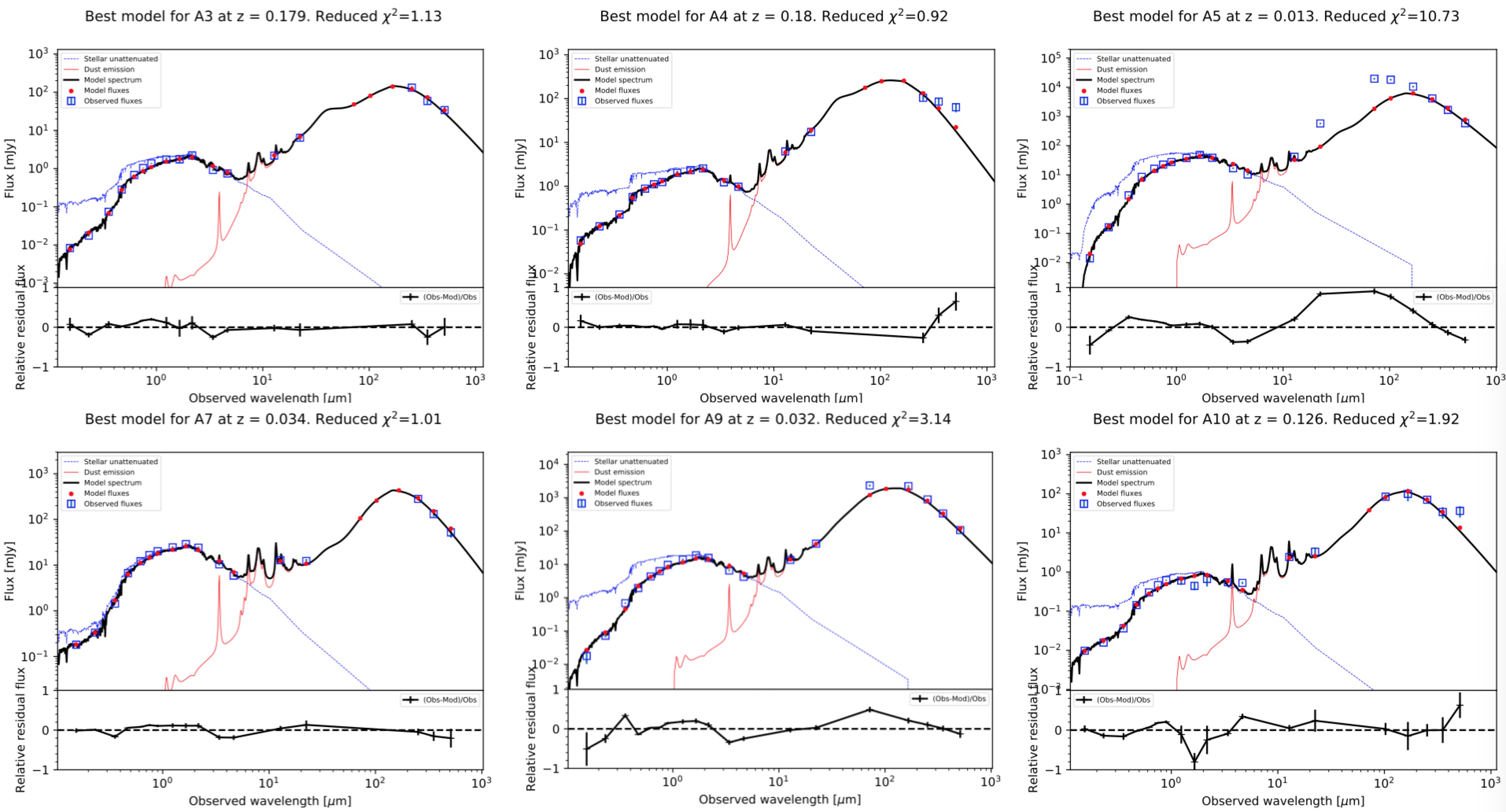}\\
\end{tabular}
\caption{Continued. The worst fit here, A5, is particularly extended ($\emph r_{50}$ = 16 arcsec) and dusty.}
\label{fig:fitsd}
\end{figure}

\begin{figure}
\figurenum{13e}
\includegraphics[width=0.68\linewidth, clip]{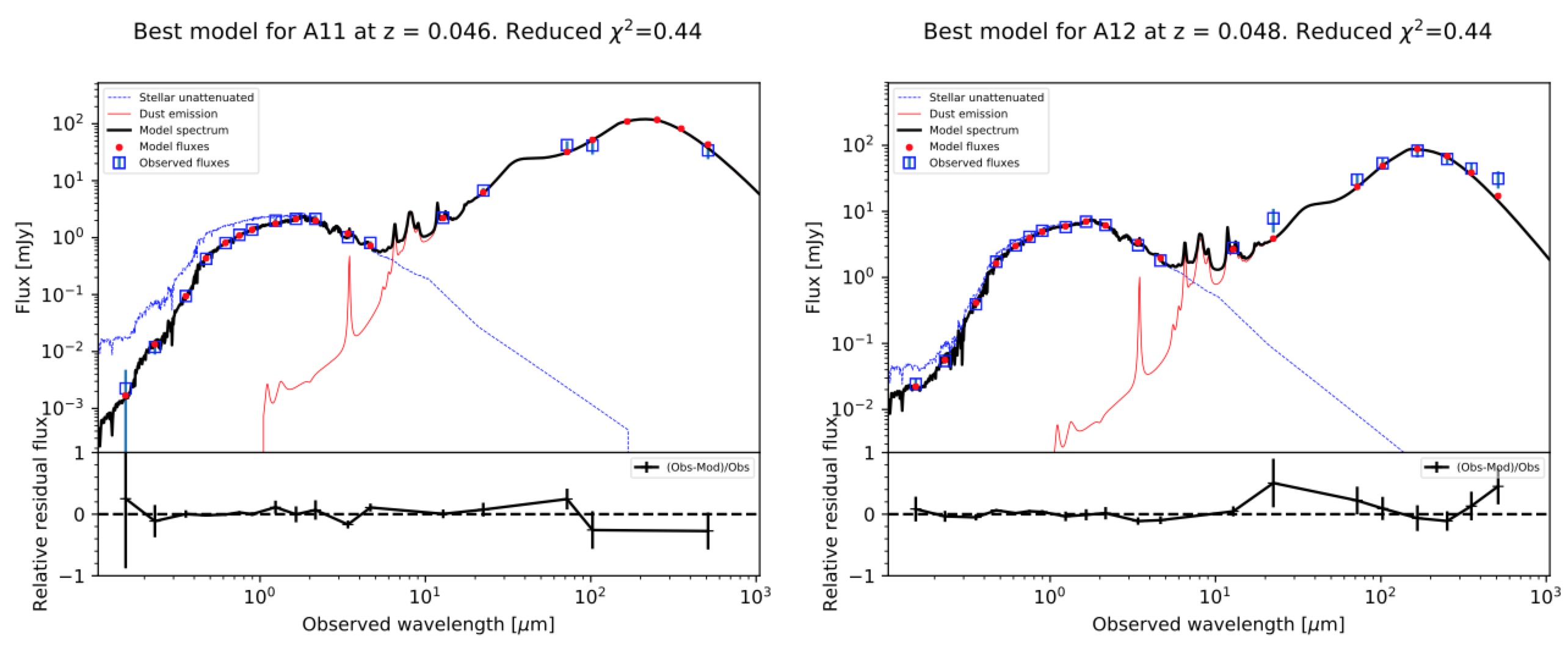}
\caption{Continued.}
\label{fig:fitse}
\end{figure}

\acknowledgments
\end{document}